%% file: mzm_bosonized.tex
\documentclass[aps,prb,reprint,groupedaddress,twocolumn,eqsecnum,footinbib]{revtex4-1}

\usepackage{color}
\usepackage[dvipsnames]{xcolor}

\input{preamble}

\pacs{}

\begin{document}

\title{Majorana zero modes and their bosonization}
\date{\today}
\author{Victor Chua}
\author{Katharina Laubscher}
\email[Corresponding author: ]{katharina.laubscher@unibas.ch} 
\author{Jelena Klinovaja}
\author{Daniel Loss}
\affiliation{Department of Physics, University of Basel, Klingelbergstrasse 82, CH-4056 Basel, Switzerland}

\begin{abstract}

The simplest continuum model of a one-dimensional non-interacting superconducting fermionic symmetry-protected topological (SPT) phase is studied in great detail using analytical methods. In a first step, we present a full exact diagonalization of the fermionic Bogoliubov-de Gennes Hamiltonian for a system of finite length and with open boundaries. In particular, we  derive exact analytical expressions for the Majorana zero modes emerging in the topologically non-trivial phase, revealing their spatial localization, their transformation properties under symmetry operations, and the exact finite-size energy splitting of the associated quasi-degenerate ground states. We then proceed to analyze the model via exact operator bosonization in both open and closed geometries. In the closed wire geometry, we demonstrate fermion parity switching from twisting boundary conditions in the topologically non-trivial phase. For the open wire, on the other hand, we first take a semiclassical approach employing the Mathieu equation to study the two quasi-degenerate ground states as well as their energy splitting at finite system sizes. We then finally  derive the exact forms of the Majorana zero modes in the bosonic language using vertex-algebra techniques. These modes are verified to be in exact agreement with the results obtained from the fermionic description. The complementary viewpoints provided by the fermionic and bosonic formulations of the superconducting SPT phase are reconciled, allowing us to provide a complete and exact account of how Majorana zero modes manifest in a bosonized description of an SPT phase.


\end{abstract}

\maketitle 

\tableofcontents


\section{Introduction}


Majorana zero modes (MZMs) may be conceptualized as ``half electrons'' with the distinguishing feature that they cost practically no additional energy to excite from the ground state. Hence their presence leads to effectively degenerate ground states which must differ by the parity (even vs.~odd) of their fermion number.\cite{kitaev2001unpaired} MZMs naturally emerge as a symptom of non-trivial topology in fermionic symmetry-protected topological (SPT) phases\cite{senthil2015symmetry} of the right symmetries. For some time now, academic interest and excitement in MZMs has been driven by the fact that they represent the simplest possible setup for the implementation of topological quantum computation, a powerful but technologically challenging paradigm for quantum computation exploiting the non-Abelian braiding properties of anyonic quasiparticles.\cite{nayak2008non} 

In this work, we revisit the simplest possible continuum model of a fermionic SPT phase in one dimension, namely that of a time-reversal symmetric topological superconductor without Kramers degeneracy. This continuum model is the low-energy linearized limit of the celebrated Kitaev Majorana chain.\cite{kitaev2001unpaired} The main results of this work are: (1) the exact analytic forms of the MZM operators, (2) their energy splitting in a system of finite length, (3) the correct bosonized Hamiltonian incorporating open boundary conditions, and (4) the exact MZM operators from constructive bosonization and vertex algebra methods. We find that with \emph{open boundaries} and \emph{finite system lengths}, the exact bosonized Hamiltonian will acquire an inhomogeneous topological superconducting mass term that vanishes at the boundaries. This insight elegantly illustrates the reason behind the MZMs behaving as approximate zero modes in a finite system. More importantly, our novel derivation of exact analytical MZM operators in both the fermionic and bosonic language not only confirms their quasi-degeneracy, but also their spatial localization and their transformation properties under time-reversal symmetry. These last two points are absolutely crucial for the demonstration of symmetry fractionalization,\cite{turner2011topological,gu2014symmetry,fidkowski2011topological} which is a defining property of fermionic SPT phases.

Thus far, there have been several actively pursued experimental proposals\cite{beenakker2013majorana} to synthesize and manipulate MZMs, the most popular of which is based on nanowires\cite{lutchyn2010majorana,oreg2010helical} and a combination of strong Rashba spin-orbit coupling, proximity-induced superconductivity as well as weak magnetic fields.~\cite{Mourik1003,Das2012,Rokhinson2012,Churchill2013,Deng2016,Prada2019} An alternative setup is based on chains of magnetic adatoms on superconducting substrates.~\cite{Nadj-Perge2013,Klinovaja2013,Pientka2013,Vazifeh2013,Braunecker2013,Poyhonen2013,Nadj-Perge2014,Ruby2014,Pawlak2016,Pawlak2019}
 It was predicted theoretically that under the right conditions, these setups could result in a topologically non-trivial superconducting wire with the tell-tale signs of localized MZMs at its ends. At the level of a `gedanken experiment',\cite{alicea2011non} topological braiding was then proposed between networks of topological nanowires that are created and manipulated by electrostatic gating. 

Extensions involving more exotic quasiparticles known as \emph{parafermions} have been recognized for some time to possess even greater potential as building blocks for topological quantum computation.\cite{nayak2008non} However,  such claims are also accompanied by more challenging experimental configurations.\cite{clarke2013exotic,lindner2012fractionalizing,vaezi2013fractional,oreg2013fractional,klinovaja2014time,klinovaja2014kramers,alicea2016topological,santos2017parafermionic,cheng2012superconducting,Mong2014,Klinovaja2014,Vaezi2014,Laubscher2019,Laubscher2020} Although these appear to be more exciting and cutting edge, they are nevertheless extremely daunting. The motivation for this work is partly driven by basic questions regarding the bosonization of this and related models such as the symmetry-enriched topological parafermionic models.\cite{bondesan2013topological,klinovaja2014kramers,zhang2014time,kane2015time,meidan2017classification,xu2018enriched} 

The paper is organized as follows. In Sec.~\ref{sec:main_results} the model is briefly introduced and our main results are summarized. This is followed by a detailed discussion of the continuum fermionic model in Sec.~\ref{sec:fermionic_model}. Topics that are touched upon are the model's symmetries, its formulation in terms of Majorana fields, the topological and trivial phases, and the treatment of open boundary conditions through unfolding. Then, in Sec.~\ref{sec:exact_fermion}, an exact diagonalization of the topological Bogoliubov-de Gennes Hamiltonian with open boundaries and finite lengths is presented. In Sec.~\ref{sec:closed_wire_bosonization} bosonization is first applied to the model with closed but twisted boundary conditions. At the semi-classical level, we demonstrate how the topological and trivial phases are distinguished by the response of the ground state(s) to twisted boundary conditions.  Thereafter, in Sec.~\ref{sec:open_wire_bosonization}, bosonization is finally applied to the model with open boundaries and finite lengths through unfolding. Here the topological and trivial phases can be distinguished at the semi-classical level by the quasi-degeneracy of the ground states. The Majorana zero modes are derived within bosonization and the resulting expressions are shown to agree with the ones obtained in Sec.~\ref{sec:exact_fermion}. Finally, in Sec.~\ref{sec:summary_discussion}, we discuss and summarize the overall results of this work. For the sake of completeness, additional material is included in the Appendices \ref{app:corr}, \ref{app:bosonic_action}, \ref{app:direct_check}, and \ref{app:Maj_chains}. 



\section{Model and summary of main results}
\label{sec:main_results}

In the interest of readability, we present a short summary of our main results without going into the minutiae of our methods.  \\


\paragraph*{\underline{Fermionic formulation:}} We specialize to the model Hamiltonian 
\begin{subequations}
\begin{align}
H  = &\int_0^\LL  \mathcal{H} \; \mathrm{d}x, \\ 
\mathcal{H} =& 
- i v_F \left( R^\dagger   \partial_x  R - L^\dagger   \partial_x  L \right)
+ i \Delta (R^\dagger L^\dagger - L R),
\end{align}\label{eq:H_intro}%
\end{subequations}
where $R(x)$ and $L(x)$ are relativistic right- and left-moving chiral fermionic fields, respectively. Here, $\Delta > 0 $ is a constant mean-field superconducting (SC) pairing potential, $v_F$ is the Fermi velocity and we work in units of $\hbar=1$. In a system of size $\LL$ with open boundary conditions, there arise two MZM operators. The enforcement of open boundary conditions is executed by the unfolding method\cite{eggert1992magnetic,fabrizio1995interacting,giamarchi2004quantum} which relates $R(x)$ and $L(x)$ according to 
\begin{align}
L(x) \equiv -R(-x)
\end{align}
in the extended domain $-\LL \leqslant x \leqslant \LL$ and with \emph{antiperiodic} boundary conditions $R(x+2\LL) = -R(x)$. The MZM operators are explicitly given as
\begin{subequations}
\begin{align}
\gamma_0 &:= \int_{-\mathrm{L}}^\mathrm{L} \mathrm{d}x \; a(x) [R(x)+ R^\dagger(x)] \equiv \gamma_0^\dagger,\\
\gamma_\mathrm{L} &:= \int_{-\mathrm{L}}^\mathrm{L} \mathrm{d}x \; b(x) [R(x)+ R^\dagger(x)] \equiv \gamma_\LL^\dagger,
\end{align}
\end{subequations}
where the real-valued wavefunctions $a(x),b(x)$ are 
\begin{subequations}
\begin{align}
a(x) &:= \mathcal{N_\kappa} \sinh(\kappa(\mathrm{L}-|x|)),   \\
b(x) & := \mathcal{N_\kappa} \sinh(\kappa x),
\end{align}
\end{subequations}
with constants $\kappa$ and $\mathcal{N}_\kappa$ defined implicitly and explicitly by 
\begin{align}
\kappa = \frac{\Delta}{v_F} \tanh (\kappa \mathrm{L}),
\quad \mathcal{N_\kappa}:={\LL}^{-\frac{1}{2}}\sqrt{\frac{\sinh(2\kappa \mathrm{L})}{2\kappa \mathrm{L}}-1}.
\end{align}
Here, $\mathcal{N_\kappa}$ is the normalization prefactor and the parameter $\kappa$ is determined by imposing boundary conditions on the decaying eigenmodes. We note that
these expressions remain valid as long as the localization length ${v_F}/{\Delta}$ is smaller than the system length $\LL$. At exactly $\LL=v_F/\Delta $ a finite-length topological transition occurs where these modes become extended,~\cite{Rainis2013} as is detailed in Appendix~\ref{app:Maj_chains}. The asymptotic behavior of 
\begin{align}
a(x) \sim \LL^{-\frac{1}{2}}\mathrm{e}^{-\kappa |x|}, \quad b(x) \sim \LL^{-\frac{1}{2}}\text{sgn}(x)\mathrm{e}^{\kappa (\mathrm{L}-|x|)} 
\end{align}
in the limit of large $\LL \gg v_F/\Delta$ exemplifies the spatial locality of these operators. Moreover, the exact fermionic quasiparticle excitation operator is given by 
\begin{align}
\psi_{E_{i\kappa}} :=  \frac{\gamma_0 - i \gamma_\LL}{2}
\end{align}
and satisfies $[H,\psi^\dagger_{E_{i\kappa}}]=E_{i\kappa}\psi_{E_{i\kappa}}^\dagger$ with the energy splitting
\begin{align}
E_{i\kappa} = \sqrt{\Delta^2 - v_F^2 \kappa^2} \sim 2\Delta \mathrm{e}^{-\frac{\Delta \mathrm{L}}{v_F}}.
\end{align}
Equivalently, the MZMs obey the commutation relations
\begin{align}
[H,\gamma_0] = i E_{i\kappa}\gamma_\mathrm{L}, \qquad [H,\gamma_\mathrm{L}] = -i E_{i\kappa}\gamma_0.
\label{eqn:check_MZMs}
\end{align}

One of the most crucial properties of $\gamma_0$ and $\gamma_\LL$ are their transformations under spinless time-reversal $\mathcal{T}$,
\begin{align}
\mathcal{T} \gamma_0 \mathcal{T}^{-1} = +\gamma_0, \quad
\mathcal{T} \gamma_\mathrm{L} \mathcal{T}^{-1} = -\gamma_\mathrm{L},
\end{align}
since $\mathcal{T} R(x) \mathcal{T}^{-1} = R(-x)$.
Here, $\mathcal{T}$ is anti-linear and satisfies $\mathcal{T}^2 = +1$. The fact that $\gamma_0$ and $\gamma_\LL$ transform oppositely in sign under $\mathcal{T}$ is a clear demonstration of symmetry fractionalization,~\cite{fidkowski2010effects,turner2011topological} since the two MZMs are spatially localized on different boundaries. 

The exact MZM solutions given above can be verified by direct substitution as is done in Appendix \ref{app:direct_check}. The main bulk of Sec.~\ref{sec:exact_fermion} is dedicated to the intuition and derivation of these solutions, as well as a full exact diagonalization of $H$. Using these eigenmode solutions, the Bardeen-Cooper-Schrieffer (BCS) quasi-degenerate ground states are constructed. \\


\paragraph*{\underline{Bosonization:}} We proceed to carry out finite-size bosonization on $H$ with closed and open boundary conditions. The case of open boundaries is analyzed using the unfolding procedure. Since unfolding reduces the number of chiral fields to just $R(x)$ in a domain of size $2\LL$, we need to bosonize only a \emph{single} chiral field with $2\LL$-antiperiodic boundary conditions
\begin{align}
R(x) &= \frac{1}{\sqrt{2\LL}}:\e^{i\phi^R(x)}:\e^{-i \frac{\pi x }{2\LL}},
\end{align}
where $\phi^R(x) \sim \phi^R(x) + 2\pi $ is a non-local compact bosonic chiral field. Here $:\;:$ denotes bosonic normal ordering. 
The bosonized Hamiltonian is then given by
\begin{align}
H =& \int_0^\LL \mathrm{d}x\; \left(\frac{v_F}{2\pi} \left( [\partial_x \varphi(x)]^2 + [\partial_x \vartheta(x)]^2 \right) \right.\nonumber \\
&\left. \hspace{1cm} -\frac{2\Delta}{\LL}\sin\left[\frac{\pi x}{\LL}\right] :\cos[2\vartheta(x)]: \right),
\end{align} 
where the local conjugate fields are
\begin{subequations}
\begin{align}
&\vartheta(x) = +\frac{\phi^R(x)+ \phi^R(-x)}{2}, \\
&\varphi(x) = -\frac{\phi^R(x)-\phi^R(-x)}{2}.
\end{align}
\end{subequations}
Importantly, we note that the SC pairing potential now has a $\sin(\pi x/\LL)$ modulation -- vanishing at $x=0,\pm\LL$ -- stemming from the fact that $R(x)$ and $L(x)$ are no longer independent due to unfolding. This is further reflected in the expressions of $\varphi(x)$ and $\vartheta(x)$ above, which show that they are \emph{non-local} superpositions of $\phi^R(x)$ in the extended domain $[-\LL,\LL]$.
At a more technical level, the compactification radii of the bosonic fields $\vartheta$ and $\varphi$ are doubled from $\pi$ to $2\pi$ as when compared to the closed wire geometry. This is absolutely crucial for the semi-classical understanding of the topological degeneracy in terms of degenerate minima.

Next, we exactly bosonize the explicit MZM operators $\gamma_0$ and $\gamma_\LL$, yielding
\begin{subequations}
\begin{align}
\gamma_0 &= \frac{1}{\sqrt{2\LL}}\int_{-\mathrm{L}}^\mathrm{L} \mathrm{d}x \; a(x) [:\mathrm{e}^{i \phi^R(x)}: + :\mathrm{e}^{-i\phi^R(x)}:]\mathrm{e}^{-i\frac{\pi x}{2\mathrm{L}}}, \\
  \gamma_\LL &= \frac{1}{\sqrt{2\LL}}\int_{-\mathrm{L}}^\mathrm{L} \mathrm{d}x \; b(x)[:\mathrm{e}^{i \phi^R(x)}: + :\mathrm{e}^{-i\phi^R(x)}:]\mathrm{e}^{-i\frac{\pi x}{2\mathrm{L}}}.
\end{align}
\label{eqn:gammas_bosonized}%
\end{subequations}
Using these expressions one can again verify the validity of Eqn.~(\ref{eqn:check_MZMs}) within the bosonized formulation. This entails using vertex algebra methods based around the exact operator identity between normal-ordered vertex operators\cite{stone1994bosonization}
\begin{align}
&:\e^{i \alpha \phi^R(x)}:\;:\e^{ i\beta \phi^R(y)}:  \nonumber \\
&= 
\left[ \e^{-i\frac{\pi x}{\LL}} - \e^{-i\frac{\pi y}{\LL}}
\right]^{\alpha \beta} :\e^{i [\alpha \phi^R(x) + \beta \phi^R(y)]}:
\end{align}
for $\alpha,\beta \in \mathbb{Z}$. 
In the extremely degenerate (flat band) limit when $v_F=0$ these bosonized MZM operators take the exactly localized forms 
\begin{subequations}
\begin{align}
\gamma_0& = \mathcal{N} \left( :\e^{i\phi^R(0)}: + :\e^{-i\phi^R(0)}:\right)
\nonumber\\&\propto R(0)+ R^\dagger(0),\\
\gamma_\LL &= -i\mathcal{N} \left( :\e^{i\phi^R(\LL)}: + :\e^{-i\phi^R(\LL)}:\right)
\nonumber\\&\propto i[R(\LL)- R^\dagger(\LL)],
\end{align}
\end{subequations}
where $\mathcal{N}$ is an infinite normalization constant. In this special limit, $H$ reduces to just the modulated cosine potential that goes as $\sin\left[\frac{\pi x}{\LL}\right]\cos[ 2\vartheta(x)]$. The fact that the SC pairing potential explicitly vanishes at the boundaries makes it abundantly clear that the above expressions for $\gamma_0$ and $\gamma_\LL$ are indeed zero-energy modes. Their particular Majorana form has to do with an inversion symmetry (present even when $v_F>0$) that we shall discuss later. Let us just mention that the corresponding superpositions are sensitive to the sign of $\Delta$, which we have so far taken to be positive. Lastly, in terms of the mode expansion of $\phi^R(x)$,
\begin{align}
\phi^R(x) = \vartheta_0 + \frac{\pi Q x}{\LL} + \sum_{n>0} 
\frac{1}{\sqrt{n}}\left(a_n \e^{i\frac{n\pi x}{\LL}} + a_n^\dagger \e^{-i\frac{n\pi x}{\LL}} \right),
\end{align}
the MZMs in the extreme limit $v_F=0$ have the simplified forms
\begin{align}
\gamma_0 \propto \e^{i\vartheta_0}, \qquad \gamma_\LL \propto -i\e^{i\vartheta_0}(-1)^Q,
\end{align}
where $\vartheta_0$ is the zero mode for the current density phase operator and $Q$ is the total charge operator conjugate to it. Even though these appear to be non-local expressions,\cite{cheng2012superconducting} they are in fact limits of the local expansions given in Eqn.~(\ref{eqn:gammas_bosonized}). This just highlights the point that non-local bosonic expressions for topological quasiparticle operators --  Majoranas in this case -- are often entirely local in terms of fermionic fields. 

The derivation of the above results and claims are the contents of Secs.~\ref{sec:closed_wire_bosonization} and \ref{sec:open_wire_bosonization}. Along the way, we shall also cover and contrast the case of closed boundary conditions while demonstrating the fermion parity switching effect\cite{fidkowski2011majorana,keselman2013inducing} under magnetic flux threading. Also, in Sec.~\ref{sec:exact_MZM_vertex}, we introduce and employ vertex algebra methods from the conformal field theory of current algebras. Finally, we should mention that although a large body of our bosonization results are exact -- and in complete agreement with the fermionic approach -- we have also utilized semi-classical methods, a point of novelty being that we use the phase-space formulation\cite{floreanini1987self} of the classical action for the zero modes and the physical observables of charge and current.



\section{Continuum fermionic model and open boundaries}\label{sec:fermionic_model}

The physical setting is that of one-dimensional spinless fermions $\Psi(x)$ with a low-energy linearized expansion
\begin{align}
&\Psi(x) \approx R(x)\e^{i k_F x} + L(x)\e^{-ik_F x},
\end{align}
where $k_F$ is a Fermi momentum and $R(x)$, $L(x)$ are independent slow-moving right and left chiral fields respectively. They satisfy the usual canonical anticommutation relations $\{R(x),R^\dagger(y) \} = \delta(x-y)$ and similarly for $L(x)$.

For additional insight into the effects of competing gapping terms, we initially consider a Hamiltonian that is slightly more general than the one given in Eqn.~(\ref{eq:H_intro}). Here, an additional $k_F$-dependent backscattering term is present. Such a term naturally competes with the SC pairing that would otherwise favor the topological MZM phase.\cite{Klinovaja2012,Klinovaja2015} Explicitly, we consider a (1+1)-dimensional continuum model with Lagrangian and Hamiltonian 
\begin{subequations}
\begin{align}
\mathcal{L} :=& 
 i R^\dagger \partial_t  R  + i L^\dagger \partial_t L - \mathcal{H}, \\ 
\nonumber \\
\mathcal{H} :=& 
- i v_F \left( R^\dagger   \partial_x  R - L^\dagger   \partial_x  L \right)\nonumber \\ 
&+ \Delta\e^{i\chi} R^\dagger L^\dagger - \mathrm{M}  \e^{-i2 k_F x} R^\dagger L + \text{h.c.},
\end{align}
\label{eqn:L_setup_kF}%
\end{subequations}
where $\mathrm{M} \geq0$ is a single-particle backscattering term and $\chi \in \mathbb{R}$ is a constant SC phase parameter. The $\e^{\pm i 2 k_F x}$ phases in the backscattering term are necessary to ensure momentum conservation. 
Such a model arises effectively in nanowires with strong spin-orbit coupling (usually Rashba-type), Zeeman coupling to a magnetic field and proximity-induced SC pairing of the $s$-wave form.\cite{lutchyn2010majorana,oreg2010helical,beenakker2013majorana} This model has been studied many times in the literature from the point of view of bosonization. \cite{fidkowski2011majorana,gangadharaiah2011majorana,sau2011number,keselman2015gapless} More directly, it may also appear in the low-energy limit of Kitaev's Majorana lattice model\cite{kitaev2001unpaired}
\begin{align}
H = &\sum_{j \in \mathbb{Z}}\left[-t c^\dagger_{j+1}c_{j} - \tfrac{\Delta}{2}\mathrm{e}^{i\chi}c^\dagger_{j+1}c^\dagger_{j}+\mathrm{M} c^\dagger_{2j+1}c_{2j} + \text{h.c.} \right] \nonumber \\
&- \mu \sum_{j\in \mathbb{Z}} c^\dagger_{j} c_j .
\label{eq:Kitaev_chain}
\end{align}
Near half-filling $\mu=0$, this model can be linearized with $k_F = \mu/(2t)$ and $v_F \approx 2t $, where the Fermi momenta are now calculated from their values at half-filling given by $\pm \pi/2$. It should be noted that the $\mathrm{M}$ process represents dimerization on the lattice and at $\Delta= \mu = 0$ we have the famed Su-Schrieffer-Heeger\cite{su1979solitons} (SSH) model. The fact that the SSH and Kitaev models share \emph{almost} the same low-energy effective description has to do with a duality transformation relating the two.\cite{cobanera2015equivalence} However, key differences lie in the implementation of symmetry transformations and open boundary conditions.\cite{unpublished} 

Interest in Eqns.~(\ref{eqn:L_setup_kF}) has to do with the emergence of a non-trivial superconducting topological phase yielding boundary Majorana zero modes whenever $0\leq \mathrm{M} < \Delta$. Intuitively speaking, large $\Delta$ favors a gapped topological phase, while large $\rm M$ favors a topologically trivial charge-density wave phase with period $1/2k_F$.\cite{Harper1955,Aubryand1980,Gangadharaiah2012,Kraus2012,Park2016,Pletyukhov2020}

It should be noted that the Kitaev chain has a predecessor in the exactly solvable Lieb-Schultz-Mattis (LSM) anti-ferromagnetic chain,\cite{lieb1961two} where boundary-localized quasi-degenerate bound states akin to MZMs were first derived exactly. At granularity of the lattice, the fermion-spin duality is supplied by the Jordan-Wigner transform. By contrast, bosonization, which will be discussed at length in Secs.~\ref{sec:closed_wire_bosonization} and \ref{sec:open_wire_bosonization}, is a fermion-boson duality that emerges in the continuum limit.


\subsection{Symmetries}\label{sec:symmetries}

For clarity, we shall assume first that the system is closed and neglect complications arising from boundaries. Without mass terms ($\Delta=\mathrm{M}=0$), there is full $\widehat{U(1)}_L \times \widehat{U(1)}_R$ conformal symmetry with central charges $c=1$ in each chirality. Thus, as Heisenberg fields, $R$ and $L$ propagate freely as
$R(x,t) = R(x-v_F t)$, $L(x,t) = L(x+v_F t)$.
There are two global symmetries, total $U(1)_{R+L}$ symmetry and chiral $U(1)_{R-L}$ symmetry, defined respectively by 
\begin{align}
&(R,L) \mapsto (R\e^{i\theta_1},L\e^{i\theta_1}), \\
&(R,L) \mapsto (R\e^{i\theta_2},L\e^{-i\theta_2})
\end{align}
for $\theta_{1,2} \in \mathbb{R}$. They are infinitesimally generated by the total charge (number) operator
\begin{align}\label{eqn:Q_fermion}
Q := \int \mathrm{d}x \; :[R^\dagger(x)R(x) + L^\dagger(x) L(x)]:
\end{align}
and the total chiral charge (current) operator
\begin{align}\label{eqn:J_fermion}
J := \int \mathrm{d}x \; :[R^\dagger(x)R(x) - L^\dagger(x) L(x)]:,
\end{align}
respectively. Here, $:\;:$ denotes fermion normal ordering with respect to the filled Fermi sea of negative-energy states. 
Non-zero values of $\mathrm{M}$ and $\Delta$ will not only break the conformal symmetries but also the global $U(1)_{R+L}$ and $U(1)_{R-L}$ symmetries. We consider these broken symmetries next on a case by case basis. \\


\noindent\underline{$\mathrm{M} \neq 0$ and $\Delta \neq 0$:} \quad In this case, only \emph{fermion parity} ($F$) and \emph{spinless time-reversal} $(\mathcal{T}$) are symmetries with actions
\begin{align}
&F R F^{-1}= -R, \quad F L F^{-1}= -L, \\
&\mathcal{T} \,R \,\mathcal{T}^{-1} = -i \mathrm{e}^{i\chi} L , \quad
\mathcal{T} \,L \,\mathcal{T}^{-1} = -i \mathrm{e}^{i \chi} R. 
\end{align}
Fermion parity is a \emph{linear} symmetry ($F i = i F$), but time-reversal is \emph{anti-linear} ($\mathcal{T}i = -i \mathcal{T}$) and $\chi$-dependent in order to compensate for the SC phase. Both are representations of $\mathbb{Z}_2$ because $F^2 = \mathcal{T}^2=1$. However, $F$ satisfies the additional identity
\begin{align}
F \equiv (-1)^Q \equiv (-1)^J.
\end{align}
It is convenient to perform a $U(1)$ phase rotation%
	\footnote{This phase rotation is not a $U(1)$ electromagnetic gauge transformation because the superconducting order parameter $\Delta\, \e^{i\chi}$ is unchanged. For a detailed discussion we recommend Ref.~\onlinecite{greiter2005electromagnetic}} 
\begin{align}
R \rightarrow R \,\e^{i\chi/2 + i\pi/4}, \qquad L \rightarrow L \,\e^{i\chi/2 + i\pi/4},
\end{align}
which effectively sets $\chi = \pi/2$. At this special point, the Hamiltonian density takes the simpler form
\begin{align}
\mathcal{H} = &-i v_F 
\left(R^\dagger \partial_x R - L^\dagger \partial_x L \right)  \nonumber \\
&+ i \Delta R^\dagger L^\dagger -\mathrm{M} \e^{-i2k_Fx}R^\dagger L  + \text{h.c.}
\label{eqn:Ham_fermion}
\end{align}
Because $F$ and $\mathcal{T}$ are the only symmetries, the model falls under the BDI-class of free fermion models.\cite{heinzner2005symmetry} This is made more apparent in the Bogoliubov-de Gennes (BdG) form of the Hamiltonian with the Nambu 4-spinor $\varPsi(x) := [R(x),L^\dagger(x),L(x),R^\dagger(x)]^T$,
\begin{subequations}
\begin{align}
H &= \frac{1}{2}\int \mathrm{d}x \; \varPsi^\dagger(x) \, \mathcal{H}_\text{BdG} \, \varPsi(x) + \text{const.}, \\ 
\mathcal{H}_\text{BdG} &= 
\begin{pmatrix}
-i v_F \sigma^z \partial_x - \Delta \sigma^y &  -\mathrm{M}\e^{-i2k_F x} \sigma^z  \\
-\mathrm{M}\e^{i2k_F x} \sigma^z & i v_F \sigma^z \partial_x + \Delta \sigma^y
\end{pmatrix}.
\end{align}
\label{eqn:BdG_Ham}%
\end{subequations}
Here $\mathcal{H}_\text{BdG}$ is traceless (particle-hole symmetric) and anticommutes with $\mathbbm{1} \otimes \sigma^x$. Time-reversal is expressed by $\mathrm{T} = (\sigma^z\otimes \sigma^x) K$ with $K$ denoting complex conjugation, giving $[\mathcal{H}_\text{BdG},\mathrm{T}]=0$. As with all BDI-class BdG Hamiltonians, there is an anti-linear symmetry corresponding to charge conjugation,
$\mathrm{C} := \mathrm{T} (\mathbbm{1}\otimes \sigma^x) \equiv K (\sigma^z\otimes \mathbbm{1})$,
such that $\{\mathrm{C},\mathcal{H}_\text{BdG}\}=0$. \\


\noindent\underline{$\mathrm{M} \neq 0$ but $\Delta =0$:} \quad Now there is $Q$ conservation but no $J$ conservation with $F$ and $\mathcal{T}$ still remaining as symmetries. Essentially, the $U(1)_{R-L}$ group has been broken down to $\mathbb{Z}_2$ while $U(1)_{R+L}$ remains unbroken. Nevertheless, there exists a linear charge-conserving spatial inversion symmetry $\mathcal{I}$ acting as 
\begin{subequations}
\begin{align}
&\mathcal{I} \, R(x)\,  \mathcal{I} =\e^{i\uptheta_x} L(-x), \\ 
&\mathcal{I} \, L(x)\,  \mathcal{I} = \e^{-i\uptheta_x} R(-x).
\end{align}
\end{subequations}
The constant phase $\uptheta_x \in [0,2\pi)$ is not set a priori because the action of $\mathcal{I}$ and chiral rotations generated by $J$ do not commute. Rather, because $J$ is odd under $\mathcal{I}$, conjugation by chiral rotations generated by $J$ changes the value of $\uptheta_x$. One can set $\uptheta_x=0$ as a choice of `chiral gauge'. However, imposing open boundary conditions will result in singling out a specific value of $\uptheta_x$, which effectively breaks the chiral symmetry at the open boundaries. \\


\noindent\underline{$\Delta \neq 0$ but $\mathrm{M} =0$:}\quad Conversely, there is now $J$ conservation at the expense of $Q$ conservation and $\mathcal{I}$ symmetry. More interestingly, there exists an additional linear $\mathbb{Z}_2$ symmetry that may be thought of as a \emph{unitary} charge conjugation transformation because it exchanges particles and holes in the manner of a unitary Bogoliubov transformation. This symmetry, which we denote by $\mathrm{I}_x$  -- to distinguish it from the true spatial inversion $\mathcal{I}$ -- is defined by  
\begin{subequations}
\begin{align}
&\mathrm{I}_x R(x) \mathrm{I}_x = -R^\dagger(x),\quad \mathrm{I}_x L(x) \mathrm{I}_x = -L^\dagger(x), \\
&\mathrm{I}_x R^\dagger(x) \mathrm{I}_x = -R(x),\quad \mathrm{I}_x L^\dagger(x) \mathrm{I}_x = -L(x),
\end{align}
\end{subequations}
where the $-1$ signs are for later convenience. In terms of the Nambu spinor field $\varPsi(x)$ we have
\begin{align}
\mathrm{I}_x \varPsi(x) \mathrm{I}_x = -(\sigma^x \otimes \sigma^x) \varPsi(x),
\end{align}
which gives another symmetry of Eqn.~(\ref{eqn:BdG_Ham}). This symmetry will allow us to attach $\pm 1$ quantum numbers to quasiparticle eigenstates of $H$ and will be exploited to exactly diagonalize the Hamiltonian with open boundaries. 


\subsection{In terms of Majoranas}\label{sec:in_terms_of_majoranas}

The Lagrangian can also be expressed in a relativistic form that will make the relativistic symmetries of the theory more apparent. A Majorana representation is obtained by defining
\begin{align}
\lambda := \frac{1}{2}\begin{pmatrix} R + R^\dagger \\ L+ L^\dagger\end{pmatrix}, \qquad 
\lambda' := \frac{1}{2i}\begin{pmatrix}R-R^\dagger \\ L-L^\dagger\end{pmatrix}.
\label{eqn:majorana_fields}
\end{align}
This implies the reality conditions
$\lambda^\dagger = \lambda^T$, $(\lambda')^\dagger = (\lambda')^T$,
and the canonical anticommutation relations 
$\{\lambda_\alpha(x),\lambda_\beta(y) \}$ = $\{\lambda'_\alpha(x),\lambda'_\beta(y) \} = \tfrac{1}{2} \delta(x-y)\delta_{\alpha \beta}.$
Under time reversal the Majorana fields transform oppositely as
\begin{align}
\mathcal{T} \lambda \mathcal{T}^{-1} = \sigma^x \lambda, \qquad 
\mathcal{T} \lambda' \mathcal{T}^{-1} = - \sigma^x \lambda',
\end{align}
whereas under spatial inversion, they transform identically as
\begin{align}
\mathcal{I}\, \lambda \, \mathcal{I} = \sigma^x \lambda, \qquad
\mathcal{I}\, \lambda' \, \mathcal{I} = \sigma^x \lambda'.  
\end{align}
Direct substitution leads to the Lagrangian 
\begin{align}
\mathcal{L} &= i \lambda^T (\partial_t + \sigma^z v_F \partial_x ) \lambda 
+ i (\lambda')^T(\partial_t + \sigma^z v_F \partial_x ) \lambda' \nonumber \\
&+ \mathrm{M} \sin(2k_F x)[\lambda^T \sigma^y \lambda + (\lambda')^T \sigma^y \lambda']
 \nonumber \\ 
&+i\mathrm{M}  \cos(2k_F x)\lambda^T \sigma^x \lambda' \nonumber \\ 
&+  \Delta [\lambda^T \sigma^y \lambda - (\lambda')^T \sigma^y \lambda'].
\end{align}
Next, we define the Dirac matrices 
\begin{align}
\gamma^0 = \sigma^x, \qquad \gamma^1 =-i \sigma^y, \qquad \gamma_5 = \gamma^0 \gamma^1 = \sigma^z,
\end{align}
such that $\{\gamma^\mu, \gamma^\nu\} = 2 g^{\mu\nu}$ in the ($+1$,$-1$) metric signature. Then, defining $\overline{\lambda} := \lambda^\dagger \gamma^ 0$, $\overline{\lambda}' := (\lambda')^\dagger \gamma^ 0$ gives
\begin{align}
\mathcal{L} &= 
i \overline{\lambda}(\gamma^0 \partial_t + v_F \gamma^1  \partial_x )\lambda 
+i \overline{\lambda}'(\gamma^0 \partial_t + v_F \gamma^1  \partial_x )\lambda'\nonumber \\
&+ i \mathrm{M} \cos(2k_F x)  \overline{\lambda}\lambda'
+  i \mathrm{M} \sin(2k_F x) \left[  \overline{\lambda}\gamma_5 \lambda - \overline{\lambda'}\gamma_5 \lambda' \right]  
\nonumber \\ 
&+ i\, \Delta [\overline{\lambda}\gamma_5 \lambda - \overline{\lambda'}\gamma_5 \lambda' ]\label{eqn:L_Majorana}.
\end{align}
Thus, only when $k_F=0$ the Lagrangian is actually relativistically invariant. We remind the reader that in the case of Kitaev's Majorana chain, we calculate the Fermi momenta from their half-filling value, see Eqn.~(\ref{eq:Kitaev_chain}). Thus, $k_F=0$ corresponds to half-filling. At that point we have a theory of \emph{two} continuum relativistic Majorana chains\cite{fidkowski2010effects} $\lambda$ and $\lambda'$ with competing masses: one \emph{scalar} with strength $\mathrm{M}$, and another \emph{pseudo-scalar} with strength $\Delta$. 

The two types of massive phases that are possible with the Lagrangian in Eqn.~(\ref{eqn:L_Majorana}) are separated by a critical point at $\Delta= \mathrm{M}$. The topologically trivial phase is the one with $\mathrm{M} > \Delta\geq 0 $ and vice versa. However confirming either type of phase is not so straightforward. The simplest method involves introducing open boundaries as will be discussed in Sec.~\ref{sec:exact_MZM}. Alternatively, one could introduce a lattice model whose low-energy continuum limit is described by either the Hamiltonian of Eqn.~(\ref{eqn:BdG_Ham}) or the Lagrangian of Eqn.~(\ref{eqn:L_Majorana}). This is carried out explicitly in Appendix \ref{app:Maj_chains}. With a lattice model, one then computes a chiral winding number defined over the first Brillouin zone.

Another more intrinsic approach that avoids open boundaries and lattice models proceeds by computing the Pauli-Villars regularized partition function in the limit $\mathrm{M}=0$ when the Majorana Lagrangians have decoupled. Then, as was argued in Ref.~\onlinecite{witten2016fermion}, the topological phase exists whenever the pseudo-scalar mass ($\Delta$) has a sign opposite to that of the regulator mass. However, since pseudo-scalar mass terms appear with both signs in the Lagrangian  Eqn.~($\ref{eqn:L_Majorana}$), either $\lambda$ or $\lambda'$ will always be topologically non-trivial while the other is trivial. Thus, the system as a whole will always remain topologically non-trivial irrespective of the sign of $\Delta$ in the decoupled chain limit where $\mathrm{M}=0$.


\subsection{Open boundaries and unfolding}\label{sec:open_bc}

We start by revisiting the analysis of 
Refs.~\onlinecite{fabrizio1995interacting,loss1992parity} and note that similar concepts appeared 
earlier in the bosonization of open spin-1/2 chains.\cite{eggert1992magnetic,lecheminant2002magnetization}
The basic idea is to impose Dirichlet boundary conditions on the fast moving fermionic field 
\begin{align}
\Psi(x) = R(x)\e^{i k_F x} + L(x)\e^{-ik_F x}.
\end{align}
The boundary conditions that can be imposed take the general form
\begin{subequations}
\begin{align}
R(0) + \e^{i \delta_0 }L(0)  &= 0, \\
R(\LL) + \e^{i(\delta_L-2k_F \LL)} L(\LL) &=0,
\end{align}
\label{eqn:Dirichlet}%
\end{subequations}
where $\LL$ is the system length, $\delta_0$ and $\delta_\LL$ are reflection phase shifts, and the additional $2k_F\LL$ phase is for later convenience. More generally, we can also impose boundary conditions that mix $R$ and $L^\dagger$ and vice-versa. These are Andreev-type boundary conditions,\cite{maslov1996josephson} which do not conserve electric charge and require superconducting reservoirs at the boundaries. 

We then proceed by an operation known as \emph{unfolding}, which entails relabeling the $L(x)$ field by $R(-x)$ defined on an extended space $[-\LL,\LL]$ that is topologically equivalent to $S^1$. Specifically, the relabeling is 
\begin{align}
L(x) \equiv - \e^{-i\delta_0} R(-x), \qquad 0 \leqslant x \leqslant \LL \, ,
\end{align}
in accordance to the boundary condition at $x=0$. Identification of the points at $x=\LL$ and $x=-\LL$ requires that the unfolded $R(x)$ field satisfies quasiperiodic boundary conditions 
\begin{align}
R(\LL) = R(-\LL)\,\e^{i(\delta_\LL -\delta_0 -2k_F \LL)},
\end{align}
where the quasiperiodic phase depends on $k_F \LL$ and the phases $\delta_0,\delta_\LL$.

It is illuminating to consider this unfolding operation from the point of view of inversion symmetry. For open boundaries it is more natural to take the center of inversion to lie at $x=\LL/2$ as opposed to $x=0$. Then, the inversion transformation is redefined by 
\begin{align}
&\mathcal{I} \, R(x)\,  \mathcal{I} =\e^{i\uptheta_x} L(\LL-x),\\ 
&\mathcal{I} \, L(x)\,  \mathcal{I} = \e^{-i\uptheta_x} R(\LL-x).
\end{align}
Applying this inversion operation to the boundary conditions (\ref{eqn:Dirichlet}) yields the condition 
\begin{align}
2\uptheta_x = \delta_\LL + \delta_0 -2k_F \LL.
\end{align}
Thus, the inversion phase $\uptheta_x$, which was formerly arbitrary because of $J$ conservation, has now been fixed (symmetry-broken) by the phases set by the boundary conditions. Incidentally, this also implies that the imposition of open boundary conditions necessarily breaks $J$ conservation, i.e., $U(1)_{R-L}$ symmetry. 

Next, the 2-spinor field $[R(x),L(x)]^T$ can be decomposed into inversion eigenstates with parities $\sigma= \pm 1$ such that
\begin{align}
L(\LL-x) = \sigma \e^{-i\uptheta_x}R(x), \quad R(\LL-x) = \sigma \e^{i\uptheta_x}L(x).
\end{align}
Applying these to the boundary conditions at $x=0,\LL$ gives
\begin{align}
&R(\LL) = -\sigma \e^{+i[\frac{\delta_\LL-\delta_0}{2}- k_F \LL]}\, R(0), \\
&L(\LL) = -\sigma \e^{-i[\frac{\delta_\LL-\delta_0}{2}- k_F \LL]}\, L(0).
\end{align}
This shows that for even parity states with $\sigma = +1$, the $R,L$ fields obey antiperiodic boundary conditions between $x=0$ and $x=\LL$ up to constant phases $\e^{\pm i [\frac{\delta_\LL-\delta_0}{2}- k_F \LL] }$. Conversely, when the inversion parity is odd, $\sigma = -1$, the $R,L$ fields will obey periodic boundary conditions up to constant phases $\e^{\pm i [\frac{\delta_\LL-\delta_0}{2}- k_F \LL]}$. We use this information to write the $R$ field as an equal superposition of periodic (Ramond) and antiperiodic (Neveu-Schwarz) right-moving fermionic fields. The names Ramond (R) and Neveu-Schwarz (NS) is the nomenclature used in conformal field theory\footnote{Here we use the cylindrical coordinates $x-iv_F \tau = \frac{i\LL }{\pi}\ln z$ so that the case of Ramond (R) fermions really does correspond to a periodic fermion. Transforming between planar $(z)$ and cylindrical $(x,\tau)$ holomorphic coordinates results in an extra $z^{-1/2}$ Jacobian which exchanges periodic and antiperiodic boundary conditions.} to describe periodic and antiperiodic fermions on the string sheet. Expanding the chiral field $R(x)$ yields 
\begin{align}
&R(x) := R^{(\mathrm{R})}(x)\e^{-i k_F' x} + R^{(\mathrm{NS})}(x)\e^{-i k_F' x} \nonumber \\
&\hspace{4mm}= \frac{1}{\sqrt{2\LL}}\sum_{n=-\infty}^\infty \left(
R_n \e^{i\frac{2\pi}{\LL}nx} + R_{n+\frac{1}{2}}\e^{i\frac{2\pi}{\LL}(n+\frac{1}{2})x}
\right)\e^{-i k_F' x},
\end{align} where 
$k_F':= k_F + (\delta_0-\delta_\LL)/(2\LL)$. 
Note that in the limit of large $\LL$, the difference between $k_F$ and $k_F'$ diminishes. The normalization constant above has been selected such that 
\begin{align}
\{R(x),R^\dagger(y) \} = \sum_{m=-\infty}^\infty \delta(x-y+  2m \LL) \,\left(\e^{-i2k_F'\LL}\right)^m.
\end{align}
This anticommutation relation suggests that the $R(x)$ field is quasiperiodic in the extended domain with period $2\LL$ and twisting $\e^{-i2k_F'\LL}$. This is nothing more than the unfolding operation that was just described. Finally, because of the requirements of inversion symmetry, we define the left-moving fields by  
\begin{subequations}
\begin{align}
L^{(\mathrm{R})}(x) &:= - R^{(\mathrm{R})}(\LL-x) = -R^{(\mathrm{R})}(-x), \\
L^{(\mathrm{NS})}(x) &:= + R^{(\mathrm{NS})}(\LL-x) = -R^{(\mathrm{NS})}(-x).
\end{align} 
This implies 
\begin{align} 
L(x) &:= L^{(\mathrm{R})}(x)\e^{i k_F' x} + L^{(\mathrm{NS})}(x)\e^{i k_F' x} = -R(-x),
\end{align}
\label{eqn:Dirichlet_final}%
\end{subequations}
where we have analytically continued the fields into $x<0$. These new unfolded chiral fields are $2\LL$-quasiperiodic such that
\begin{subequations}
\begin{align}
&R(x+2\LL) = \e^{-i2 k_F' \LL}R(x),\\
&L(x+2\LL) = \e^{i2 k_F' \LL}L(x).
\end{align}
\end{subequations}
In this manner, a non-zero $k_F$ and reflection phases acquired at the open boundaries can influence the continuum model through a quasiperiodic twisting after unfolding. 

The above quasiperiodic boundary conditions for the unfolded chiral fields differ from the ones stated in the standard literature such as Refs.~\onlinecite{fabrizio1995interacting,gogolin1999bosonization,giamarchi2004quantum}. They only agree when $\e^{i2k_F' \LL}=1$, which corresponds to an implicit assumption that $k_F'$ is exactly quantized in units of $\pi/\LL$ such that $R(x)\equiv -L(-x)$ is periodic with periodicity $2\LL$. 
Nevertheless, the chiral field may also be analytically continued to an antiperiodic field with $R(x+2\LL) = - R(x)$. In this case, $k_F'$ is a half-integer multiple of $\pi/\LL$ such that $\e^{i 2k_F' \LL} = -1$ and the boundary condition at $x=\LL$ in Eqn.~(\ref{eqn:Dirichlet}) remains satisfied. Often, these differences become less important -- in regard to bulk properties -- in the limit of large $\LL$.

One important point to appreciate is that the Dirichlet boundary condition Eqns.~(\ref{eqn:Dirichlet}) selects inversion-symmetric solutions from the periodic and antiperiodic modes for the closed version of the wire. Half are from the NS and the other half from the R types, and this depends on their properties under inversion. In fact, the exact eigenstates of an open system will alternate between even and odd inversion parity states. This observation is also suggestive of the connection between the fermion parity switching effect and twisted boundary conditions in an open wire with quasi-zero modes.



\section{Exact fermionic eigenstates with open boundaries}\label{sec:exact_fermion}

We shall now carry out a full diagonalization of the continuum fermionic Hamiltonian of Eqn.~(\ref{eqn:Ham_fermion}) in the limit $\mathrm{M}=0$ such that
\begin{align}
H &= \int_0^\LL \mathrm{d}x\big[
-iv_F R^\dagger \partial_x R + iv_F L^\dagger \partial_x L \nonumber\\&\hspace{15mm}+ i \Delta(R^\dagger L^\dagger -LR)\big]
\label{eqn:Ham_fermion2}
\end{align}
at finite $\LL$ and with open boundaries through unfolding. Our approach significantly extends the analyses of Refs.~\onlinecite{gangadharaiah2011majorana,fabrizio1995interacting} as it does not make an approximation for large system lengths $\LL$. As was noted in Sec.~\ref{sec:symmetries}, there is an enhanced linear $\mathbb{Z}_2$ symmetry $\mathrm{I}_x$ whenever $\mathrm{M}=0$, and this will serve as a guide for us in carrying out an exact analytic diagonalization. 

First, we define a two-component Nambu spinor field 
\begin{align}
\psi(x) := \begin{pmatrix} R(x) \\ L^\dagger(x) \end{pmatrix},
\end{align}
which is an effective linearized representation of the microscopic field $\Psi(x)\approx R(x)\e^{ik_F x} + L(x)\e^{-ik_F x}.$ 
For reasons that will become clearer later, we will demand now that $\e^{2ik'_FL}=-1$, which leads to an antiperiodic (NS) unfolded $R(x)$ fermion. The simplest way to satisfy this condition as well as $\Psi(0)=\Psi(\LL) =0$ is to have
\begin{align}
\delta_0=\delta_\LL =0, \qquad k_F = \left(n+\frac{1}{2}\right)\frac{\pi}{\LL}
\end{align}
for $n \in \mathbb{Z}$. This leads to $L(x) \equiv -R(-x)$, implying
\begin{align}
R(0)+L(0)= 0, \quad R(\LL)- L(\LL) = 0.
\end{align}
Therefore, we find $R(x+2\LL) = -R(x)$ and $\psi(x+2\LL) = -\psi(-x)$. Moreover, we must also have that 
\begin{align}
\psi(x) \equiv -\sigma^x [\psi^\dagger(-x)]^T,
\label{eqn:psi_conditions}
\end{align}
which is a reality condition that translates to a condition on the Majorana fields $\lambda,\lambda'$ as
\begin{align}
\lambda(x) = -\sigma^x \lambda(-x), \qquad \lambda'(x) = -\sigma^x \lambda'(-x).
\end{align}
The Hamiltonian in terms of $\psi(x)$ is 
\begin{align}
H = \int_0^\LL \mathrm{d}x\; \psi^\dagger(x) \left[
-iv_F \sigma^z \partial_x -\Delta \sigma^y
\right]\psi(x),
\end{align}
which after unfolding corresponds to
\begin{align}\label{eqn:Ham_unfolded}
H =\frac{1}{2}\int_{-\LL}^\LL \mathrm{d}x\; \psi^\dagger(x) \left[
-iv_F \sigma^z \partial_x -\Delta \,s(x)\sigma^y 
\right] \psi(x).
\end{align}
Here, we have introduced the $2\LL$-periodic square wave form 
\begin{align}
s(x) := \text{sgn}\left[\sin (\tfrac{\pi x}{\LL})\right] =  2\sum_{n \text{ odd}} \frac{\e^{i \frac{n \pi x}{\LL}}}{i n \pi },
\end{align}
which serves as a profile function for a mass domain wall. Hence we have the BdG mode Hamiltonian\cite{fabrizio1995interacting,gangadharaiah2011majorana}
\begin{align}
\mathcal{H}_\mathrm{BdG} = -iv_F \sigma^z \partial_x -\Delta \,s(x)\,\sigma^y,
\label{eqn:Ham_BdG_unfolded}
\end{align}
which contains in it a mass domain wall at $x=0,\pm \LL$. 

To diagonalize $H$, one needs to find operators 
\begin{align}
\psi_E^\dagger := \int_{-\LL}^\LL \mathrm{d}x\; \psi^\dagger(x)\cdot \phi_E(x)
\end{align}
that satisfy
\begin{align}
[H,\psi_E^\dagger] = E \psi_E^\dagger, \quad \mathcal{H}_\text{BdG} \,\phi_E(x) = E \phi_E(x)
\end{align}
such that $\phi_E$ is a normalized BdG eigenmode. Due to fermionic Nambu doubling, the negative-energy eigenmodes are redundant because it can be shown that $\psi_{-E} \propto \psi_E^\dagger$. Tentatively, this yields the diagonalization 
\begin{align}
H &= \sum_{E \geqslant 0} E \, \psi_E^\dagger \psi_E\;\;  + E_0 \mathbbm{1},
\end{align}
where $E_0$ is the formally infinite ground state energy.

Although superficially similar to Eqn.~(\ref{eqn:BdG_Ham}), $\mathcal{H}_\text{BdG}$ in Eqn.~(\ref{eqn:Ham_BdG_unfolded}) has its symmetries expressed differently. Specifically, time reversal $\mathrm{T}$ and unitary charge conjugation $\mathrm{I}_x$ have the following actions on 2-spinor wavefunctions $\chi(x)$ defined over $[-\LL,\LL]$:
\begin{subequations}
\begin{align}
&\mathrm{T}:\chi(x) \mapsto [\chi(-x)]^*, \\ \nonumber \\
&\mathrm{I}_x:\chi(x) \mapsto \sigma^x \chi(-x).
\end{align}
\label{eqn:Z2_spinor}%
\end{subequations}
In this sense, $\mathrm{I}_x$ now plays the role of an inversion symmetry in the unfolded domain. This is because $R(-x) = -L(x)$ such that coordinate inversion in the unfolded domain $x \mapsto -x$ automatically exchanges left and right movers. For this reason, we will from now on refer to $\mathrm{I}_x$ as an inversion symmetry.

Linear momentum is not conserved because $\mathcal{H}_\text{BdG}$ is \emph{not} invariant under arbitrary finite amounts of spatial translations. Nevertheless, $\mathcal{H}_\text{BdG}$ is \emph{locally} translationally invariant in the sense that $\mathcal{H}_\text{BdG}$ commutes with the infinitesimal generator $-i\partial_x$ everywhere except where $s(x)$ is discontinuous. This suggests diagonalizing the BdG Hamiltonian in $(0,\LL)$ and $(-\LL,0)$ separately at first and then forming superpositions. The relevant BdG equations are 
\begin{align}
\begin{pmatrix}
-iv_F \partial_x & i \Delta \\
-i \Delta & i v_F \partial_x
\end{pmatrix} w_>(x) &= E w_>(x), \quad x \in (0,\LL), \\
\begin{pmatrix}
-iv_F \partial_x & -i \Delta \\
i \Delta & i v_F \partial_x
\end{pmatrix} w_<(x) &= E w_<(x), \quad x \in (-\LL,0).
\end{align}


\subsection{Extended states}\label{sec:ext_states}

We focus first on states above the SC gap ($E^2 \geqslant \Delta^2$) in $0<x<\LL$. The obvious eigenfunction candidates are plane waves of the form
\begin{align}
w_>(x) = \begin{pmatrix} u_k \\ v_k \end{pmatrix}\e^{ikx}, \quad k>0.
\end{align}
Direct substitution then leads to 
\begin{align}
\begin{pmatrix}
\xi_k & i \Delta \\ -i \Delta & -\xi_k
\end{pmatrix} \begin{pmatrix} u_k \\ v_k \end{pmatrix} 
= E \begin{pmatrix} u_k \\ v_k \end{pmatrix} ,
\end{align}
where $\xi_k := v_F k$ is the normal state dispersion. The resulting BdG mean-field eigenvalue equation is readily diagonalized with solutions
\begin{align}
w^{(+)}_{>,k}(x) &= \begin{pmatrix} \cos \theta_k \\ -i\sin \theta_k \end{pmatrix} \e^{ikx}, \qquad E = +E_k,
\label{eq:BdG_sol1}\\
w_{>,k}^{(-)}(x) &= \begin{pmatrix} \sin \theta_k \\ i\cos \theta_k \end{pmatrix}\e^{ikx} 
, \qquad E = -E_k,
\label{eq:BdG_sol2}
\end{align}
where the BdG dispersion is given by
\begin{align}
E_k := \sqrt{\Delta^2 + \xi_k^2}
\label{eqn:Ek}
\end{align}
and the angle $\theta_k \in [0, \pi/4)$ is defined by 
\begin{align}
\tan 2\theta_k = \frac{\Delta}{\xi_k}.
\label{eqn:thetak}
\end{align}
This also implies the `reflection' property 
\begin{align}
\cos \theta_k = \sin \theta_{-k}.
\end{align}

The eigenmodes in the negative domain $-\LL < x < 0$ can be obtained by replacing $\Delta \rightarrow - \Delta$, which simply corresponds to multiplying the eigenmodes given in Eqns.~(\ref{eq:BdG_sol1}) and (\ref{eq:BdG_sol2}) by $\sigma^z$. This yields 
\begin{align}
w^{(+)}_{<,k}(x) &= \begin{pmatrix} \cos \theta_k \\ i\sin \theta_k \end{pmatrix} \e^{ikx}, \qquad E = +E_k,  \\
w_{<,k}^{(-)}(x) &= \begin{pmatrix} \sin \theta_k \\ -i\cos \theta_k \end{pmatrix}\e^{ikx}, \qquad E = -E_k.
\end{align} 
However, $w^{(\pm)}_{>,k}(x)$ and $w^{(\pm)}_{<,k}(x)$ do not agree at the boundary points $x=0,\pm \LL$. Nonetheless, we can exploit the fact that $E_k = E_{-k}$ and consider \emph{superpositions} of degenerate energy eigenstates with opposite $k$ wavenumbers. 

Because $\mathcal{H}_\text{BdG}$ is invariant under inversion as defined by $\mathrm{I}_x$, we can attempt to determine continuous eigenfunctions from $\mathrm{I}_x$-symmetric eigenstates, i.e., from eigenstates with even (+1) and odd (-1) $\mathrm{I}_x$ parities. The superpositions that achieve this for $E = +E_{\pm k}$ are exactly
\begin{subequations}
\begin{align}
\phi^{(+)}_{>,k}(x) &= 
\frac{1}{\sqrt{4\LL \mathrm{N}_k}} 
\left[
\e^{i\theta_k}w_{>,k}^{(+)}(x) - i \e^{-i\theta_k}w^{(+)}_{>,(-k)}(x) 
\right] \nonumber \\
&=\frac{1}{\sqrt{4\LL \mathrm{N}_k}}
\begin{pmatrix}
\cos(kx + 2\theta_k)+ i \sin (kx) \\ -\cos(kx+2\theta_k)+i\sin(kx)
\end{pmatrix},
\\
\varphi^{(+)}_{>,k}(x) &= 
\frac{1}{\sqrt{4\LL \mathrm{N}_k}} 
\left[
\e^{-i\theta_k}w_{>,k}^{(+)}(x) + i \e^{i\theta_k}w^{(+)}_{>,(-k)}(x) 
\right] \nonumber \\
&=\frac{1}{\sqrt{4\LL \mathrm{N}_k}}
\begin{pmatrix}
\cos(kx -2\theta_k)+ i \sin (kx) \\ \cos(kx-2\theta_k)-i\sin(kx)
\end{pmatrix},
\end{align}
\label{eqn:magik}%
\end{subequations}
and their analytic continuations through 
\begin{align}
\phi^{(+)}_{<,k}(x) = \phi^{(+)}_{>,k}(-x)^*, \quad 
\varphi^{(+)}_{<,k}(x) = \varphi^{(+)}_{>,k}(-x)^* 
\end{align}
as required by time-reversal symmetry. The $k$-dependent $O(1)$ constant $\mathrm{N}_k$ is chosen such that the eigenfunctions are properly normalized over $[-\LL,\LL]$. Finally, we arrive at the normalized entire eigenfunctions
\begin{align}
\phi^{(+)}_{k}(x) &= \frac{1}{\sqrt{2\LL \mathrm{N}_k}}
\begin{pmatrix}
\cos(k|x| + 2\theta_k)+ i \sin (kx) \\ -\cos(k|x|+2\theta_k)+i\sin(kx)
\end{pmatrix}, \label{eqn:phi_soln} \\
\varphi^{(+)}_k (x) &=\frac{1}{\sqrt{2\LL \mathrm{N}_k}}
\begin{pmatrix}
\cos(k|x|-2\theta_k)+ i \sin (kx) \\ \cos(k|x|-2\theta_k)-i\sin(kx)
\end{pmatrix}, \label{eqn:varphi_soln}
\end{align}
valid for $-\LL \leqslant x \leqslant \LL$ and a $k >0$ which is yet undetermined. Note that $\phi^{(+)}_0(x) \equiv 0$ and $\varphi^{(+)}_0(x)\equiv 0$, which excludes the $k=0$ solution entirely. Moreover, the negative $k$ values are excluded since they have already been used in forming the superpositions. 

It is easy to verify that these states are $\mathrm{I}_x$ eigenstates with 
\begin{align}
\mathrm{I}_x \phi_k^{(+)} = - \phi_k^{(+)} , \qquad
 \mathrm{I}_x \varphi_k^{(+)} = + \varphi_k^{(+)}.
\end{align}
More importantly, they are continuous at $x=0$. However, at $x=\pm \LL$, enforcing the antiperiodic boundary condition leads to quantization of $k$ through the conditions
\begin{align}
&\cos(k\LL+2\theta_k) = 0 \quad \text{for} \quad \phi^{(+)}_k, \\ 
&\cos(k\LL-2\theta_k) = 0 \quad \text{for} \quad \varphi^{(+)}_k,
\end{align}
which give the self-consistent equations
\begin{subequations}
\begin{align}
&k_{n,1} = \left(n +\frac{1}{2}\right)\frac{\pi}{\LL} - \frac{2\theta_{k_{n,1}}}{\LL}  \quad \text{for} \quad \phi^{(+)}_{k_{n,1}}, \quad n \in \mathbb{N}, \\
&k_{n,2} = \left(n +\frac{1}{2}\right)\frac{\pi}{\LL} + \frac { 2\theta_{k_{n,2}}} {\LL} \quad \text{for} \quad \varphi^{(+)}_{k_{n,2}} , \quad n \in \mathbb{N}\cup \{0\}.
\end{align}
\label{eqn:kn12_quantization}%
\end{subequations}
In general, these have to be solved numerically for each $n$. However, in the limit of large $n$ either $k_{n,1}$ or $k_{n,2}$ approaches a half-integer multiple of $\pi/\LL$ because $\theta_k \rightarrow 0$ in the limit of large $k$. One should also note that while $\varphi^{(+)}_{k_{0,2}}$ is a viable solution irrespective of $\Delta$, its counterpart $\phi^{(+)}_{k_{0,1}}$ is not whenever $v_F/\Delta < \LL$. Nevertheless, solutions to $k_{0,1}$ in Eqn.~(\ref{eqn:kn12_quantization}) will exist when $v_F/\Delta \geqslant \LL$, which is the regime where the coherence length is longer than or equal to the system length. In fact, the mode $\phi^{(+)}_{k_{0,1}}$ is continuously connected to the topological MZM eigenfunction that we will describe in the next subsection. 

Finally, the solutions for the quantized values of $k=k_{n,1},k_{n,2}$ lead to the fermionic Bogoliubon operators
\begin{align}
\phi_{n} &:= \int_{-\LL}^\LL \mathrm{d}x\; [\phi^{(+)}_{k_{n,1}}(x)]^\dagger \psi(x), \\
\varphi_{n} &:= \int_{-\LL}^\LL \mathrm{d}x\; [\varphi^{(+)}_{k_{n,2}}(x)]^\dagger \psi(x),
\end{align}
that diagonalize the part of $H$ above the SC gap.
It is convenient to define the functions
\begin{align}
u_k(x) &:= \frac{1}{\sqrt{\LL \mathrm{N}_k}}\left[
\cos(k|x| + 2\theta_k)+ i \sin(kx) 
\right], \\
v_k(x) &:= \frac{1}{\sqrt{\LL\mathrm{N}_k}}\left[
\cos(k|x| - 2\theta_k)+ i \sin(kx) 
\right],
\end{align}
which allows us to express our eigenfunction solutions more neatly as
\begin{align}
\phi^{(+)}_{k_{n,1}}(x) &= \frac{1}{\sqrt{2}}
\begin{pmatrix} u_{k_{n,1}}(x) \\ -[u_{k_{n,1}}(x)]^* \end{pmatrix}, \\
\varphi^{(+)}_{k_{n,2}}(x) &= \frac{1}{\sqrt{2}}
\begin{pmatrix} v_{k_{n,2}}(x) \\ [v_{k_{n,2}}(x)]^* \end{pmatrix}.
\end{align}
In addition, these functions satisfy $[u_k(-x)]^* = u_k(x)$ and $[v_k(-x)]^* = v_k(x)$
as is required by time-reversal symmetry $\mathrm{T}$ and inversion symmetry $\mathrm{I}_x$. This then yields the following equivalent expressions for the Bogoliubon operators:
\begin{align}
\phi_n = \int_{-\LL}^\LL \mathrm{d}x\; \frac{[u_{k_{n,1}}(x)]^*}{\sqrt{2}}\left[R(x) + R^\dagger(x)\right], \\
\varphi_n = \int_{-\LL}^\LL \mathrm{d}x\; \frac{[v_{k_{n,2}}(x)]^*}{\sqrt{2}}\left[R(x) - R^\dagger(x)\right].
\end{align}
When expressed in terms of the Majorana field operators $\lambda(x)$, $\lambda'(x)$ defined in Eqns.~(\ref{eqn:majorana_fields}), these expressions translate to
\begin{align}
\phi_n &= 2 \int_0^\LL \mathrm{d}x\,  [\phi_{k_{n,1}}^{(+)}(x)]^\dagger \lambda(x), \\
\varphi_n &= 2i\, \int_0^\LL \mathrm{d}x\, [\varphi_{k_{n,2}}^{(+)}(x)]^\dagger\sigma^z \lambda'(x).
\end{align}
This just reflects the fact that the theory is comprised of two decoupled Majorana field theories with opposite pseudo-scalar masses as described by the Lagrangian in Eqn.~(\ref{eqn:L_Majorana}) in the limit $\mathrm{M}=0$. One may also interpret each Majorana model as being an independent Kitaev Majorana chain as is explicitly discussed in Appendix \ref{app:Maj_chains}.


\subsection{In-gap states and Majorana zero modes}\label{sec:exact_MZM}

We now move on to the localized states that have their energies in the SC gap ($E^2 < \Delta^2$). It is best to work directly in the unfolded $[-\LL,\LL]$ domain. Recall the BdG Hamiltonian 
\begin{align}
\mathcal{H}_\text{BdG} = -iv_F \sigma^z \partial_x - \Delta \, \text{sgn}[\sin (\tfrac{\pi x}{\LL})]\sigma^y,
\label{eqn:Ham_BdG_Open_L}
\end{align}
the eigenvalue equation of which we can rearrange as 
\begin{align}
\left[
\partial_x - {l} \,\text{sgn}[\sin (\tfrac{\pi x}{\LL})] \, \sigma^x - i \sigma^z \epsilon 
\right] w(x) =0 \label{eqn:1st_diff_w}
\end{align}
with ${l} := \frac{\Delta}{v_F}$ and $\epsilon := \frac{E}{v_F}$. For the moment, we will not fix the $\mathrm{I}_x$ parity of the spinor-valued function $w(x)$. 

Now applying the linear differential operator on the LHS of Eqn.~(\ref{eqn:1st_diff_w}) twice to $w(x)$ yields 
\begin{align}
&\left[
-\partial_x^2 + 2 {l} \sum_{m=-\infty}^\infty (-1)^m \delta(x-m\LL)\sigma^x
\right] w(x) \nonumber \\
&\hspace{3.9cm}= - ({l}^2 - \epsilon^2)w(x)
\end{align}
after using 
\begin{align}
\partial_x \left[ \text{sgn}(\sin (\tfrac{\pi x}{\LL}))\right]
= 2 \sum_{m=-\infty}^\infty (-1)^m \delta(x-m\LL).
\end{align}
This then suggests that we expand $w(x)$ in the eigenbasis of $\sigma^x$ as
\begin{align}
w(x) = a(x)\begin{pmatrix} 1 \\-1 \end{pmatrix} + i b(x)\begin{pmatrix} 1 \\ 1\end{pmatrix},\label{eqn:w_a_b}
\end{align}
which yields the following decoupled Kronig-Penney like time-independent Schr\"odinger equations with delta-function potentials
\begin{subequations}
\begin{align}
\left[-\partial_x^2 - 2 {l}  \sum_{m=-\infty}^\infty (-1)^m \delta(x-m\LL) \right] &a(x) \nonumber\\= -({l}^2 - \epsilon^2) &a(x), \\
\left[-\partial_x^2 + 2 {l} \sum_{m=-\infty}^\infty (-1)^m \delta(x-m\LL) \right]&b(x)  \nonumber\\= - ({l}^2 - \epsilon^2) &b(x).
\end{align}\label{eqn:Kronig_Penney}%
\end{subequations}
This is a somewhat unusual Kronig-Penney model because of the alternating sign of the delta-function potential. However, because $a(x)$ and $b(x)$ are required to be antiperiodic as opposed to being generally quasiperiodic, the `crystal-momentum' is located at the boundary of the first Brillouin zone. Moreover, because the `Hamiltonian' operator is purely real -- as opposed to just being Hermitian -- the functions $a(x)$ and $b(x)$ can be chosen to be purely real. It is then easy to appreciate that $a(x)$ and $b(x)$ must vanish where the singular potential is positive in order to be $2\LL$-antiperiodic. Thus we have the nodal conditions
\begin{align}
a(\LL) = a(-\LL)= 0, \qquad b(0) = 0,
\end{align}
which accompany the kinks at $a'(0)$ and $b'(\LL)$. In the domain $-\LL<x<\LL$, the normalized functions with these properties are
\begin{align}
a(x) &= \mathcal{N_\kappa} \sinh(\kappa(\LL-|x|)),\label{eqn:a}\\
b(x) &= \mathcal{N_\kappa} \sinh(\kappa x), \label{eqn:b}
\end{align}
with inverse decay length $\kappa := +\sqrt{{l}^2 - \epsilon^2}$ and  
\begin{align}
\mathcal{N_\kappa}=\frac{1}{\sqrt{\LL}}\left(\frac{\sinh(2\kappa \LL)}{2\kappa \LL}-1\right)^{-\frac{1}{2}}
\end{align}
as a normalization constant. Here $a(x)$ is allowed to be even because it vanishes at the boundaries $x=\pm \LL$, whilst $b(\pm \mathrm{L})$ is allowed to take non-zero values because it is an odd function. Because $a(x)$ and $b(x)$ are really required to be $2\LL$-antiperiodic, these functions 
can be described as Fourier series:
\begin{align}
a(x) &= \mathcal{N}_\kappa \sum_{n \text{ odd}} \frac{4 \kappa \LL\, \cosh(\kappa \LL)}{(n \pi)^2 + (2\kappa \LL)^2}\; \e^{i\frac{n\pi x}{2\LL}}, \\
b(x) &= \mathcal{N}_\kappa \sum_{n \text{ odd}} \frac{4 \kappa \LL\, \cosh(\kappa \LL)}{(n \pi)^2 + (2\kappa \LL)^2}\; i^{-n}\;\e^{i\frac{n\pi x}{2\LL}}.
\end{align}
In fact, it is easy to see from the Fourier series representations that $b(x) = a(x-\LL)$.

Finally, we can derive the exact `gap equation' from the kink conditions at $x=0,\LL$ which read as 
\begin{align}
-a'(0^+)+ a'(0^-) &= 2 {l} a(0), \\
-b'(-\LL+0^+) + b'(\LL^-) &= 2 {l} b(\LL).
\end{align}
These are identical conditions which translate to\footnote{It is amusing to point out that this gap equation relating $\kappa$ and ${l}$ is identical to the self-consistent Weiss mean-field equation with $\kappa$ as the order parameter and $\LL^{-1}$ as the temperature.} 
\begin{align}
\kappa = {l} \tanh (\kappa \LL)  \quad \Longleftrightarrow \quad \frac{i v_F \kappa}{\Delta} = \tan\left(i\kappa \LL \right), 
\end{align}
or, in terms of the original parameters,
\begin{align}
\sqrt{1-\left(\tfrac{E}{\Delta}\right)^2}
= \tanh \left(\frac{\Delta \LL}{v_F}\sqrt{1-\left(\tfrac{E}{\Delta}\right)^2}\right).
\label{eqn:gap_eqn}
\end{align}
By identifying $v_F /\Delta$ with the SC coherence length, we see that as $|E| \rightarrow \Delta$ we have $\LL \rightarrow v_F /\Delta$. Likewise, as $\LL \rightarrow \infty$ with $v_F /\Delta$ finite, then $|E|\rightarrow 0$. Thus we have the correct limits describing the extremely small and infinite $\LL$ limits, respectively, as compared to the SC coherence length. Moreover it is easy to see from Eqns.~(\ref{eqn:a}) and (\ref{eqn:b}) that as $\LL \rightarrow \infty$, $a(x)$ and $b(x)$ approach the expected exponentially localized forms. By contrast, as $\Delta \rightarrow v_F/\LL$ from above (while all other parameters remain finite) we have $\kappa \rightarrow 0$, meaning that these solutions cease to be normalized eigenfunctions at points beyond $\Delta < v_F/\LL$. 

\begin{figure}
\begin{center}
\includegraphics[width=0.45\textwidth]{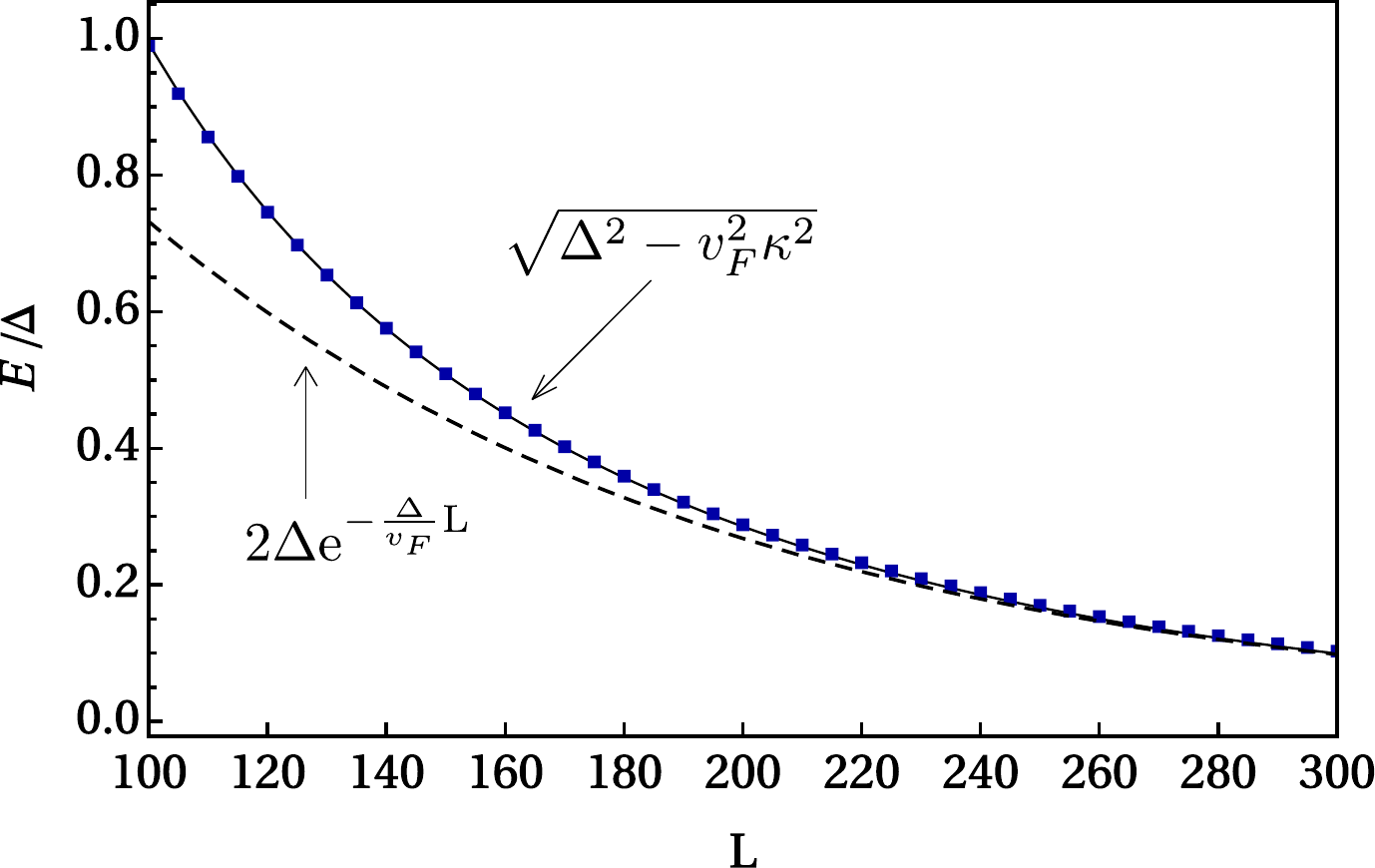}
\end{center}
\caption{Comparison of the in-gap energy $E$ using Eqn.~(\ref{eqn:gap_eqn}) [solid line] and numerically computed eigenvalues of the lattice model given in Eqn.~(\ref{eqn:H_Kit_wire}) [square points]. The numerical parameters are $t'=0.99$ and $t=1.0$, where $v_F = (t+t')a/2$ and $\Delta = t-t'$. The coherence length is given by $v_F/\Delta = 99.5$ in units of $a$. The dashed line is the asymptotic form given in Eqn.~(\ref{eqn:Egap_asymp}), which underestimates the energy gap for short lengths.}
\label{fig:Egap}
\end{figure}

To arrive at an estimate of $|E|$ in the large $\LL$ limit, we use the asymptotic expansion
\begin{align}
\tanh x = 1 - 2\e^{-2x} + O(\e^{-4x}) 
\end{align}
to arrive at 
\begin{align}
\kappa \approx {l} (1- 2\e^{-2\kappa \LL}) \quad \Rightarrow
\quad 
|E|\sim 2\Delta \e^{-\frac{\Delta \LL}{v_F}}.
\label{eqn:Egap_asymp}
\end{align}
We note that here the usual oscillating prefactor to the finite-size energy splitting is absent (equal to one) due to our choice that $k_F$ is a half-integer multiple of $\pi/\mathrm{L}$. With this choice of $k_F$, we find a constraint from above on the value of the finite-size energy splitting. Shown in Fig.~\ref{fig:Egap} is a comparison between this asymptotic estimate, Eqn.~(\ref{eqn:gap_eqn}) and the numerically computed gaps for a Kitaev lattice model realization. As is discussed in Appendix \ref{app:Maj_chains}, the continuum SC wire model may be realized as two Kitaev Majorana chains of opposite SC ``mass terms" in accordance with the discussion in Sec.~\ref{sec:in_terms_of_majoranas}.

We can now write down the exact finite $\LL$ eigenstates of the original BdG equation. These are not surprisingly given by the equal superpositions
\begin{align}
\Phi^{(\pm)}(x) :=& \frac{a(x)}{2}\begin{pmatrix} 1 \\ -1\end{pmatrix}
\pm  \frac{i b(x)}{2} \begin{pmatrix} 1 \\ 1 \end{pmatrix} \nonumber \\
\equiv& \frac{\mathcal{N}_\kappa}{2}\begin{pmatrix} \sinh(\kappa(\LL-|x|)) \pm i \sinh(\kappa x) \\ -\sinh(\kappa(\LL-|x|)) \pm i \sinh(\kappa x ) 
\end{pmatrix}
\label{eqn:Phi_soln}
\end{align}
with energies $\pm E_{i\kappa}=\pm \sqrt{\Delta^2 - v_F^2 \kappa^2}$. This specific linear combination respects time-reversal symmetry $[\Phi^{(\pm)}(-x)]^* = \Phi^{(\pm)}(x)$. 
To check that these are indeed eigenfunctions of Eqn.~(\ref{eqn:Ham_BdG_Open_L}), one can first verify that for $0<x<\LL$ we have
\begin{subequations}
\begin{align}
\partial_x a(x) + {l} a(x) &= -\sqrt{{l}^2 -\kappa^2}\,b(x), \\
\partial_x b(x) - {l} b(x) &= + \sqrt{{l}^2 -\kappa^2} a(x),
\end{align}
\end{subequations}
using the gap equation Eqn.~(\ref{eqn:gap_eqn}) and hyperbolic trigonometric identities. Even more generally, we can derive similar equations for $-\LL<x<0$ and taken together produce
\begin{subequations}
\begin{align}
[v_F\partial_x  + \Delta s(x)] a(x) &= - E_{i\kappa} b(x),\\
[v_F\partial_x  - \Delta s(x)] b(x) &= + E_{i\kappa}  a(x)
\end{align}
\label{eqn:BdG_a_b}
\end{subequations}
in $-\LL < x <\LL $. Using these one readily obtains
\begin{subequations}
\begin{align}
[-iv_F \sigma^z \partial_x- \Delta s(x)\sigma^y] \begin{pmatrix}a(x) \\ -a(x)\end{pmatrix} &= iE_{i\kappa} \,  \begin{pmatrix} b(x) \\ b(x) \end{pmatrix}, \\
[-iv_F \sigma^z \partial_x- \Delta s(x)\sigma^y] \begin{pmatrix} ib(x) \\ ib(x)\end{pmatrix} &= E_{i\kappa} \,  \begin{pmatrix} a(x) \\ -a(x) \end{pmatrix}.
\end{align}
\end{subequations}
Together, these relations can be used to show that 
\begin{align}
[-iv_F \sigma^z \partial_x - \Delta s(x) \sigma^y ]\Phi^{(\pm)}(x) = \pm E_{i\kappa}  \Phi^{(\pm)}(x).
\end{align}
Defining the fermionic eigenmode operator 
\begin{align}
\psi^\dagger_{E_{i\kappa}}:= \int_{-\LL}^\LL \mathrm{d}x\ \psi^\dagger(x)\, \Phi^{(+)}(x) 
\end{align}
then implies that $[H,\psi^\dagger_{E_{i\kappa}}]=E_{i\kappa}\psi_{E_{i\kappa}}^\dagger$.  

It will furthermore prove useful to define the Majorana operators
\begin{subequations}
\begin{align}
\gamma_0 &:= \int_{-\LL}^\LL \mathrm{d}x \; a(x) [R(x)+ R^\dagger(x)], \\
\gamma_\LL &:= \int_{-\LL}^\LL \mathrm{d}x \; b(x) [R(x)+ R^\dagger(x)],
\end{align}
\label{eqn:gamma_0L}%
\end{subequations}
which are localized at $x=0$ and $x=\pm\LL$ respectively. This leads to 
\begin{align}
\psi_{E_{i\kappa}} = \int_{-\LL}^{\LL} \mathrm{d}x\; [\Phi^{(+)}(x)]^\dagger \, \psi(x) \equiv \frac{\gamma_0 - i \gamma_
\LL}{2}
\end{align}
and 
\begin{align}
[H,\gamma_0] = i E_{i\kappa}\gamma_\LL, \qquad [H,\gamma_\LL] = -i E_{i\kappa}\gamma_0.  
\end{align}
The first of these commutation relations is explicitly verified in Appendix \ref{app:direct_check}.


\subsection{Full diagonalization and symmetry fractionalization} \label{sec:fermion_diag}

The extended and localized eigenmodes lead to the final diagonalization of $H$ as 
\begin{align}
H =& - \frac{i}{2}E_{i\kappa}\gamma_0 \gamma_\LL + E_{k_{0,2}}\varphi^\dagger_0 \varphi_0 \nonumber \\
&+\sum_{n >0}  \left[ E_{k_{n,1}}\phi^\dagger_n \phi_n 
+ E_{k_{n,2}}\varphi^\dagger_n \varphi_n\right] + \text{const.} 
\label{eqn:Ham_final_diag}
\end{align}
 In second-quantized form, time-reversal $\mathcal{T}$ corresponds to the anti-linear transformation\footnote{This differs by an irrelevant sign from our previous definition of $\mathcal{T}$.}
\begin{align}
\mathcal{T} R(x) \mathcal{T}^{-1} = R(-x), \quad
\mathcal{T} R^\dagger(x) \mathcal{T}^{-1} = R^\dagger(-x)
\end{align}
because of unfolding. Thus, while the excitation operators $\varphi_n,\phi_n$ are time-reversal invariant, we have that 
\begin{align}
\mathcal{T} \gamma_0 \mathcal{T}^{-1} = +\gamma_0, \quad
\mathcal{T} \gamma_\LL \mathcal{T}^{-1} = -\gamma_\LL,
\end{align}
as required for time-reversal symmetry of $H$. Since the overlap $\int_0^\LL a(x)b(x)\,\mathrm{d}x \sim \e^{-\kappa \LL}$ is exponentially suppressed, we have the delocalization of symmetry properties. This is the hallmark of fractionalization of symmetry transformations\cite{turner2011topological,fidkowski2010effects,fidkowski2011topological} that is a defining property of SPT phases. 

In the short length limit where $\LL < v_F/\Delta \leqslant \infty$, the solution $\Phi^{(+)}{(x)}$ is no longer a valid eigenfunction. Rather, the extended state solution $\phi^{(+)}_{k_{0,1}}(x)$ replaces it as the lowest positive-energy eigenfunction. Physically speaking, in the short $\LL$ limit, the quantization scale of the kinetic energy exceeds the SC gap $\Delta$ such that only extended state solutions exist. The change from localized to extended state occurs at exactly $\mathrm{L} = v_F/\Delta$, where the functions $a(x)$ and $b(x)$ become piecewise linear functions with $\kappa=0$.

It is interesting to note that under the change of sign $\Delta \rightarrow -\Delta$, the roles of $k_{n,1}$ and $k_{n,2}$ in Eqn.~(\ref{eqn:kn12_quantization}) are exchanged because the relation $2\theta_k = \tan^{-1}(\frac{\Delta}{v_F k})$ is odd in $\Delta$. In this instance, $\varphi^{(+)}_{k_{0,2}}(x)$ is no longer a viable eigenfunction but instead turns into a topological localized boundary mode with $\mathrm{I}_x= +1.$ At the same time, $\Phi^{(+)}(x)$ as defined above gets replaced by $\phi^{(+)}_{k_{0,1}}(x)$ with $\mathrm{I}_x= -1$. In this way, we always have a single topological boundary mode irrespective of the sign of $\Delta$. This is one of the biggest differences with the SSH model, where the sign of the `mass' plus a choice of open boundary conditions dictates whether or not topological boundary modes will be present in the open wire energy spectrum. 

\begin{figure}
\includegraphics[width=0.5\textwidth]{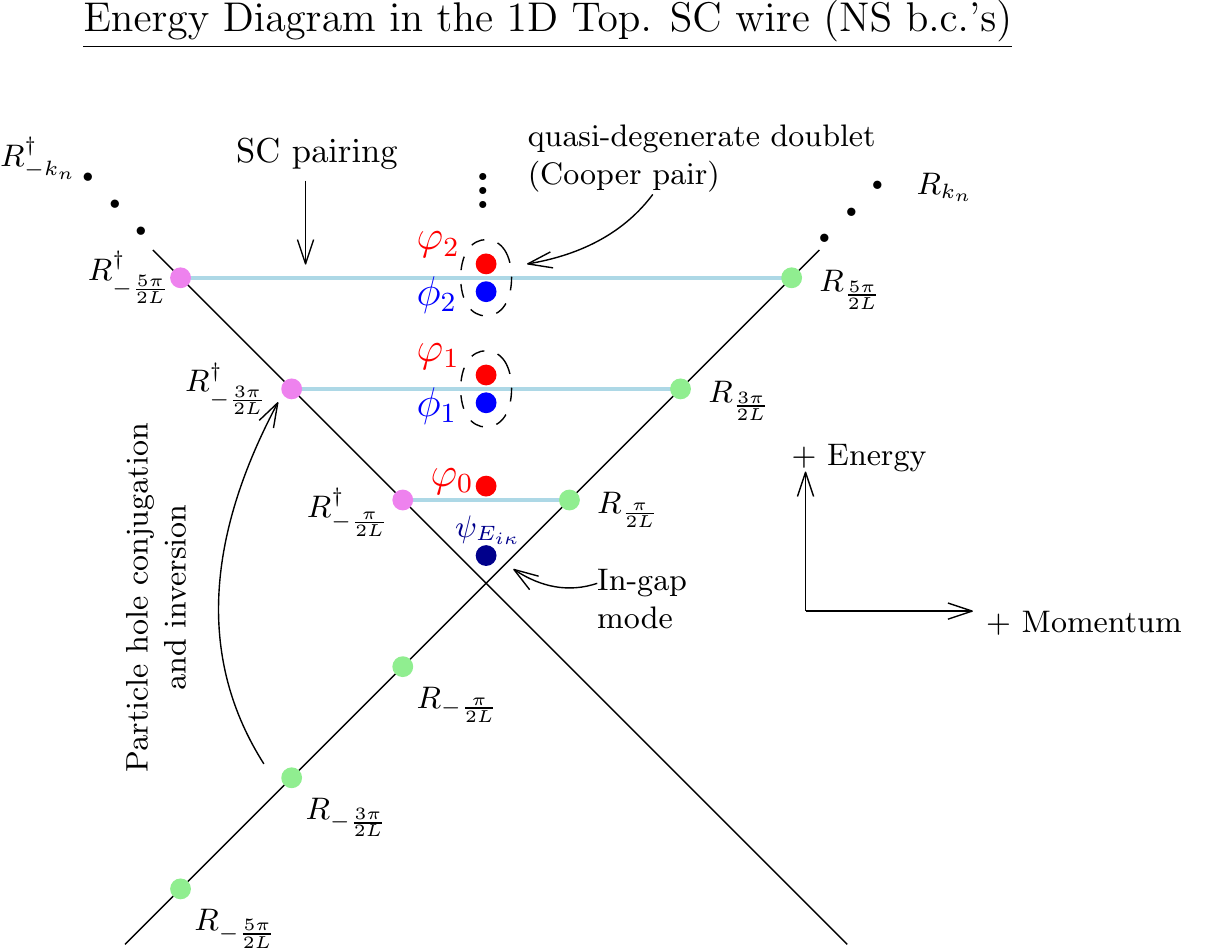}
\caption{Schematic energy diagram of the diagonalized linearized model for an open wire of length $\LL$ obeying $2\LL$-antiperiodic (NS) boundary conditions after unfolding. Note that with the SC pairing interaction turned on, energy levels are pushed up (pushed down if we are counting states into the filled Fermi sea), which is not shown to scale here for the sake of clarity. Also, the SC pairing function having the square-wave profile $\Delta s(x)$ couples all momentum eigenstates amongst themselves. The solid light blue line only denotes the most dominant channel which couples a particle-hole pair with opposite momenta.}
\label{fig:energy_diagram}
\end{figure}


\subsection{BCS wavefunction}\label{sec:BCS}

With the explicit 1-body eigensolutions to the BdG Hamiltonian at hand, we can express the exact many-body BCS wavefunction for a finite system of length $\LL$ with open boundaries. 
This is surprisingly not as straightforward as it seems. Typically, the ground state BCS wavefunction $|\text{BCS}\rangle$ should be defined such that 
\begin{align}
&\varphi_n|\text{BCS}\rangle = \phi_n|\text{BCS}\rangle = 0 && \forall \;\; n \in \mathbb{N}, \\
&\varphi_0|\text{BCS}\rangle = \psi_{E_{i\kappa}}|\text{BCS}\rangle = 0, &&{v_F \kappa} =  \Delta \tanh(\kappa \LL).
\end{align}
Naively, one would then take 
\begin{align}
|\text{BCS}\rangle =  \psi_{E_{i\kappa}} \varphi_0 \prod_{n>0} \phi_n\varphi_{n} |\text{vac} \rangle,
\end{align}
where $|\text{vac}\rangle$ is the particle-number vacuum which is annihilated by all $R(x)$. There are, however, problems with this definition: In the limit $\Delta = 0$ we have that 
\begin{align}
\phi_n\varphi_n  = R^\dagger_{-{k_n}}R_{k_n}, \qquad k_n = \left(n+\frac{1}{2}\right)\frac{\pi}{\LL},
\end{align}
which then annihilates the vacuum state $|\text{vac}\rangle$. Nevertheless, in the limit $\Delta=0$, we expect that the state $|\text{BCS}\rangle$ will become  $|0\rangle$, i.e., the filled Fermi sea. It turns out that the relevant state to act on is the `anti-Fermi sea' $|\bar{0}\rangle$ defined by 
\begin{align}
|\bar{0}\rangle := \prod_{n\geqslant 0 }R_{k_n}^\dagger |\text{vac}\rangle.
\end{align}
This is chosen such that 
\begin{align*}
\prod_{n\geqslant 0} (R^\dagger_{-{k_n}}R_{k_n})|\bar{0}\rangle &= \prod_{n\geqslant 0}R^\dagger_{-{k}_n}|\text{vac}\rangle =|0\rangle.
\end{align*}
Hence we define the BCS ground state to be
\begin{align}|\text{BCS}\rangle :=  \psi_{E_{i\kappa}} \varphi_0\prod_{n>0}\phi_n\varphi_n |\bar{0} \rangle.
\end{align}
By analogy with the usual $s$-wave BCS wavefunctions, we see that the (hard-core) bosonic operators $b_n := \phi_n \varphi_n$ are the ``Cooper pair'' annihilation operators for an open wire. However, the Bogoliubons $\phi_n$ and $\varphi_n$ are only approximately degenerate because $k_{n,1}\neq k_{n,2}$. 
Also implicit is the recognition that in the small $\Delta$ limit, where $v_F/\Delta \geqslant \LL$, the mode $\psi_{E_{i\kappa}}$ is replaced continuously by the extended mode $\phi_{0}$.
Hence, we see that the boundary fermionic mode $\psi_{E_{i\kappa}}$ was once part of a Cooper pair $b_0 = \phi_0 \varphi_0$ that was `broken off' and localized when the system length $\LL$ exceeded the SC coherence length $v_F/\Delta$. Moreover, by changing the sign of $\Delta$, we can make $\varphi_0$ evolve into the topological boundary mode when $|\Delta| >v_F/\LL$. A schematic energy diagram of the diagonalized open wire is shown in Fig.~\ref{fig:energy_diagram}.

Relative to $|\text{BCS}\rangle$, we can define two quasi-degenerate ground states
\begin{align}
|\Omega_0\rangle := |\text{BCS}\rangle, \qquad |\Omega_1\rangle := \psi_{E_{i\kappa}}^\dagger |\text{BCS}\rangle.\label{eq:omega}
\end{align}
This gives $\gamma_0 = \sigma^x$ and $\gamma_\LL = \sigma^y$ when projected onto the quasi-degenerate ground state manifold. Moreover, $|\Omega_0\rangle$ and $|\Omega_1\rangle$ only differ in their correlation functions due to the different occupations of the boundary mode $\psi_{E_{i\kappa}}$. Otherwise, contributions to correlation functions from the bulk extended modes $\phi_n$ and $\varphi_n$ are always identical. In Appendix~\ref{app:corr}, we will discuss in detail the ``topological'' contribution to the correlation functions due to $\psi_{E_{i\kappa}}$.


\subsection{Extreme superconducting limit}\label{sec:fermion_extreme}

It is clear that the topological boundary mode $\psi_{E_{i\kappa}}$ cannot appear in the normal metal phase when $\Delta=0$. In fact, the localized in-gap states already cease to exist whenever $v_F/\Delta \geqslant \LL$. 
Let us now consider the other extreme limit when the SC interaction is infinitely strong with $v_F/\Delta =0$ and $\kappa \rightarrow \infty$. For the in-gap state wavefunctions this gives
\begin{align*}
&a(x)^2 \rightarrow \sum_{m=-\infty}^\infty\delta(x-2m\LL), \\  
&b(x)^2 \rightarrow \sum_{m=-\infty}^\infty \delta(\LL-x-2m\LL).
\end{align*}
Taking the square root, we formally arrive at
\begin{align*}
&a(x) \rightarrow \sum_{m=-\infty}^\infty (-1)^m\sqrt{\delta(x-2m\LL)}, \\
& b(x) \rightarrow \sum_{m=-\infty}^\infty (-1)^m\sqrt{\delta(\LL-x-2m\LL)}.
\end{align*}
The notion of a square root to the Dirac delta function needs some clarification because these are not generalized functions in the conventional Schwartz theory of distributions.%
\footnote{Naively one could try taking the definition of $\sqrt{\delta(x)}$ to be the square root of the limit of appropriately normalized Gaussian functions. However, one would find that the normalization of such functions diminishes too rapidly in the limit such that they (weakly) converge to the zero function within the space of Schwartz generalized functions.}
Rather they require a construction known as non-linear generalized functions in order to have a rigorous definition.\cite{colombeau2011elementary,colombeau2013nonlinear} With this caveat in mind, we recognize that in this artificial limit there exist perfectly localized exact Majorana zero modes at the boundary points $x=0,\LL$. Moreover, these infinitely localized Majorana operators commute exactly with the Hamiltonian $H$ in Eqn.~(\ref{eqn:Ham_unfolded}) in the limit $v_F=0$. This is easy to appreciate since the SC pairing potential exactly vanishes at $x=0,\LL$ but is constant and non-zero everywhere else. 

We should point out that this extreme limit is not technically accessible in a lattice model. This is because the zero correlation length is smaller than the finite lattice spacing, which invalidates the continuum approximation. Nonetheless, there are similarities with the flat band limit of a lattice model, namely that the dispersion is exactly zero and if the model possesses topological boundary modes, these are also exactly lattice-localized zero-energy modes. 

From the point of view of the bulk eigenmodes, one has in this limit $\theta_{k_{n,1}}=\theta_{k_{n,2}} = \pi/4$ for all $n$. This means that
\begin{align*}
k_{n,1} &= \left(n+\frac{1}{2}\right)\frac{\pi}{\LL} - \frac{\pi}{2\LL } = \frac{n\pi}{\LL}, \quad n=1,2,\ldots \\
k_{n,2} &= \left(n+\frac{1}{2}\right)\frac{\pi}{\LL} + \frac{\pi}{2\LL } = \frac{(n+1)\pi}{\LL}, \quad n=0,1,\ldots
\end{align*}
so that 
$k_{n,1} = k_{n-1,2}= n\pi/\LL$ for $n \in \mathbb{N}$.
One can think of this as a rearrangement of modes such that there is a preference to form a Cooper pair out of $\phi_{n}\varphi_{n-1}$. 



\section{Bosonization with closed boundaries}\label{sec:closed_wire_bosonization}

In this section, we bosonize the continuum model with closed periodic and antiperiodic boundary conditions. Contrasting periodic vs. antiperiodic boundary conditions is important because of the intimate relationship between 1D topological superconducting phases and fermion parity switches:\cite{fidkowski2011majorana,keselman2013inducing} Namely, the number of fermion number parity switches (modulo 2) induced by adiabatic flux threading -- which twists boundary conditions --  is the very definition of a $\mathbb{Z}_2$ SPT topological index. Such a fermion parity switching property is also responsible for the $4\pi$ Josephson junction period in the topological phase. 


\subsection{Bosonization preliminaries}

We consider two sources of twisted boundary conditions. The first one is a threaded magnetic flux through the closed wire of length $\LL$, which acts \emph{equally} on left- and right-moving fields, while the second one is a chiral magnetic flux with \emph{opposite} action on left and right movers. 
We write our bosonized fermionic fields in normal-ordered form as\cite{haldane1981luttinger}
\begin{subequations}
\begin{align}
R(x) 
&= \frac{\eta^R}{\sqrt{\LL} } :\e^{i\phi^R(x)} :\e^{i\frac{\pi x}{\LL}(\delta_J + \delta_Q-1)} \nonumber \\ 
&= \frac{\eta^R}{\sqrt{\LL} }:\e^{i[\vartheta(x)-\varphi(x)]}:\e^{i\frac{\pi x}{\LL}(\delta_J + \delta_Q-1)}, \\ \nonumber \\
L(x) 
&= \frac{\eta^L}{\sqrt{\LL} } :\e^{i\phi^L(x)}: \e^{i\frac{\pi x}{\LL}(\delta_J - \delta_Q+1)} \nonumber \\
&= \frac{\eta^L}{\sqrt{\LL} }:\e^{i[\vartheta(x)+\varphi(x)]}:\e^{i\frac{\pi x}{\LL}(\delta_J - \delta_Q+1)},
\end{align}
\label{eqn:bosonized_RL_normal}%
\end{subequations}
where $\phi^{R,L}(x)$ are non-local bosonic chiral fields and $\vartheta(x),\varphi(x)$ are their corresponding local conjugate fields. The $\eta^{R,L}$ are the Klein factors, which we take here to be Hermitian Majorana operators, and $\delta_Q, \delta_J$ represent the effect of twisted boundary conditions.
As Klein factors, $\eta^L=(\eta^L)^\dagger$ and $\eta^R=(\eta^R)^\dagger$ must satisfy
\begin{align}
&\{\eta^R,\eta^L\} = 0,&  &(\eta^R)^2 = (\eta^L)^2 =1.
\end{align}
Their role is to ensure the correct anticommutation statistics between $R$ and $L$ operators.
Normal ordering $:\;:$ is taken with respect to the reference non-interacting Fermi sea (vacuum) $|0\rangle$ that is annihilated by the topological number operators $Q,J$ and positive-frequency bosonic excitation operators defined in the free limit. 
For reference, the unnormal-ordered forms are 
\begin{subequations}
\begin{align}
R(x) 
&= \frac{\eta^R}{\sqrt{2\pi \epsilon} } \e^{i\phi^R(x)} \e^{i\frac{\pi x}{\LL}(\delta_J + \delta_Q)} \nonumber \\
&= \frac{\eta^R}{\sqrt{2\pi \epsilon} }\e^{i[\vartheta(x)-\varphi(x)]}\e^{i\frac{\pi x}{\LL}(\delta_J + \delta_Q)}, \\ \nonumber \\
L(x) 
&= \frac{\eta^L}{\sqrt{2\pi \epsilon} } \e^{i\phi^L(x)} \e^{i\frac{\pi x}{\LL}(\delta_J - \delta_Q)} \nonumber \\
&= \frac{\eta^L}{\sqrt{2\pi \epsilon} }\e^{i[\vartheta(x)+\varphi(x)]}\e^{i\frac{\pi x}{\LL}(\delta_J - \delta_Q)},
\end{align}
\label{eqn:bosonized_RL}%
\end{subequations}
where $\epsilon>0$ is the UV cutoff taken to zero at the end of the calculation. It is important to point out that normal ordering yields additional $\e^{\pm i \frac{\pi x}{\LL}}$ phase factors, and that neglecting these phase factors often leads to the wrong identification of boundary condition type.

In both normal- and unnormal-ordered forms, periodic (R) boundary conditions correspond to 
\begin{align}
\delta_Q + \delta_J = 1 \; \mod 2,
\end{align}
whereas antiperiodic (NS) boundary conditions correspond to
\begin{align}
\delta_Q + \delta_J = 0 \; \mod 2.
\end{align}
For convenience, we limit $\delta_Q$ and $\delta_J$ to integral values, but this is not a physical requirement. 
	\footnote{At special integral values of $\delta_Q$ and $\delta_J$ [e.g., $\delta_Q=1$ and $\delta_J = 2$], the spectrum of momentum states is shifted by quantized integral amounts. One can then singularly `gauge' away the effect of the twisted boundary conditions, essentially by relabeling momentum states. However, because the bosonized Hamiltonian is normal-ordered with respect to a reference vacuum {$|0\rangle$} that remains fixed, the Hamiltonian will not appear to be invariant under the singular gauge transformation.}
Hence we have 
\begin{align}
R(x+\LL) &= R(x)\e^{i\pi(\delta_J+\delta_Q-1)}, \\
L(x+\LL) &= L(x)\e^{i\pi(\delta_J-\delta_Q+1)}.
\end{align}
Note that $\delta_Q$ has an alternative interpretation as a quantized shift in the Fermi momentum $k_F$ in units of $\pi/\LL$. Absorbing the boundary condition shifts $\delta_{Q,J}$ into $Q$ and $J$ is a convention often used by many authors.\cite{von_Delft_Schoeller,loss1992parity} However, because $\delta_Q$ and $\delta_J$ commute with the zero modes of $\varphi(x)$ and $\vartheta(x)$, they do not fluctuate quantum-mechanically, nor do they contain data about actual particle occupation numbers. For this reason, we prefer to keep them separate from $Q$ and $J$. 

The conjugate bosonic fields $\varphi(x)$ and $\vartheta(x)$ satisfy the equal-time commutator
\begin{align}
[\varphi(x),\vartheta(y)] &= -i \pi \varepsilon_1(x-y) 
\label{eqn:phi_theta_com}
\end{align} with
\begin{align}
\varepsilon_1(x-y) &:= \frac{1}{\LL}(x-y) + \frac{i}{2\pi}
\ln \left(\frac{1-\e^{+\frac{i2\pi}{\LL}(x-y)}}{1-\e^{-\frac{i2\pi}{\LL}(x-y)}}\right)\nonumber \\
& = \left\lceil \frac{x-y}{\LL} \right\rceil  -\frac{1}{2},
\end{align}
where $\lceil * \rceil$ denotes the integer ceiling.
The function $\varepsilon_1(x)$ is $\LL$-periodic in $x$, possesses a branch cut at $x=0$ and is a Green's function with respect to $\partial_x$,
\begin{align}
\varepsilon_1'(x-y)= \sum_{m \in \mathbb{Z}}\delta(x-y-m\LL).
\end{align}
Thus, Eqn.~(\ref{eqn:phi_theta_com}) is the finite-length version of the often-quoted bosonized commutation relation\cite{giamarchi2004quantum,gogolin1999bosonization}
\begin{align}
[\varphi(x),\vartheta(y)] &= -\frac{i\pi}{2}\text{sgn}(x-y)
\end{align} 
for $|x-y|<\LL$. The conjugate fields $\vartheta(x), \varphi(x)$ have the mode expansion 
\begin{subequations}
\begin{align}
\varphi(x) &= \varphi_0 - \frac{\pi Q}{\LL} x - \sum_{q \neq 0} \sqrt{\frac{\pi}{2|q|\LL}} (a_q + a_{-q}^\dagger)\e^{iq x}, \\
\vartheta(x) &= \vartheta_0 + \frac{\pi J}{\LL}x + \sum_{q\neq 0} \sqrt{\frac{\pi}{2|q|\LL}}\,\text{sgn}(q)\, (a_q -a_{-q}^\dagger)\e^{iqx},
\end{align}
\label{eqn:phi_theta_mode_expansion}%
\end{subequations}  
where the sums run over $q \in (2\pi/\LL)\mathbb{Z}$ and $[a_q,a_{q'}^\dagger]=\delta_{qq'}$. Here, $\vartheta_0$ and $\varphi_0$ are zero modes such that 
\begin{subequations}
\begin{align}
&[\varphi_0,\vartheta_0]= 0 ,\quad
[\varphi_0,J] = -[\vartheta_0,Q] = i, \\ \nonumber \\  
&[Q,J]=0, \quad [\varphi_0,Q]= [\vartheta_0,J]=0.
\end{align}
\label{eqn:boson_zero_comm}%
\end{subequations}
It is important to emphasize that the integer-quantized operators $Q$ and $J$ are not completely independent and must satisfy the parity selection rule 
\begin{align}
F:= (-1)^{Q} \equiv (-1)^J \quad \Longleftrightarrow \quad Q = J \text{ mod }2,
\label{eqn:parity_rule}
\end{align}
where $F$ is the fermion number parity operator. This is a necessary constraint given the fermionic field representations of $Q$ and $J$ introduced in Eqns.~(\ref{eqn:Q_fermion}) and (\ref{eqn:J_fermion}).

More importantly, the angular bosonic fields $\varphi$ and $\vartheta$ have to be compactified\cite{klassen1993sine} with radius $\pi$,
\begin{align}
\varphi(x) \sim \varphi(x) + \pi, \qquad 
\vartheta(x) \sim \vartheta(x) + \pi,
\label{eqn:compact_radius}
\end{align} 
in line with the mode expansion in Eqns.~(\ref{eqn:phi_theta_mode_expansion}). 

Before bosonizing the Hamiltonian in Eqn.~(\ref{eqn:Ham_fermion}), we introduce the following equivalent expressions of the bosonic fields
\begin{subequations}
\begin{align}
\varphi(x) &= \varphi_0 - \frac{\pi Q }{\LL}x - \pi \int_{-\LL/2}^{\LL/2}\zeta_1(x-y) \rho(y)\,\mathrm{d}y, \\
\vartheta(x) &= \vartheta_0 + \frac{\pi J }{\LL}x + \pi \int_{-\LL/2}^{\LL/2} \zeta_1(x-y) j(y)\,\mathrm{d}y, 
\end{align}
\label{eqn:phi_theta_rho_j}%
\end{subequations}
where $\zeta_1$ defined as
\begin{align}
\zeta_1(x-y) &:= \varepsilon_1(x-y) - \frac{x-y}{\LL} \equiv \sum_{n>0} \frac{\sin\left(\frac{2\pi n}{\LL}(x-y)\right)}{n\pi}
\end{align}
is the descending saw-tooth function of period $\LL$ such that
\begin{align}
\zeta_1'(x) = -\frac{1}{\LL}+\sum_{m\in \mathbb{Z}} \delta(x+m\LL).
\end{align}
The physical charge and current densities are given by
\begin{subequations}
\begin{align}
\rho_\text{tot}(x) &= -\tfrac{1}{\pi}\partial_x \varphi(x) = \tfrac{Q}{\LL} + \rho(x), \\
j_\text{tot}(x) &= +\tfrac{1}{\pi}\partial_x \vartheta(x) = \tfrac{J}{\LL} + j(x).
\end{align}
\end{subequations}
The derivation of these uncommon but useful bosonization expressions is extensively described in Appendix \ref{app:bosonic_action}. There, it is also shown that the fields $\rho(x)$ and $j(x)$ satisfy the finite-length current algebra commutation relation
\begin{align}
[\rho(x),j(y)] = -\frac{i}{\pi} \sum_{m \in \mathbb{Z}}\delta'(x-y-m\LL) = -\frac{i}{\pi}\varepsilon''(x-y).
\end{align}

Finally, we must emphasize a redundant $\mathbb{Z}_2$ `gauge symmetry' that is always present with this type of bosonization convention. Indeed, in the bosonization expressions of Eqns.~(\ref{eqn:bosonized_RL_normal}), the Klein factors and zero modes always appear in the combinations\footnote{To add confusion to matters, these products of operators are sometimes also referred to as Klein factors (see, e.g., Ref.~\onlinecite{von_Delft_Schoeller})} 
\begin{align}
\eta^R \e^{i (\vartheta_0 + \varphi_0)}, \qquad
\eta^L \e^{i (\vartheta_0 - \varphi_0)}.
\label{eqn:Z2_vertex}
\end{align}
These combinations are \emph{invariant} under the discrete transformation
\begin{align*}
&\eta^{R} \rightarrow - \eta^{R}, \qquad \eta^{L} \rightarrow - \eta^{L}, \\
&\varphi_0 \rightarrow \varphi_0, \quad \qquad \vartheta_0 \rightarrow \vartheta_0 + \pi. 
\end{align*}
Roughly speaking, the combinations of operators in Eqn.~(\ref{eqn:Z2_vertex}) are $\mathbb{Z}_2$ slave-particle fractionalizations\cite{fradkin2013field,wen2004quantum} of the fermionic operators that raise or lower the fermion number. This internal $\mathbb{Z}_2$ gauge symmetry is generated by the $\mathbb{Z}_2$ charge operator
\begin{align}
\xi := i \eta^L \eta^R (-1)^Q, \qquad \xi^2 = 1. 
\end{align}
Hence, the nominal Hilbert space is always twice as big as compared to the physical Hilbert space of the underlying chiral fermions. Fortunately, we can project onto the physical Hilbert space by selecting a gauge fixing choice. In this instance, a convenient choice is just to fix the gauge to $\xi=1$ such that \begin{align}
F \equiv (-1)^Q = i\eta^L\eta^R.
\label{eqn:Z2_gauge_fix}
\end{align}


\subsection{Bosonized Hamiltonian and action}

Applying this bosonization dictionary to the fermionic Hamiltonian given in Eqn.~(\ref{eqn:Ham_fermion}) in the case of coincident Fermi points $k_F=0$ gives
\begin{subequations}
\begin{align}
H &= \int_{-\LL/2}^{\LL/2} \left(\mathcal{H}_0 + \mathcal{H}_1 \right)\mathrm{d}x , 
\\ \nonumber \\
\mathcal{H}_0 &= \frac{v_F}{2\pi}:\left\{[\partial_x \varphi(x)]^2+[\partial_x \vartheta(x)]^2\right\}: \nonumber \\ &\quad+ \frac{\pi v_F}{\LL^2}[ Q \, (\delta_Q-1)  +  J \, \delta_J], \\ \nonumber \\
\mathcal{H}_1 
&=\left(\frac{2\mathrm{M}}{\LL}\right) F
:\sin\left(2\varphi(x) -\left(\tfrac{2\pi}{\LL}x\right)\delta_Q \right): \nonumber \\
&\quad- \left(\frac{2\Delta}{\LL}\right) F :\cos\left(2\vartheta(x) + \left(\tfrac{2\pi}{\LL}x\right)\delta_J\right):.
\end{align}
\label{eqn:Ham_boson_PBC}%
\end{subequations}
We have chosen to focus on the $k_F=0$ case since the competition between backscattering and superconducting interactions is most intense at half-filling. However, this comment only applies to the Kitaev wire interpretation of the model which remains the most convenient for our purpose. By contrast, the realization in Rashba nanowires is more complicated and involves charge/spin density order set by $k_F$.~\cite{Klinovaja2012,Klinovaja2015}

Very often, authors\cite{giamarchi2004quantum,gogolin1999bosonization} neglect the effect of Klein factors by setting $F\equiv i \eta^L \eta^R=1$. However, this can lead to mistakes because of the parity switching effect. In fact, one can already appreciate\cite{fidkowski2011majorana} from Eqn.~(\ref{eqn:Ham_boson_PBC}c) the role that twisted boundary conditions ($\delta_Q,\delta_J$) and $F$ will play in deciding the pinned values of the fields $\varphi$ and $\vartheta$. Thus we must allow $F$ to be a degree of freedom even though it is an integral of motion.

Although exactly solvable in the fermionic representation, the bosonized Hamiltonian in Eqn.~(\ref{eqn:Ham_boson_PBC}) cannot be so transparently dealt with. Nevertheless, one can still gain a great deal of intuition from the effective low-energy approach. This is best carried out with an action/functional approach. As is derived in Appendix \ref{app:bosonic_action}, the real-time classical action associated to the model for fixed fermion parity $F=\pm 1$ can be expressed as $S_\text{tot}[\varphi,\vartheta] := S_c[\rho,j] + S_0[\varphi,\vartheta] + S_1[\varphi,\vartheta]$, where
\begin{widetext}
\begin{subequations}
\begin{align}
S_c[\rho,j] &:= -\pi \int \mathrm{d}t \int_{-\LL/2}^{\LL/2} \mathrm{d}x \int_{-\LL/2}^{\LL/2} \mathrm{d}y \; \zeta_1(x-y) \,j(x)\, \partial_t \rho(y) -\int \mathrm{d}t\int_{-\LL/2}^{\LL/2} \mathrm{d}x \;\frac{\pi v_F}{2}[\rho(x)^2 + j(x)^2], \\
S_0[\varphi,\vartheta] &:= \int \mathrm{d}t  \left(
J\, \partial_t \varphi_0 - Q \, \partial_t \vartheta_0 - \frac{\pi v_F }{2\LL} (J^2 + Q^2) - \frac{\pi v_F}{\LL}[ Q (\delta_Q-1)  +  J \, \delta_J]
\right), \\
S_1[\varphi,\vartheta] &:= \int \mathrm{d}t \int_{-\LL/2}^{\LL/2} \mathrm{d}x\left\{
\left(\tfrac{2\Delta}{\LL}\right) F \cos\left[2\vartheta(x) + \left(\tfrac{2\pi}{\LL}x\right)\delta_J\right]-\left(\frac{2\mathrm{M}}{\LL}\right) F
\sin\left[2\varphi(x) -\left(\tfrac{2\pi}{\LL}x\right)\delta_Q \right] 
\right\},
\end{align}
\label{eqn:S_total_boson_PBC}%
\end{subequations}
\end{widetext}
and the local conjugate fields $\varphi(x)$ and $\vartheta(x)$ are related to $\rho$, $j$, $\varphi_0$, $\vartheta_0$, $Q$, and $J$ by Eqns.~(\ref{eqn:phi_theta_rho_j}). Compared to Eqn.~(\ref{eqn:S_usual}a), we have now included additional cosine potentials that arise from the fermionic mass terms. 
Note that in order to derive the correct classical action from the quantum Hamiltonian and the canonical commutation relations, one must first decide on an operator ordering convention. As detailed in Appendix \ref{app:bosonic_action}, we choose to work in the convention that the non-zero bosonic modes $a^{(\dagger)}_q$ are Wick normal-ordered, while the zero modes $\vartheta_0,\varphi_0$ and their charges $J,Q$ are $qp$-ordered, where the angular operators act as the $q$'s and the charges as the $p$'s. This yields cosine potentials that are independent of the UV cutoff. 

We now proceed by semi-classically examining the action $S_\text{tot}$ when either one of the cosine terms dominates. By seeking homogeneous saddle points of the action, we set $\rho=0$ and $j=0$, which gives the simplified effective Lagrangian
\begin{align}
L_\text{eff} &= 
J\, \partial_t \varphi_0 - Q \, \partial_t \vartheta_0 \nonumber\\&\hspace{4mm}- \tfrac{\pi v_F }{2\LL} [J(J+2\delta_J) + Q(Q+2\delta_Q-2)] \nonumber \\
& \hspace{4mm}- F\left\{
2\mathrm{M}\left(\frac{\sin[\pi(Q+\delta_Q)]}{\pi(Q+\delta_Q)}\right) \sin(2\varphi_0) \right. \nonumber \\
&\hspace{6mm}\left.\qquad-2\Delta \left(\frac{\sin[\pi(J+\delta_J)]}{\pi(J+\delta_J)}\right)\cos(2\vartheta_0)
\right\},
\label{eqn:L_eff_ring}
\end{align}
where we have performed the finite integrals in the cosine terms. In general, for simultaneously non-zero couplings, both potentials will compete and we can expect tunneling between saddle-point minima of the two potentials. In principle, once the saddle points have been determined, one could reincorporate fluctuations
by expanding about the saddle-point solution in $\rho$ and $j$ perturbatively.\cite{zinn1996quantum,fradkin2013field} 

Next, we would like to define the fermion parity quantum eigenstates which will correspond to possible saddle points of pinned states. These are conventionally defined as
\begin{subequations}
\begin{align}
&|\vartheta_0 = \theta\rangle_+\, := \sum_{m \text{ even}}\e^{i m \theta}|Q=m\rangle, \\
&|\vartheta_0 = \theta\rangle_-\, := \sum_{m \text{ odd}}\e^{i m \theta}|Q=m\rangle,   \\
&|\varphi_0 = \phi\rangle_+\, := \sum_{m \text{ even}}\e^{-i m \phi}|J=m\rangle, \\
&|\varphi_0 = \phi\rangle_-\, := \sum_{m \text{ odd}}\e^{-i m \phi}|J=m\rangle,
\end{align}
\end{subequations}
and satisfy 
\begin{align}
&|\vartheta_0 = (\theta+ \pi)\rangle_\pm = \pm |\vartheta_0 = \theta\rangle_\pm, \\
&|\varphi_0 = (\phi+ \pi)\rangle_\pm = \pm |\varphi_0 = \phi\rangle_\pm.
\end{align}
Hence these quantum states obey the compactification radius conditions of Eqn.~(\ref{eqn:compact_radius}). Moreover, 
\begin{align}
&\e^{i\vartheta_0}|\vartheta_0 = \theta\rangle_\pm = \e^{i\theta}|\vartheta_0 = \theta\rangle_\mp, \\
&\e^{-i\varphi_0}|\varphi_0 = \phi\rangle_\pm = \e^{-i\phi}|\varphi_0 = \phi\rangle_\mp,
\end{align}
since raising and lowering obviously changes fermion parity. More importantly, this means that it is \emph{possible} to pin both $\varphi_0$ and $\vartheta_0$ simultaneously [since they commute from Eqns.~(\ref{eqn:boson_zero_comm})] but only in the specific tensor product combinations
\begin{align}
|\vartheta_0 = \theta\rangle_+  \otimes|\varphi_0 = \phi\rangle_+, \quad 
|\vartheta_0 = \theta\rangle_-  \otimes|\varphi_0 = \phi\rangle_-,
\end{align}
in accordance to the fermion parity rule. We should emphasize that these $\vartheta_0,\varphi_0$ `vacuum states' are meant to represent variational ground states in the limit of infinitely strong pinning. 

In the following subsections, we shall examine the topologically trivial and non-trivial phases separately. We do so by considering the cases where $\mathrm{M} \neq 0$, $\Delta =0 $ and $\Delta \neq 0$, $\mathrm{M} =0 $ in turn.

\subsubsection{Trivial phase}

The saddle point of the trivial phase can be characterized by an effective theory when $\Delta=0$, leaving the potential derived from fermionic backscattering as the only interaction term in Eqn.~(\ref{eqn:L_eff_ring}). In this limit, $Q$ itself becomes an integral of motion and hence a conserved quantum number in the effective quantum description. Hence in minimizing the potential energy we can set $Q= -\delta_Q$ with $\delta_Q=0,1$, which also implies $F=(-1)^{\delta_Q}$. This yields the effective Hamiltonian
\begin{align}
H_\text{eff} &= 
\frac{\pi v_F}{2\LL}\left[J(J+2\delta_J)+\delta_Q(2-\delta_Q)\right]
\nonumber\\&\hspace{5mm}+2\mathrm{M}(-1)^{\delta_Q}\sin(2\varphi_0).
\end{align}
Varying $\delta_Q$ can be seen to be equivalent to varying the chemical potential (Fermi momentum) by non-quantized values. For this reason, we shall set $\delta_Q = 1$ and let the twisted boundary conditions merely specify changes in $\delta_J$. Explicitly, we have periodic (antiperiodic) boundary conditions for $\delta_J=0$ ($\delta_J=1$). 
Note that this automatically determines the fermion parity as $F=(-1)^{\delta_Q}=-1$ and the charge as $Q=-1$. Nevertheless, it is important to remember that the fermion parity has not actually been fixed. Rather, the low-energy sector just prefers to have an even parity in the absence of twisting by a chiral magnetic field.  \\

\paragraph*{\underline{Periodic boundary conditions:}}

In the Ramond (R) case we have $\delta_J=0$, leading to the effective Hamiltonian
\begin{align}
H^{(R)}_\text{eff} &= 
\frac{\pi v_F}{2\LL}\left(J^2+1\right)
-2\mathrm{M}\sin(2\varphi_0),
\label{eqn:Ham_trivial_PBC}
\end{align}
which resembles the Hamiltonian of a Josephson junction. It is minimized by $J=0$ and $\varphi_0 = \pi/4$ mod $\pi$. From the effective Lagrangian, we still have the Poisson bracket $\{\varphi_0,J\}=1$ and hence in the quantum theory we must have that $[\varphi_0,J]=i$. In the limit of strong pinning, where $\mathrm{M}  \gg v_F / \LL$, the variationally optimized non-degenerate pinned ground state is then
\begin{align}
|\Psi^{(R)}_\mathrm{M}\rangle =|Q=-1\rangle  \otimes |\varphi_0 = \tfrac{\pi}{4}\rangle_-.
\end{align}

\paragraph*{\underline{Antiperiodic boundary conditions:}} 

Turning to the Neveu-Schwarz (NS) case, we now have the condition $\delta_J=1$. This then results in the effective Hamiltonian
\begin{align}
H^{(NS)}_\text{eff} &= 
\frac{\pi v_F}{2\LL}\left[J(J+2)+1\right]
-2\mathrm{M}\sin(2\varphi_0)
\label{eqn:Ham_trivial_ABC}
\end{align}
with the same variational ground state in the strongly pinned limit,
\begin{align}
|\Psi^{(NS)}_\mathrm{M}\rangle =|Q=-1\rangle  \otimes |\varphi_0 = \tfrac{\pi}{4}\rangle_- ,
\end{align}
and the same fermion parity $F=-1$.
In short, we do not see a switch in fermion number parity of the variational ground state whenever the quasiperiodic boundary conditions are twisted by a magnetic field threaded through the closed 1D system. 

\subsubsection{Topological phase}

The saddle point of the topologically non-trivial phase can be characterized by an effective theory at $\mathrm{M}=0$. In this limit, $J$ becomes an integral of motion and hence a conserved quantum number in the effective quantum description. From minimizing the potential energy, we set $J= -\delta_J$, where $\delta_J=0,1$. In contrast to the trivial phase, taking $\delta_Q=1$ does not fix the fermion parity. In fact, the fermion parity of the effective low-energy sector is now explicitly set by the choice of twisted boundary condition with $F=(-1)^{\delta_J}$. The effective Hamiltonian that results is 
\begin{align}
H_\text{eff} &= 
\frac{\pi v_F}{2\LL}\left(\delta_J^2+Q^2\right)
-2\Delta(-1)^{\delta_J}\cos(2\vartheta_0).
\label{eqn:Ham_top_PBC_gen}
\end{align}\\

\paragraph*{\underline{Periodic boundary conditions:}} 

Setting $J=\delta_J=0$ and hence $F=1$, we arrive at another Josephson-junction type Hamiltonian
\begin{align}
H^{(R)}_\text{eff} &= 
\frac{\pi v_F}{2\LL}Q^2-2\Delta\cos(2\vartheta_0),
\label{eqn:Ham_top_PBC}
\end{align}
minimized by $Q=0$ and $\vartheta_0 = 0$ mod $\pi$. Again, in the strong pinning limit, we have the variational ground state
\begin{align}
|\Psi^{(R)}_\Delta\rangle = |\vartheta_0 = 0\rangle_+ \otimes |J=0\rangle.
\end{align}

\paragraph*{\underline{Antiperiodic boundary conditions:}}

Now instead setting $\delta_J=1$ and hence $F=-1$, we arrive at a different Josephson-junction type Hamiltonian
\begin{align}
H^{(NS)}_\text{eff} &= 
\frac{\pi v_F}{2\LL}(Q^2+1)+2\Delta\cos(2\vartheta_0),
\label{eqn:Ham_top_ABC}
\end{align}
minimized by $Q=0$ and $\vartheta_0 = \pi/2$ mod $\pi$. In the strongly pinned limit, the variational ground state is 
\begin{align}
|\Psi^{(NS)}_\Delta\rangle = |\vartheta_0 = \frac{\pi}{2}\rangle_- \otimes |J=-1\rangle.
\end{align}
Physically, the non-zero $J$ is a result of attempting to `screen' the applied magnetic flux, much like in the Luttinger-liquid problem.\cite{loss1992parity} Hence, there is now fermion number parity switching which is indicative of a topologically non-trivial phase. 


\subsection{Mathieu function solutions to the effective Hamiltonians}\label{sec:mathieu_functions}

In principle, we should include the effects of fluctuations from the ``kinetic terms'' that are proportional to $Q^2$ and $ J^2$. This could be carried out to lowest order by using the harmonic approximation on $\sin (2\varphi_0)$ and $\cos(2\vartheta_0)$ about their pinning values. In the lowest non-trivial order, one then obtains a simple harmonic oscillator ground state eigenfunction as the variational ground state. More accurately, we can use solutions of the Mathieu equation to obtain exact normalizable eigenstates\cite{devoret2004superconducting} to the Josephson-junction like Hamiltonians of Eqns.~(\ref{eqn:Ham_trivial_PBC}), (\ref{eqn:Ham_trivial_ABC}), (\ref{eqn:Ham_top_PBC}) and (\ref{eqn:Ham_top_ABC}). The Mathieu equation is\cite{abramowitz1965formulas} 
\begin{align}
\frac{\mathrm{d}^2 y(z)}{\mathrm{d}z} +[a-2q \cos (2z)]y(z) = 0,
\label{eqn:Mathieu}
\end{align}
where $a$ and $q$ are constants. It can be seen that all of the Eqns.~(\ref{eqn:Ham_trivial_PBC}), (\ref{eqn:Ham_top_PBC}) and (\ref{eqn:Ham_top_ABC}) may be brought into this form, where $z = \varphi_0$ or $z=\vartheta_0$, $a$ is proportional to the energy, and $q$ is proportional to $\mathrm{M}$ or $\Delta$. Equation~(\ref{eqn:Ham_trivial_ABC}) can also be treated this way but requires a Floquet-type solution to the Mathieu equation which we will not discuss here. For our purposes we require the periodic solutions $y(u+\pi) = +y(u)$ (even fermion parity) and the doubly-periodic solutions $y(u+\pi) = -y(u)$ (odd fermion parity). These are real eigensolutions for real $q$ and depend smoothly on $q$. They are denoted by $ce_r(z,q)$, $se_{r+1}(z,q)$ where $r = 0,1,2,\ldots$. Their eigenvalues (characteristics) are conventionally denoted by $a_r(q)$ and $b_r(q)$ such that 
\begin{align}
&\partial_z^2 ce_r(z,q) +[a_r(q)-2q \cos (2z) ] ce_r(z,q) =0, \\
&\partial_z^2 se_r(z,q) +[b_r(q)-2q \cos (2z) ] se_r(z,q) =0.
\end{align}

Now specializing to the topologically non-trivial phase, the low-energy effective (unnormalized) ground state wavefunction and energy for the periodic (R) case is
\begin{align}
|\Psi^{(R)}_\Delta\rangle &= \sqrt{\frac{2}{\pi}}\int_0^\pi \mathrm{d}\vartheta \; ce_{0}(\vartheta-\tfrac{\pi}{2},\tfrac{2\Delta \LL}{\pi v_F})\;|\vartheta \rangle_+ \otimes |J=0\rangle, \\
E^{(R)}_\Delta &=  \frac{\pi v_F}{2\LL}a_0(\tfrac{2\Delta \LL}{\pi v_F}).
\end{align}
For the antiperiodic (NS) case we instead have
\begin{align}
|\Psi^{(NS)}_\Delta\rangle &= \sqrt{\frac{2}{\pi}}\int_0^\pi \mathrm{d}\vartheta \; se_{1}(\vartheta,\tfrac{2\Delta \LL}{\pi v_F})\; |\vartheta \rangle_- \otimes |J=-1\rangle, \\
E^{(NS)}_\Delta &= \frac{\pi v_F}{2 \LL} +  \frac{\pi v_F}{2\LL}b_1(\tfrac{2\Delta \LL}{\pi v_F}).
\end{align}
Their difference is 
\begin{align}
\delta E_{\Delta} &= E^{(NS)}_\Delta - E^{(R)}_\Delta \nonumber \\
&=\frac{\pi v_F }{2\LL}\left(1 + b_1(\tfrac{2\Delta \LL}{\pi v_F}) - a_0(\tfrac{2\Delta \LL}{\pi v_F})\right) \nonumber \\ 
&\sim  \frac{\pi v_F }{2\LL} \left(1 + 2^5 \sqrt{\frac{2}{\pi}}\left(\frac{2\Delta \LL}{\pi v_F}\right)^{3/4}\e^{-4 \sqrt{\frac{2\Delta \LL}{\pi v_F}}}\right)
\end{align}
as $\LL \rightarrow \infty$, where we have used the asymptotic relation\cite{abramowitz1965formulas} 
\begin{align}
b_{r+1}(q) - a_r(q) \sim \frac{2^{4r+5}}{r!} \sqrt{\frac{2}{\pi}} \,q^{\frac{r}{2}+\frac{3}{4}}\e^{-4\sqrt{q}}.
\label{eqn:b_a_asymp}
\end{align}
This demonstrates the fermion parity switching effect between NS and R variational solutions to the topologically non-trivial phase, with an energy splitting of $\pi v_F/ (2\LL)$ to leading order.  
Furthermore, it is interesting to note that as $\LL \rightarrow \infty$, the eigenfunctions $|\Psi^{(NS)}_\Delta\rangle$ and $|\Psi^{(R)}_\Delta\rangle$ turn into square-root Dirac delta functions localized at their respective pinning values, meaning that their respective squared amplitudes -- in angle space -- approach that of a Dirac delta function in the infinitely pinned limit.



\section{Bosonization with open boundaries}\label{sec:open_wire_bosonization}

In this section, we bosonize the fermionic model with open boundaries using the unfolding procedure of Sec.~\ref{sec:open_bc}. 


\subsection{Bosonization preliminaries}

Recall the linearization 
\begin{align}
\Psi(x) = R(x)\e^{ik_F x} + L(x)\e^{-ik_F x} 
\end{align}
and the number-conserving boundary conditions  
\begin{align}
R(0) + \e^{i\delta_0 }L(0) &= 0, \\
R(\LL) + \e^{i(\delta_\LL-2k_F \LL) }L(\LL) &= 0. 
\end{align}
We take $\delta_0 = \delta_\LL =0$ and $k_F \in \frac{\pi}{2\LL}+ \frac{\pi}{\LL}\mathbb{N}$, which gives us Dirichlet boundary conditions $\Psi(0)=\Psi(\LL)=0$ and an antiperiodic (NS) $R(x)$ field in $[-\LL,\LL]$, 
\begin{align}
&R(x+2\LL) = -R(x), \\
&L(x) \equiv - R(-x). 
\end{align} 
Then we bosonize the \emph{single} chiral fermionic field $R(x)$, such that
\begin{subequations}
\begin{align}
R(x) &= \frac{\eta^R}{\sqrt{2\LL}}:\e^{i\phi^R(x)}:\e^{-i \frac{\pi x }{2\LL}}, \\
R^\dagger(x) &= \frac{\eta^R}{\sqrt{2\LL}}:\e^{-i\phi^R(x)}: \e^{- i \frac{\pi x }{2\LL}}.
\end{align}
\label{eqn:bosonized_RL_open}%
\end{subequations}
The additional phase twisting of $\e^{-i \frac{\pi x }{2\LL}}$ that is equal for both $R(x)$ and $R^\dagger(x)$ is a result of normal ordering the zero mode. The chiral field $\phi^R(x)$ has the mode expansion\cite{fabrizio1995interacting,mattsson1997properties}
\begin{align}
\phi^R(x) := \vartheta_0 + \frac{\pi Q x}{\LL} + \sum_{n>0} 
\frac{1}{\sqrt{n}}\left(a_n \e^{i\frac{n\pi x}{\LL}} + a_n^\dagger \e^{-i\frac{n\pi x}{\LL}} \right)
\end{align}
with $[\vartheta_0,Q]=-i$ and $[a_n,a^\dagger_{n'}] = \delta_{nn'}$. The compactification radius of $\phi^R$ remains to be $2\pi$,
\begin{align}
\phi^R(x) \sim \phi^R(x) + 2\pi,
\end{align}
which leads to the quantization of the eigenvalues of $Q$ in $\mathbb{Z}$. The mode expansion yields the usual non-local commutation relation for a system of length $2\LL$, i.e.
\begin{align}
[\phi^R(x),\phi^R(y)] 
&= i2 \pi\,  \varepsilon_2(x-y) \nonumber \\ 
&\equiv i2\pi \left\lceil \frac{x-y}{2\LL} \right\rceil  -{i \pi}\nonumber \\
&= i \pi \,\text{sgn}(x-y) \quad \text{for } |x-y| < 2\LL.
\end{align}
The kernel function $\varepsilon_2(x-y)$ is defined analogously to $\varepsilon_1(x-y)$, except that it has a period of $2\LL$ instead of $\LL$. 

A crucial exact operator identity is the product of vertex functions
\begin{align}
&:\e^{i \alpha \phi^R(x)}:\;:\e^{ i\beta \phi^R(y)}: \nonumber \\
&= \left[ \e^{-i\frac{\pi x}{\LL}} - \e^{-i\frac{\pi y}{\LL}}
\right]^{\alpha \beta} :\e^{i [\alpha \phi^R(x) + \beta \phi^R(y)]}:
\label{eqn:OBC_OPE} 
\end{align}
for $\alpha,\beta\in\mathbb{Z}$. This is just the conformal field theory operator product expansion (OPE) for normal-ordered vertex functions of a chiral field $\phi^R(x)$ in a system of length $2\LL$. It can be used to verify many other bosonization identities like the canonical anticommutation relations for an antiperiodic $R(x)$,
\begin{align}
\{R(x),R^\dagger(y)\} = \sum_{m=-\infty}^\infty (-1)^m \delta(x-y-2m\LL).
\end{align}
Furthermore, the unfolding relation between $R(x)$ and $L(x)$ gives 
\begin{subequations}
\begin{align}
L(x) &= -\frac{\eta^R}{\sqrt{2\LL}}:\e^{i\phi^R(-x)}: \e^{i \frac{\pi x}{2\LL}},\\
L^\dagger(x)
&= -\frac{\eta^R}{\sqrt{2\LL}}:\e^{-i\phi^R(-x)}:\, \e^{i \frac{\pi x }{2\LL}}.
\end{align}
\end{subequations}
It is crucial to realize that both $R$ and $L$ are the \emph{same} field such that $\eta^L \equiv - \eta^R$.  

Next, we define the angular bosonic fields
\begin{align}
\vartheta(x) := \frac{\phi^R(x)+ \phi^R(-x)}{2}, \quad
\varphi(x) := -\frac{\phi^R(x)-\phi^R(-x)}{2},
\end{align}
which are local, $[\varphi(x),\varphi(y)] = [\vartheta(x),\vartheta(y)] = 0$, since $\varepsilon_2$ is an odd function. However, they remain mutually non-local because
\begin{align}
[\varphi(x),\vartheta(y)]  &= -i\pi \left[
\varepsilon_2(x-y) + \varepsilon_2(x+y)
\right]   \\
&= -i \pi \,\Theta(x-y)\quad \text{for}\;\; 0< x,y<\LL.\nonumber 
\end{align}
Their mode expansions take the form 
\begin{subequations}
\begin{align}
\vartheta(x) &= \vartheta_0 + \sum_{n>0} \frac{1}{\sqrt{n}}\left(
a_n + a_n^\dagger
\right) \cos\left(\frac{n \pi x}{\LL}\right),\\
\varphi(x) &= -\frac{\pi Q}{\LL}x +i \sum_{n>0}\frac{1}{\sqrt{n}}\left(a_n - a_n^\dagger \right) \sin\left(\frac{n\pi x}{\LL}\right).
\end{align}
\label{eqn:phi_theta_mode_expansion_OBC}%
\end{subequations}
Hence, $\vartheta$ $(\varphi)$ satisfies homogeneous Neumann (Dirichlet) boundary conditions at $x = 0,\LL$. Both fields remain periodic in $x\in [-\LL,\LL]$ modulo $2\pi$ such that their compactification radius is $2\pi$ and not $\pi$:
\begin{align}
\varphi(x) \sim \varphi(x) + 2\pi, \qquad \vartheta(x) \sim \vartheta(x) + 2\pi.
\end{align}
The Neumann condition on $\vartheta(x)$ means that the current density 
\begin{align}
j(x) := \frac{1}{\pi} \partial_x \vartheta(x)
\end{align}
vanishes at the boundaries, $j(0)=j(\LL)=0$. Likewise the fluctuating charge density defined by
\begin{align}
\rho(x) := -\frac{1}{\pi}\partial_x \varphi(x) - \frac{Q}{\LL} 
\end{align}
obeys the integral sum rule $\int_0^\LL \rho(x)\, \mathrm{d}x = 0$. It is clear that there cannot be any topological windings in $\vartheta(x)$ leading to the conclusion that $J=0$ always. This is entirely consistent with our earlier assertions in Sec.~\ref{sec:open_bc} regarding the incompatibility of $U(1)_{R-L}$ symmetry and open boundary conditions.  

Next, we have the mode expansions
\begin{align}
\rho(x) &= -\frac{i}{\LL} \sum_{n>0} \sqrt{n}(a_n - a_n^\dagger )\cos\left(\frac{n\pi x}{\LL}\right), \label{eqn:rho_mode}\\
j(x) &= -\frac{1}{\LL}\sum_{n>0}\sqrt{n}(a_n + a_n^\dagger )\sin\left(\frac{n\pi x}{\LL}\right),
\label{eqn:j_mode}
\end{align}
from which we can directly verify that they satisfy the modified current algebra
\begin{align}
&[\rho(x),j(y)] \nonumber \\
&= -\frac{i}{\pi}\sum_{m=-\infty}^\infty\left[
\delta'(x-y+ 2m\LL) + \delta'(x+y+2m\LL)
\right], 
\end{align}
which now has \emph{two} Schwinger terms. 
Alternatively, we can express $\phi^R$ in terms of $\rho$ and $j$:
\begin{subequations}
\begin{align}
&\frac{1}{\pi}\partial_x \phi^R(x) = j(x) + \rho(x) + \frac{Q}{\LL}, \\
&\frac{1}{\pi}[\partial_x \phi^R](-x) = -j(x) +\rho(x) + \frac{Q}{\LL}.
\end{align}
\label{eqn:jrhophiR}%
\end{subequations}
Accompanying this is the functional relation
\begin{align}
&\phi^R(x) = \vartheta_0 + \frac{\pi Q}{\LL}x + \pi \int^{\LL}_{-\LL} \zeta_2(x-y) [j(y)+\rho(y)]\mathrm{d}y, 
\end{align}
where 
\begin{align}
\zeta_2(x-y) &:= \varepsilon_2(x-y) - \frac{x-y}{2\LL}\nonumber \\ 
&\equiv \sum_{n>0}\frac{\sin\left[\frac{n \pi }{\LL}(x-y)\right]}{\pi n}
\end{align}
is the descending saw-tooth function of period $2\LL$. These expressions also lead to the following equivalent expressions for the angular fields:
\begin{subequations}
\begin{align}
\vartheta(x) &= \vartheta_0 + \pi \int_{-\LL}^\LL \zeta_2(x-y)j(y)\, \mathrm{d}y, \\
\varphi(x) &= -\frac{\pi Q}{\LL}x - \pi \int_{-\LL}^\LL \zeta_2(x-y) \rho(y)\,\mathrm{d}y.
\end{align}
\label{eqn:theta_phi_rho_j}%
\end{subequations}
The convolutions with the saw-tooth function $\zeta_2$ effectively implement anti-derivatives. In short, three things to note are: 
\begin{itemize}
\item[(1)]{$J$ and $\varphi_0$ are now absent, }
\item[(2)]{the parity rule $(-1)^Q = (-1)^J$ is no longer relevant,}
\item[(3)]{the compactification radii of the $\vartheta,\varphi$ fields are now doubled to $2\pi$ from $\pi$ as previously in Sec.~\ref{sec:closed_wire_bosonization}.}
\end{itemize}
These new aspects make the open system drastically different from the closed one from the point of view of bosonization. 


\subsection{Bosonized Hamiltonian and action}\label{sec:open_wire_Ham}

Now we apply the open boundary bosonization dictionary to the fermionic Hamiltonian in Eqn.~(\ref{eqn:Ham_fermion}). The normal-ordered kinetic term is given by
\begin{align}
H_0 &= \frac{ v_F}{4\pi }\int_0^\LL \mathrm{d}x :\left([\partial_x \phi^R(x)]^2 + [\partial_x \phi^R(-x)]^2\right): \nonumber \\
&=\frac{\pi v_F}{2}\int_0^\LL \mathrm{d}x :\left( j(x)^2 + \rho(x)^2 \right):
+ \frac{\pi v_F }{2\LL}Q^2,
\end{align}
and is missing the inductive $J^2$ term. 

Next, we consider the backscattering terms like $R^\dagger L$. An application of the product formula Eqn.~(\ref{eqn:OBC_OPE}) yields
\begin{align}
R^\dagger(x) L(x) &= -\frac{i}{4\LL}\left[ \frac{:\e^{i2\varphi(x)}:}{\sin(\tfrac{\pi x}{\LL})}\right].
\end{align}
Taking the limits $x\rightarrow 0,\LL$ produces divergences and suggests the need to fermion normal order [cf. ~Eqn.~(\ref{eqn:point_split})]. Doing so gives
\begin{align}
:R^\dagger(x) L(x): &=\lim_{\epsilon\rightarrow 0}\left[R^\dagger(x+\epsilon) L(x) - \langle 0|R^\dagger(x+\epsilon) L(x)|0\rangle\right] \nonumber \\
&= -\frac{i}{4\LL}\left[ \frac{:\e^{i2\varphi(x)}:-1}{\sin(\tfrac{\pi x}{\LL})}\right]
\end{align}
and an analogous expression for its Hermitian conjugate. Note that for $x \rightarrow 0^+$, we have
\begin{align}
:R^\dagger(0^+) L(0^+): \;=  \frac{1}{2\pi}\partial_x \varphi(0),
\end{align}
as can be checked by applying L'H\^opital's rule. Further manipulation with trigonometric identities then gives the final normal-ordered bosonized form for the backscattering term
\begin{align}
&:\e^{-i2k_F x}R^\dagger(x) L(x) + \e^{i2k_F x}L^\dagger(x)R(x): \nonumber \\
&= \frac{1}{2\LL} \left(
\frac{:\sin[2\varphi(x)-2k_F x]:}{\sin(\tfrac{\pi x}{\LL})} + \frac{\sin[2k_F x]}{\sin(\tfrac{\pi x}{\LL})}
\right).
\end{align}
Importantly, we notice that: 
\begin{itemize}
\item[(a)]{there is an additional `form factor' that is sensitive to how near $x$ is to either boundary,}
\item[(b)]{the $2k_F x$ shift in the sine potential promotes charge-density order in the phase field $\varphi(x)$.}
\end{itemize}
This result comes from the fact that $R(x)$ and $L(x)$ are no longer independent because $\phi^R(x)$ and $\phi^L(y) \equiv \phi^R(-y)$ no longer commute. This leads to additional factors when fusing the vertex operators in $R^\dagger(x) L(y)$. Nevertheless, deep in the bulk when $x \approx \LL/2$ and $\sin[\frac{\pi x }{\LL}] \approx 1$, one recovers the translationally invariant form of this term. Lastly, the second term in the parentheses is not an operator but a pure function and integrates to a constant,
\begin{align}
\frac{1}{2\LL} \int_0^\LL \mathrm{d}x\, \frac{\sin(k_F x)}{\sin(\tfrac{\pi x}{\LL})} = \frac{1}{2},
\end{align}
because $k_F \in \frac{\pi}{2\LL}+(\frac{\pi}{\LL})\mathbb{Z}$.
Hence we may replace the above expression by
\begin{align}
&:\e^{-i2k_F x}R^\dagger(x) L(x) + \e^{i2k_F x}L^\dagger(x)R(x): \nonumber \\
&= \frac{1}{2\LL} \left(
\frac{:\sin[2\varphi(x)-2k_F x]:}{\sin(\tfrac{\pi x}{\LL})} + 1
\right)
\end{align}
without loss of accuracy.

By contrast, bosonizing the SC pairing potential gives
\begin{align}
R^\dagger(x)L^\dagger(x) - L(x)R(x) = \frac{2i}{\LL}\sin\left(\frac{\pi x}{\LL}\right) :\cos[2\vartheta(x)]:.
\end{align}
The remarkable thing about this expression is that it vanishes at the boundary, consistent with $R^\dagger(0)L^\dagger(0) = -R^\dagger(0)R^\dagger(0)=0$. Nevertheless, deep in the bulk where $x \approx \LL/2$ we recover the translationally invariant form of Sec.~\ref{sec:closed_wire_bosonization}.

Finally, the total bosonized Hamiltonian is then given by
\begin{widetext}
\begin{align}
H &= \int_0^\LL \mathrm{d}x \left\{
\frac{\pi v_F}{2}:\left(j(x)^2+\rho(x)^2\right): + \frac{\pi v_F}{2\LL^2}Q^2
-\left(\frac{\mathrm{M}}{\LL}\right)\left(\frac{:\sin[2\varphi(x)-2k_F x]:}{2\sin[\tfrac{\pi x}{\LL}]}+\frac{1}{2}\right) 
-\left(\frac{2\Delta}{\LL}\right) \sin[\tfrac{\pi x}{\LL}] :\cos[2\vartheta(x)]: \right\}. 
\label{eqn:Ham_boson_open}
\end{align}
This bosonized Hamiltonian corresponds to the following real-time classical action that is obtained by the same methods as in Appendix \ref{app:bosonic_action},
\begin{subequations}
\begin{align}
S &= \int\mathrm{d}t(L_0 + L_1) \\
L_0 &= -\pi \int_0^\LL  \int_0^\LL j(x)[\zeta_2 (x-y)+\zeta_2(x+y)]\partial_t \rho(y) 
\, \mathrm{d}x\, \mathrm{d}y - Q \partial_t \vartheta_0 -\int_0^\LL \frac{\pi v_F}{2}(j(x)^2+ \rho(x)^2)\, \mathrm{d}x - \frac{\pi v_F}{2\LL}Q^2,  \\ \nonumber \\
L_1 &= \int_0^\LL \mathrm{d}x\left\{
\left(\frac{\mathrm{M}}{\LL}\right)\left(\frac{\sin[2\varphi(x)-2k_F x]}{2\sin[\tfrac{\pi x}{\LL}]}+\frac{1}{2}\right) + \left(\frac{2\Delta}{\LL}\right) \sin[\tfrac{\pi x}{\LL}] \cos[2\vartheta(x)]
\right\},
\end{align}
\label{eqn:S_total_boson_OBC}%
\end{subequations}
\end{widetext}
subject to the constraints that $\vartheta(x)$ and $\varphi(x)$ are related to $j(x),\rho(x)$, $\vartheta_0$, and $Q$ by Eqns.~(\ref{eqn:theta_phi_rho_j}).
From this action, we will perform a saddle-point analysis in the trivial and topological phases just like in Sec.~\ref{sec:closed_wire_bosonization}.

\subsubsection{Trivial phase}

\begin{figure}
\includegraphics[width=0.4\textwidth]{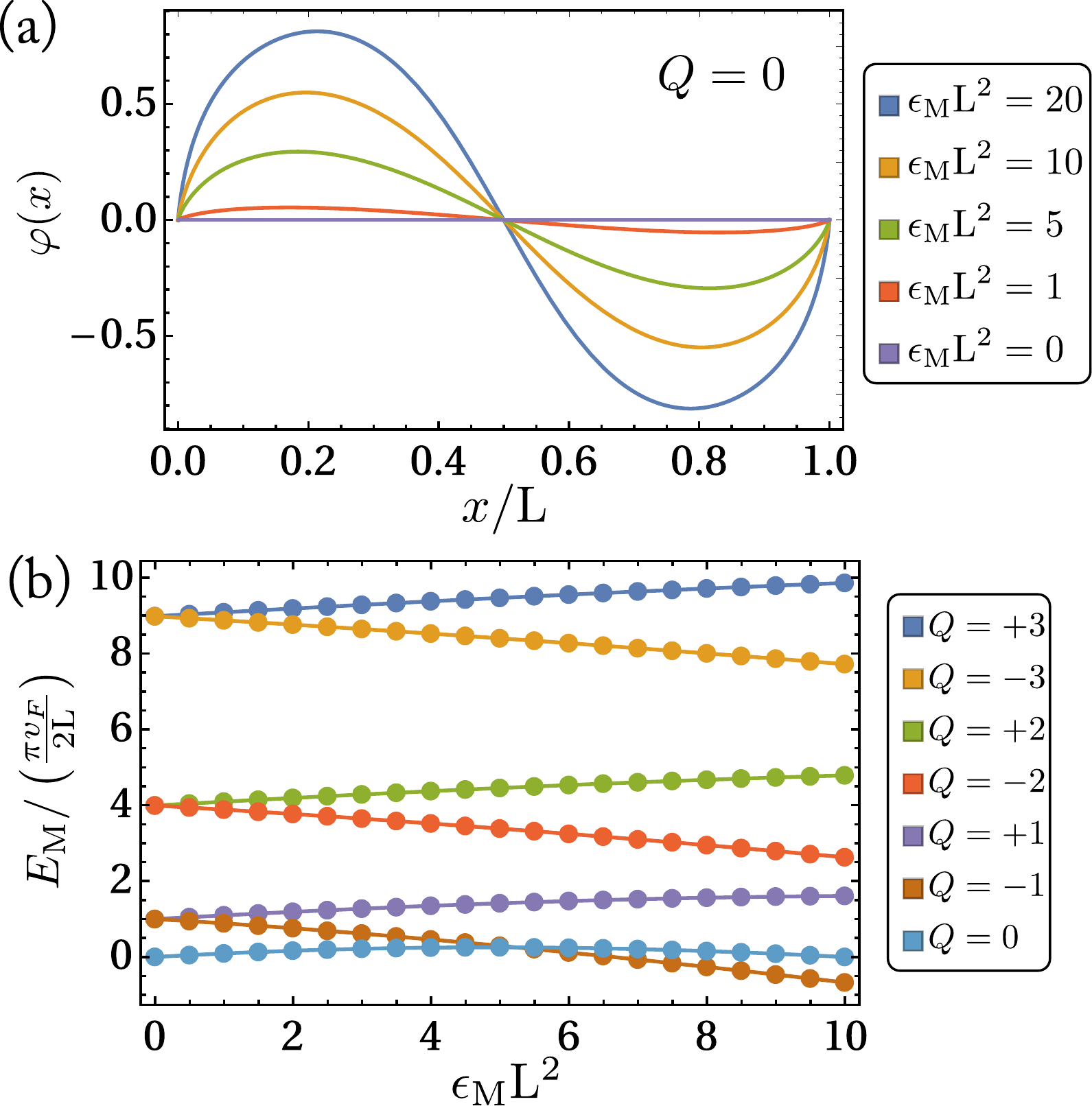}
\caption{Numerical solutions to the non-linear saddle-point ODE of Eqn.~(\ref{eqn:saddle_trivial_ODE}), where we have set $k_F = \pi/(2\LL)$. (a) Solutions in the $Q=0$ sector at varying pinning potential strength $\epsilon_\mathrm{M}$. (b) Saddle-point energies at varying pinning potential strength $\epsilon_\mathrm{M}$ for different $Q$ sectors. The sectors corresponding to $Q=0$ and $Q=-1$ happen to cross at large interaction strengths.}
\label{fig:saddle_trivial}
\end{figure}

Setting $\Delta=0$, we may integrate out the field $\vartheta(x)$ in favor of $\varphi(x)$. Since the degrees of freedom contained in $\vartheta(x)$ -- namely $\vartheta_0$ and $j(x)$ -- appear in Gaussian form, we can do this exactly. This performs a Legendre transformation from a phase-space action to a configuration-space action in terms of $\varphi(x)$. By standard arguments we arrive at the total Lagrangian density
\begin{align}
\mathcal{L}_\mathrm{M} &= \frac{1}{2\pi}\left( \frac{1}{v_F}[\partial_t \varphi(x)]^2 - v_F [\partial_x \varphi(x)]^2\right) \nonumber \\
&\quad + \left(\frac{\mathrm{M}}{2\LL}\right) \frac{\sin[2\varphi(x)-2k_F x]}{\sin(\tfrac{\pi x}{\LL})} ,
\end{align}
where for convenience we have dropped the constant term which arises from fermion normal ordering. The influence of $Q$ now manifests as a topological winding condition
\begin{align}
\varphi(0) - \varphi(\LL) = \pi Q \in \pi\mathbb{Z}.
\end{align}

Seeking saddle-point solutions, the Euler-Lagrange equation of motion is 
\begin{align}
[\partial_x^2 - v_F^{-2} \partial_t^2]\,\varphi(x,t) + \left(\frac{\pi \mathrm{M}}{v_F \LL}\right) \frac{\cos[2\varphi(x,t)-2k_F x]}{\sin(\tfrac{\pi x}{\LL})} = 0.
\end{align}
Focusing first on static solutions, we set $\partial_t \varphi =0$ and note that the homogeneous solutions $\varphi(x) = 0,\pi$ are no longer viable solutions. This is in stark contrast to the closed  wire and can directly be attributed to the presence of open boundaries. 

Nevertheless we can obtain numerical solutions (see Fig.~\ref{fig:saddle_trivial}) to the Euler-Lagrange equations, which may be re-expressed as the non-linear ODE
\begin{align}
\varphi_{xx}(x) + \epsilon_\mathrm{M} \frac{\cos[2\varphi(x)-2k_F x]}{\sin(\tfrac{\pi x}{\LL})} = 0
\label{eqn:saddle_trivial_ODE}
\end{align}
such that  $\varphi(0) = 0, \;\varphi(\LL) = -\pi Q$ and the newly introduced non-linearity parameter $\epsilon_\mathrm{M} := \pi \mathrm{M}/(v_F \LL)$ has dimensions of (length)$^{-2}$. Hence, at this level of approximation, the variational energies in the different $Q$ sectors are generically split and the saddle-point ground state is \emph{non-degenerate}. 

\subsubsection{Topological phase}\label{sec:top_open_wire}

Next we will focus on the topologically non-trivial phase and set $\mathrm{M}=0$ in Eqn.~(\ref{eqn:S_total_boson_OBC}c). Since $[\vartheta_0,Q]=-i$, $Q$ is no longer an integral of motion. We integrate out $\rho$ and $Q$ in favor of $j$ and $\vartheta_0$. Again because $\rho$ and $Q$ appear only in Gaussian contributions in the action, their functional integration amounts to a Legendre transformation. Consequently, this gives the simplified Lagrangian density in terms of $\vartheta(x)$
\begin{align}
\mathcal{L} &= \frac{1}{2\pi}\left\{\frac{1}{v_F}[\partial_t \vartheta(x)]^2 
-v_F [\partial_x \vartheta(x)]^2
\right\} \nonumber \\
&\quad + \left(\frac{2\Delta}{\LL}\right)\sin\left[\frac{\pi x}{\LL}\right] \cos[2\vartheta(x)].
\label{eqn:Lagrange_density_topological_OBC}
\end{align}
In this situation, the field $\vartheta(x)$ is constrained not to have any topological windings ($J\equiv 0$) and must satisfy homogeneous Neumann boundary conditions $\vartheta'(0) = \vartheta'(\LL)=0$. We are then led to the equation of motion 
\begin{align}
[\partial_x^2 -v_F^{-2}\partial_t^2]\vartheta(x,t) - \left(\frac{4\pi \Delta}{v_F \LL}\right)\sin\left[\frac{\pi x}{\LL}\right] \sin[2\vartheta(x,t)] =0,
\end{align}
and focusing on static saddle-point solutions yields
\begin{align}
\vartheta_{xx}(x) - \epsilon_\Delta \sin\left[\frac{\pi x}{\LL}\right] \sin (2\vartheta(x)) =0.
\end{align}
Here, $\epsilon_\Delta := 4\pi\Delta/(v_F\LL) >0 $ is a non-linearity parameter with dimensions of $(\text{length})^{-2}$. In contrast to the trivial case, homogeneous saddle-point solutions are now permissible and the ones with minimum energy correspond to 
\begin{align}
\vartheta(x) = \vartheta_0 = 0,\pi \mod 2\pi. 
\end{align}
The situation is similar to the variational solutions to the topological phase in the closed wire, with the exception that there are no parity constraints and $\vartheta_0$ is unique mod $2\pi$ instead of $\pi$. 

Nonetheless, to obtain finite variational energies we need to include kinetic fluctuations to $\vartheta_0$. Again, this is easily achieved by the semi-classical approximation $\rho =0$ and  $j =0$, which we know is consistent with saddle-point solutions. 
This yields the effective Lagrangian
\begin{align}
L_\text{eff} = - Q \partial_t \vartheta_0 -\frac{\pi v_F}{2\LL}Q^2 + \left(\frac{4\Delta}{\pi}\right)\cos 2\vartheta_0,
\end{align}
where we have performed the spatial integral over the finite interval. This then yields the effective quantum Hamiltonian 
\begin{align}
H_\text{eff} = \frac{\pi v_F}{2\LL}Q^2 - \left(\frac{4\Delta}{\pi}\right)\cos (2\vartheta_0),
\label{eqn:H_eff_top}
\end{align}
where $[\vartheta_0 ,Q] =-i$. The kinetic term now splits the degeneracy between even and odd fermion parity states. As a basis of fermion parity eigenstates, we can define the zero-mode kets
\begin{align}
|\vartheta_0=\theta\rangle &:= \sum_{m \in \mathbb{Z}} \e^{i m \theta}|Q= m\rangle, \\
|\vartheta_0=\theta \rangle_+  &:= \sum_{m \text{ even}} \e^{i m \theta}|Q= m\rangle,\label{eqn:zero_mode_ket_even}   \\
|\vartheta_0=\theta\rangle_-  &:= \sum_{m \text{ odd}} \e^{i m \theta}|Q= m\rangle,
\label{eqn:zero_mode_ket_odd}
\end{align}
and express the two lowest eigensolutions to $H_\text{eff}$ as Mathieu functions [see Sec.~\ref{sec:mathieu_functions}]:
\begin{align}
|\Psi^{(0)}_{\Delta}\rangle = \sqrt{\frac{1}{\pi}}\int_{-\pi}^{\pi}\mathrm{d}\theta\, ce_0 (\theta -\tfrac{\pi}{2},\tfrac{4\Delta\LL}{\pi^2 v_F})|\vartheta_0 = \theta \rangle_+,  \\
|\Psi^{(1)}_{\Delta}\rangle = \sqrt{\frac{1}{\pi}}\int_{-\pi}^{\pi}\mathrm{d}\theta\, se_1 (\theta -\tfrac{\pi}{2},\tfrac{4\Delta\LL}{\pi^2 v_F})|\vartheta_0 = \theta \rangle_- .
\end{align}
These have the respective energies 
\begin{align}
E^{(0)}_\Delta = \frac{\pi v_F}{2\LL}a_0(\tfrac{4\Delta\LL}{\pi^2 v_F}), \quad E^{(1)}_\Delta = \frac{\pi v_F}{2\LL}b_1(\tfrac{4\Delta\LL}{\pi^2 v_F}).
\end{align}
In the strongly pinned regime where $\Delta \gg v_F/\LL $, these wavefunctions are strongly localized about $|\vartheta_0= 0\rangle$ and $|\vartheta_0=\pi\rangle$ in equal measure. The splitting in energy between these states is\cite{abramowitz1965formulas}
\begin{align}
\delta E_\Delta &= \frac{\pi v_F}{\LL}\left[
b_1(\tfrac{4\Delta\LL}{\pi^2 v_F})- a_0(\tfrac{4\Delta\LL}{\pi^2 v_F})
\right] \nonumber \\
&\sim 2^5 \sqrt{\frac{2}{\pi}}\left(\frac{\pi v_F}{\LL} \right)\left(\frac{4\Delta\LL}{\pi^2 v_F}\right)^{3/4}
\e^{-4 \sqrt{\frac{4\Delta\LL}{\pi^2 v_F}}},
\end{align}
demonstrating the asymptotically exact degeneracy in the limit of large $\LL$. However, this estimate for the energy splitting is not quite right because the decay appears to take the form $\sim \LL^{-\frac{1}{4}}\e^{- \beta \sqrt{\LL}}$, where $\beta>0$ is a constant. This asymptotic is much slower than the exponential form $\sim \e^{-\alpha \LL}\; (\alpha>0)$ obtained from solving the free fermionic model [see Sec.~\ref{sec:exact_MZM}]. This implies that spatially homogeneous tunneling events whose kinetic energetics are captured solely by the capacitive term $\frac{v_F Q^2}{2 \LL}$ are insufficient to arrive at the correct asymptotic estimate for the quasi-degenerate energy splitting. To do so, we argue that \emph{localized} instanton kinks between the quasi-degenerate `vacua' $|\Psi^{(0)}_{\Delta}\rangle \pm |\Psi^{(1)}_{\Delta}\rangle$ are necessary at the semi-classical level. These kinks are tunneling events due to the transversal of localized $\pi$-kinks in the $\vartheta(x)$ field through the bulk. Physically, they are localized superconducting phase slips connecting degenerate semi-classical ground states. From the Lagrangian Eqn.~(\ref{eqn:Lagrange_density_topological_OBC}), we see that at low energy, these kinks are predisposed to nucleate near the open boundaries where the potential energy barrier is low due to the $\sin[\pi x/\LL]$-modulated cosine potential. Although their effective semi-classical equations of motion in the bulk should be affected by the changing strength of the cosine potential, we can expect on purely qualitative grounds that these $\pi$-kinks should have a rest mass that scales, to lowest order, like $m_\text{kink} \sim  v_F /{\xi_\Delta}$ with $\xi_\Delta = \frac{v_F}{\Delta}$. Here $\xi_\Delta$ is just the superconducting correlation length of the system. 

This is in contrast to the analyses of Refs.~\onlinecite{cheng2015fractional,fidkowski2011majorana}, where the correct exponential law for the quasi-degenerate energy splitting was also derived. However, those studies are based on a homogeneous cosine potential and spatially uniform instanton kinks to the lowest order in the semi-classical approximation. In principle, the exact eigenstates of the Mathieu equation incorporate all higher-order spatially uniform instanton kink fluctuations. Yet, the resulting degeneracy splitting does not diminish quickly enough with $\LL$ to reflect the local energy gap of the system. That is, one that is characterized by a correlation length scale and localized semi-classical fluctuations. In Sec.~\ref{sec:MZM_extreme}, we will discuss the relationship between these boundary-localized kinks and local Majorana zero mode operators.  


\subsection{Majorana zero modes in the extreme SC limit}\label{sec:MZM_extreme}

The semi-classical analysis of the previous subsection yielded quasi-degenerate ground states of opposite fermion number parity that are present only in the topologically non-trivial phase. Unfortunately, the bosonic semi-classical approach was still unable to verify that this quasi-degeneracy is associated with boundary-localized MZMs. However, in the extreme superconducting limit $v_F=0$, we can indeed derive the MZM operators analytically. In this case, the total Hamiltonian is just the SC pairing potential in an open wire. Despite being a singular Hamiltonian, it remains possible to capture many of the qualitative properties that are protected by the SC excitation gap.

First, we have to consider the effects of a charge-conjugation symmetry that is present in the model. Recall that the Lagrangian in the topological phase with $\mathrm{M}=0$ is
\begin{align}
\mathcal{L} &= \frac{1}{2\pi}\left[
\frac{1}{v_F}(\partial_t \vartheta)^2 - v_F (\partial_x \vartheta)^2
\right] \nonumber \\ & \quad +\left(\frac{2\Delta}{\LL}\right)\sin\left(\frac{\pi x}{\LL}\right)\cos(2\vartheta).
\end{align}
Recall also that the $\mathbb{Z}_2$ inversion symmetry $\mathrm{I}_x$ which was defined in Sec.~\ref{sec:exact_fermion} acts \emph{linearly} by 
\begin{align}
\mathrm{I}_x : R(x) \mapsto -R^\dagger(x) ,\qquad 
\mathrm{I}_x : R^\dagger(x) \mapsto -R(x),
\end{align}
because it acts on $\psi(x)=[R(x),-R^\dagger(-x)]^T$ by
\begin{align}
\mathrm{I}_x :\psi(x) \mapsto \sigma^x \psi(-x). 
\end{align}
When translated to the bosonic fields, this transformation acts linearly as 
\begin{align}
\mathrm{I}_x : \phi^R(x) \mapsto \pi - \phi^R(x), \qquad
\mathrm{I}_x : \eta^R \mapsto \eta^R. 
\end{align}
This means that 
\begin{align}
\mathrm{I}_x : \vartheta(x) \mapsto \pi - \vartheta(x), \qquad 
\mathrm{I}_x : \varphi(x) \mapsto -\varphi(x),
\end{align}
such that the density and current are odd under this transformation, i.e.,  
\begin{align}
j(x) \mapsto -j(x), \quad \rho(x) \mapsto -\rho(x), \quad Q \mapsto -Q.
\end{align}
Likewise, the zero mode transforms as 
\begin{align}
\vartheta_0 \mapsto \pi -\vartheta_0. 
\end{align}
These properties prompt us to identify $\mathrm{I}_x$ as a \emph{unitary} charge conjugation operation. 

The crucial point to note is that the cosine in the Lagrangian is invariant under this transformation because
\begin{align}
\cos(2\vartheta(x)) \mapsto \cos(2\pi-2\vartheta(x)) \equiv \cos(2\vartheta(x)).
\end{align}
The kinetic term is trivially invariant because it depends only on the square of gradient terms. Next, it is also important to note that this symmetry respects the required boundary conditions and constraints that are
\begin{align*}
&j(0)= j(\LL) =0,& &\int_0^\LL \rho(x)\,\mathrm{d}x =0, \\ 
&\varphi(0)=0,&  &\varphi(\LL)= -\pi Q.  
\end{align*}

Furthermore, spinless time-reversal symmetry $\mathcal{T}$ acts on fermions by 
\begin{align}
i \mapsto -i ,\qquad R^{(\dagger)}(x) \mapsto L^{(\dagger)}(x) \equiv-R^{(\dagger)}(-x).  
\end{align}
This translates to the chiral bosons as the anti-linear transformations
\begin{align}
\phi^R(x) \mapsto \pi - \phi^R(-x), \qquad \eta^R \mapsto \eta^R,
\end{align}
\begin{align}
\vartheta(x) \mapsto \pi - \vartheta(x), \qquad \varphi(x) \mapsto \varphi(x),
\end{align}
and gives
\begin{subequations}
\begin{align}
&\rho(x) \mapsto \rho(x) ,\quad j(x)\mapsto -j(x), \\ 
&\vartheta_0 \mapsto \pi - \vartheta_0, \qquad Q \mapsto Q.  
\end{align}
\end{subequations}
It is straightforward to verify that the Lagrangian, the Hamiltonian, and the constraints/boundary conditions on $\rho(x)$ and $j(x)$ are invariant under $\mathcal{T}$. Because $\mathcal{T}^2=1 $ there are no degenerate Kramers pairs and we can always choose a basis of energy eigenstates that are Kramers singlets. In particular, if an eigenstate is non-degenerate, then it must be a Kramers singlet.

Now we specialize to the $v_F=0$ limit in the Hamiltonian formulation. Notice that the local vertex operators $:\e^{\pm i\phi^R(0)}:$ and $:\e^{\pm i\phi^R(\LL)}:$ exactly commute with the Hamiltonian 
\begin{align}
H= H_1=   -\left(\frac{2\Delta}{\LL}\right) \int_0^\LL \mathrm{d}x\;\sin[\tfrac{\pi x}{\LL}] :\cos[2\vartheta(x)]:
\end{align}
precisely because the Hamiltonian density is exactly zero at the boundaries. Moreover, for $\Delta>0$ the cosine potential is exactly minimized by 
\begin{align}
\vartheta(x) = \theta_\text{min}, \qquad \quad \theta_\text{min} =0,\pi \mod 2\pi. 
\end{align}
Furthermore, under $\mathrm{I}_x$, we have that $\theta_\text{min} \rightarrow \pi - \theta_\text{min}$ such that the degenerate `vacua' transform non-trivially under $\mathrm{I}_x$. This suggests that we should express the Hermitian Majorana zero mode operators in the combinations of vertex operators
\begin{subequations}
\begin{align}
\tilde{\gamma}_0 &= \mathcal{N} \left( :\e^{i\phi^R(0)}: + :\e^{-i\phi^R(0)}:\right),  \\
\tilde{\gamma}_\LL &= -i\mathcal{N} \left( :\e^{i\phi^R(\LL)}: + :\e^{-i\phi^R(\LL)}:\right),
\end{align}
\label{eqn:MZM_extreme}%
\end{subequations}%
where $\mathcal{N}$ is a formal normalization constant that is fixed by the conditions $\tilde{\gamma}_0^2 = \tilde{\gamma}_\LL^2=1$. The phase factor $-i$ in $\tilde{\gamma}_\LL$ is necessary to make it Hermitian.
The particular equal superposition of $:\e^{\pm i \phi^R(x)}:$ is crucial such that both $\tilde{\gamma}_0$ and $\tilde{\gamma}_\LL$ transform non-trivially under $\mathrm{I}_x$,
\begin{align}
\tilde{\gamma}_0 \mapsto - \tilde{\gamma}_0, \qquad
\tilde{\gamma}_\LL \mapsto - \tilde{\gamma}_\LL. 
\end{align}
However, the complementary case where $\Delta<0$ would have led to $\theta_\text{min}=-\frac{\pi}{2},\frac{\pi}{2}$, which is invariant under $\mathrm{I}_x$. In this instance we would have chosen the other linear combination of boundary vertex operators that would have made $\tilde{\gamma}_0$ and $\tilde{\gamma}_\LL$ invariant under $\mathrm{I}_x$. In this sense the sign of $\Delta$ dictates the specific $\mathrm{I}_x$ sector of the ground state, which is in complete agreement with our prior analysis in Sec.~\ref{sec:exact_MZM}. Furthermore, the expressions for $\tilde{\gamma}_0$ and $\tilde{\gamma}_\LL$ agree with those of our exact treatment of the same extreme limit in Sec.~\ref{sec:fermion_extreme}. Finally, under time-reversal $\mathcal{T}$ we have 
\begin{align}
\widetilde{\gamma}_0 \mapsto + \widetilde{\gamma}_0, \qquad
\widetilde{\gamma}_\LL \mapsto  -\widetilde{\gamma}_\LL , 
\end{align}
consistent with symmetry fractionalization. 

It is important to appreciate that the Majorana zero mode operators in Eqns.~(\ref{eqn:MZM_extreme}) are purely local fermionic operators, though singular in their normalization. An alternative but non-local expression that has been suggested by Cheng\cite{cheng2012superconducting} and Mazza et~al.~\cite{mazza2018poor} is 
\begin{align}
\tilde{\gamma}_0 = \e^{i\vartheta_0}, \qquad \tilde{\gamma}_\LL = -i\e^{i\vartheta_0}(-1)^Q.
\label{eqn:MZM_nonlocal}
\end{align}
However these operators are only Hermitian and square to 1 in the low-energy subspace where $\vartheta_0 = \theta_\text{min}$. They are clearly non-local because neither $Q$ nor $\vartheta_0$ are local operators. The non-locality of these Majorana zero modes derives from the application of a low-energy projection onto the subspace of degenerate ground states. Moreover, we will next demonstrate that the expressions in Eqns.~(\ref{eqn:MZM_extreme}) reduce to the above forms under the appropriately defined low-energy projection. 
We should remark that the main interest of Refs.~\onlinecite{cheng2012superconducting,mazza2018poor} lies with parafermionic zero modes derived from a generalization of the Hamiltonian in Eqn.~(\ref{eqn:Ham_boson_open}) where $\mathrm{M}=0$ but $\cos[2\vartheta(x)]$ is replaced by $\cos[2M\vartheta(x)]$ with $M\in \mathbb{N}$. The case of $M=1$ directly corresponds to SC Majorana zero modes. 

Next, recall that the phase field $\vartheta(x)$ has the operator expansion 
\begin{align}
\vartheta(x) = \vartheta_0 + \pi \int_{-\LL}^\LL \zeta_2(x-y)j(y) \, \mathrm{d}y.
\end{align}
Thus, demanding that $\vartheta(x)$ is uniform is tantamount to requiring that $j(x) =0$. From the mode expansion of $j(x)$ in Eqn. (\ref{eqn:j_mode}), this is equivalent to the condition that $a_n = -a_n^\dagger$ for all positive integers $n$. Hence, we define a variational ground state manifold spanned by the tensor product of states
\begin{align}
|\vartheta_0=\theta \rangle \otimes |\widetilde{\Omega}\rangle
\end{align}
with 
\begin{align}
\e^{i\vartheta_0} |\vartheta_0=\theta \rangle \;&=\; \e^{i\theta} |\vartheta_0=\theta \rangle, \\  
\langle\vartheta_0= \theta' |\vartheta_0= \theta \rangle \;&=\; 2\pi\sum_{m\in \mathbb{Z}}\delta(\theta' - \theta-2m \pi )
\end{align}
for $\theta,\theta' \in (-\pi,\pi]$. Note that as opposed to the zero-mode kets in Eqns.~(\ref{eqn:zero_mode_ket_even}) and (\ref{eqn:zero_mode_ket_odd}), the ket $|\vartheta_0=\theta \rangle$ does not have a definite fermion parity.
The state $|\widetilde{\Omega}\rangle$ is characterized by the condition
\begin{align}
a_n |\widetilde{\Omega}\rangle = - a_n^\dagger |\widetilde{\Omega}\rangle.
\label{eqn:a_constraint_extreme}  
\end{align}
This makes $|\widetilde{\Omega}\rangle$ quite different from the filled Fermi sea $|0\rangle$ which is annihilated by all such $a_n$. Finally, the choice of sign of $\Delta>0$ leads to $|\vartheta_0 = 0\rangle$ and  $|\vartheta_0 = \pi\rangle$ as being the only two possible pinning minima. Projecting onto the subspace spanned by the states 
\begin{align}
|\vartheta_0=0 \rangle \otimes |\widetilde{\Omega}\rangle, \qquad |\vartheta_0=\pi \rangle \otimes |\widetilde{\Omega}\rangle
\end{align}
yields the effective non-local Hamiltonian
\begin{align}
H_\text{eff} =   - \left(\frac{4\Delta}{\pi}\right)\cos (2\vartheta_0),
\end{align}
which is the previously considered effective Hamiltonian without the fluctuation or charge capacitive term $\propto v_F Q^2$. 

More importantly, the vertex operators acting on this subspace simplify to
\begin{align}
&:\e^{i\alpha \phi^R(x)}: |\widetilde{\Omega}\rangle \nonumber \\
&= 
\e^{i \alpha \vartheta_0} \e^{i \alpha \frac{\pi Q x}{\LL}} 
\exp\left(2 \alpha \sum_{n>0} \frac{a_n^\dagger}{\sqrt{n}}\sin\left(\frac{n \pi x}{\LL}\right)\right)|\widetilde{\Omega}\rangle,
\end{align}
which is derived using Eqn.~(\ref{eqn:a_constraint_extreme}). These relations can then be used to derive the correlation function 
\begin{align*}
&\langle R(-x)R^\dagger(y) \rangle \nonumber \\
&=\left(\langle \vartheta_0=\theta| \otimes \langle \widetilde{\Omega} |\right) R(-x)R^\dagger(y)\left( |\vartheta_0=\theta \rangle \otimes |\widetilde{\Omega}\rangle \right) \\
&= \frac{1}{2\LL}\e^{\frac{i\pi}{2}(x-y)}\e^{i\frac{\pi}{\LL}x}
\langle \vartheta_0 =\theta|\e^{i2\vartheta_0}\e^{-\frac{i\pi Q}{\LL}(x-y)}|\vartheta_0=\theta \rangle \\ 
&\quad  \times \langle \widetilde{\Omega} | 
\exp\left(2\sum_{n>0}\frac{a_n}{\sqrt{n}}[\sin(\tfrac{n\pi x}{\LL})-\sin(\tfrac{n\pi y}{\LL})]\right)
|\widetilde{\Omega} \rangle
\end{align*}
for $\theta=0,\pi$. It is only non-zero whenever $x=y$ because of the delta-function normalization of the $|\vartheta_0=\theta \rangle$ kets. Thus, we are led to 
\begin{align}
\langle R(-x)R^\dagger(y) \rangle = \e^{i\frac{\pi x}{\LL}} \delta(x-y).
\end{align}

Lastly, we have that 
\begin{align}
:\e^{\pm i \phi^R(0)}:|\widetilde{\Omega}\rangle &= \e^{\pm i \vartheta_0}|\widetilde{\Omega}\rangle, \\ \nonumber \\
:\e^{\pm i \phi^R(\LL)}:|\widetilde{\Omega}\rangle &= \e^{\pm i \vartheta_0}(-1)^Q|\widetilde{\Omega}\rangle.
\end{align}
This means that for $|\vartheta_0 = \theta_\text{min}\rangle \otimes |\widetilde{\Omega}\rangle $ with  $\theta_\text{min} =0 ,\pi$, we have the eigenvalue equations
\begin{align}
&\tilde{\gamma}_0\, \left( |\vartheta_0 = \theta_\text{min}\rangle \otimes |\widetilde{\Omega}\rangle \right) \nonumber \\
&= +2 \mathcal{N} \cos \theta_\text{min} \left(|\vartheta_0 = \theta_\text{min}\rangle \otimes |\widetilde{\Omega}\rangle\right), \\ \nonumber \\
&i\tilde{\gamma}_\LL\, \left(|\vartheta_0 = \theta_\text{min}\rangle \otimes |\widetilde{\Omega}\rangle\right) \nonumber \\
&= -2 \mathcal{N} \cos \theta_\text{min} \left(|\vartheta_0 = \theta_\text{min}\rangle \otimes |\widetilde{\Omega}\rangle\right).
\end{align}
Thus we recover the non-local forms of Eqn.~(\ref{eqn:MZM_nonlocal}) since $\pi = -\pi$ mod $2\pi$. The singular normalization of $\tilde{\gamma}_0$ and $\tilde{\gamma}_\LL$ derives from the fact that the eigenkets of $\vartheta_0$ are delta-function normalized. 

It should be emphasized that by neglecting the kinetic contributions entirely, we have an artificial degeneracy of the model with singularly localized Majorana zero mode operators. The previous analysis of Sec.~\ref{sec:top_open_wire} in fact does a better job of lifting this degeneracy by providing some fluctuations about the pinned minima. However, as was already mentioned, the exact energy splitting derived from that effective Hamiltonian is not sufficiently suppressed with increasing $\LL$ because the fluctuations in $\vartheta_0$ and $Q$ are non-local by nature. To remedy this situation we really do need to diagonalize the model in local bosonic degrees of freedom. This is the topic of the next subsection. In a sense, we intend to smear the operators $\tilde{\gamma}_0 ,\tilde{\gamma}_\LL$ in Eqn.~(\ref{eqn:MZM_extreme}) into the bulk and thereby produce Majorana `wavefunctions'. This smearing is essentially driven by the non-zero kinetic term. 


\subsection{Majorana zero modes from the vertex algebra}\label{sec:exact_MZM_vertex}

While Secs.~\ref{sec:top_open_wire} and \ref{sec:MZM_extreme} provided complementary results on MZMs and the quasi-degenerate ground states, we shall now explicitly derive the exact bosonized MZMs. The approach taken here uses the machinery of vertex algebras.\cite{kac1998vertex}

First, recall the bosonization identity for the open wire given in Eqns.~(\ref{eqn:bosonized_RL_open}),
\begin{align}
R(x) &= \frac{1}{\sqrt{2\LL}}:\e^{i\phi^R(x)}: \e^{-i\frac{\pi x}{2\LL}},
\end{align}
where we can safely drop the Klein factor $\eta^R$ since it does not play a role after unfolding. The normal-ordered Hamiltonian Eqn.~(\ref{eqn:Ham_boson_open}) then bosonizes to 
\begin{subequations}
\begin{align}
H &= H_0 + H_1, \\
H_0 &= \frac{v_F}{4\pi} \int_{-\LL}^\LL \mathrm{d}x\; :[\partial_x \phi^R(x)]^2:, \\ 
%
H_1 &= \frac{i\Delta}{4\LL}\int_{-\LL}^\LL \mathrm{d}x\; s(x) \left[
: \e^{i\phi^R(-x)} : : \e^{i\phi^R(x)} : \right. \nonumber \\
&\left.\hspace{2.6cm}- : \e^{-i\phi^R(x)} : : \e^{-i\phi^R(-x)} :
\right],
\end{align}
\end{subequations}
where we have set $\mathrm{M}=0$ to focus only on the topological phase, and $s(x):=\text{sgn}(\sin[\tfrac{\pi x}{\LL}])$ is the square wave form. 
The goal is to determine a fermionic operator $\widetilde{\psi}_E$ that has the expansion
\begin{align}
\widetilde{\psi}_E = \int_{-\LL}^\LL \frac{\mathrm{d}x}{2\LL} \; \,\widetilde{\psi}(x) [:e^{i\phi^R(x)}:+ :e^{-i\phi^R(x)}:] \e^{-i\frac{\pi x}{2\LL}}
\label{eqn:vertex_fermion}
\end{align}
such that it is an exact quasiparticle excitation satisfying
\begin{align}
[H_0+H_1,\widetilde{\psi}_E]  = - E \widetilde{\psi}_E
\label{eqn:H_tilde_psi_com}
\end{align}
with energy $0<E<\Delta$. Note that the superposition of vertex operators in Eqn.~(\ref{eqn:vertex_fermion}) was chosen such that $\mathrm{I}_x=-1$ for this operator. This is based on the previous subsection's analysis regarding the consequences of $\Delta>0$ when choosing the $\mathrm{I}_x$ symmetry of the ground state. 
The additional phase factor $\e^{-i\frac{\pi x}{2\LL}}$ is to ensure that antiperiodic boundary conditions are satisfied, i.e., $\widetilde{\psi}(x+2\LL)=-\widetilde{\psi}(x)$, while maintaining a single-valued integrand. 
Also, being non-degenerate, $\widetilde{\psi}_E$ must be a Kramers singlet under time-reversal symmetry $\mathcal{T}$. This leads to the symmetry condition on the wavefunction 
\begin{align}
\widetilde{\psi}(x) =[ \widetilde{\psi}(-x)]^*.   
\end{align}
One should think of the expansion Eqn.~(\ref{eqn:vertex_fermion}) as a superposition of Mandelstam kinks\cite{stone1994bosonization} in the chiral field $\phi^R$.
For $\tfrac{v_F}{\LL} < \Delta $, we expect to find two Majorana zero modes of the form
\begin{align}
\tilde{\gamma}_0 = \frac{\widetilde{\psi}_E + \widetilde{\psi}_E^\dagger}{2}, \qquad
\tilde{\gamma}_\LL = i\frac{\widetilde{\psi}_E - \widetilde{\psi}_E^\dagger}{2}.
\end{align}
However, the spatial localization profiles of $\tilde{\gamma}_0$ and $\tilde{\gamma}_\LL$ remain to be determined. 

The solution to this problem in terms of fermionic operators, as we have seen in Sec.~\ref{sec:exact_fermion}, is fairly straightforward and entails working with Nambu space and the Bogoliubov-de Gennes Hamiltonian. However, in this subsection, we choose to work entirely in the bosonic language. 

First, we require expressions for commutators of vertex operators with $H_0$ and $H_1$. The computation of such commutators is most naturally done using operator product expansions (OPEs) of the conformal field theory defined by $H_0$ alone. Although the total Hamiltonian does not describe a conformally invariant theory, the computation of \emph{equal-time} commutation relations using OPEs remains valid. Note that this involves analytically continuing $x \rightarrow x + i v_F \tau$ for short times $\tau$ in order to evaluate equal-time commutators using the well-known relationship between radially ordered products and equal-time commutators.\cite{ginsparg1989applied,belavin1984infinite,blumenhagen2009basics}

We begin by defining the holomorphic coordinate in radial quantization in the $\mathbb{C}$ plane
\begin{align}
z = \e^{-i\frac{\pi}{\LL}(x+iv_F \tau)} \in \mathbb{C}, \qquad \tau \in [-\infty,\infty].
\label{eqn:z_holo}
\end{align} 
The chiral fields then take the holomorphic form\footnote{In some more common conventions of the holomorphic expansion of the chiral field (see, e.g., Refs.~\onlinecite{gogolin1999bosonization,fradkin2013field}) $z$ is replaced by $1/z$ below. This has to do with our convention for the Fourier series expansion such that $a_n$ is accompanied by the phase $\e^{+i\frac{\pi x}{\LL}}$.}
\begin{align}
\phi^R(z) = \vartheta_0 + i Q \ln z + \sum_{n>0}\frac{1}{\sqrt{n}}\left(a_nz^{-n} + a_n^\dagger z^n \right),
\end{align}
which can be regarded as the $H_0$-evolved Heisenberg operator
\begin{align}
\phi^R(z(x,\tau)) &\equiv \phi^R(z(x+iv_F\tau,0)) \nonumber \\ 
&\equiv \e^{H_0 \tau} \phi^R(z(x,0)) \e^{-H_0 \tau}.
\end{align}
As such, $x$ has been holomorphically continued to complex values. 
In the new coordinates, the vertex operator takes the elegant form\cite{goddard1986kac} 
\begin{align}
:e^{i \alpha \phi^R(z)}: \;= \; \e^{i \alpha \vartheta_0 }z^{-\alpha Q} \e^{i \alpha \phi^R_+(z)} \e^{i \alpha \phi_-^R(z)},
\end{align}
where 
\begin{align}
\phi^R_+(z) := \sum_{n>0}\frac{a_n^\dagger z^n}{\sqrt{n}} , \quad 
\phi^R_-(z) := \sum_{n>0}\frac{a_n z^{-n}}{\sqrt{n}}.   
\end{align}
The set of vertex operators themselves satisfy the fundamental product identities
\begin{subequations}
\begin{align}
:e^{i \alpha \phi^R(z)}: \, :e^{i \beta \phi^R(w)}: 
&= (z-w)^{\alpha \beta} :e^{i [\alpha \phi^R(z)+\beta \phi^R(w)]}: \\
&= (-1)^{\alpha \beta}  :e^{i \beta \phi^R(w)}:\, :e^{i \alpha \phi^R(z)}:
\end{align}
\label{eqn:vertex_alg}%
\end{subequations}
for $|z|>|w|$.

Changing coordinate system from $(x,\tau)$ to $z$ at zero time $\tau=0$ gives $\bar{z} z =1$ with
\begin{align}
\mathrm{d}x = \frac{i \LL}{\pi} \frac{\mathrm{d}z}{z}, \qquad \partial_x = -\frac{i\pi}{\LL}\, z \partial_z. 
\end{align}
Thus the kinetic Hamiltonian $H_0$ in holomorphic coordinates is 
\begin{align}
H_0 &= - \frac{\pi v_F}{2\LL} \oint_{|z|=1} \frac{\mathrm{d}z}{i2\pi}\, z\, :[\partial_z \phi^R(z)]^2:\nonumber \\
&= \frac{\pi v_F}{\LL} \oint_{|z|=1}\frac{\mathrm{d}z}{i2\pi}\, z \,T(z),
\end{align}
where $T(z)$ is the stress-energy tensor of a free chiral boson,
\begin{align}
T(z) &= -\frac{1}{2}:\partial_z \phi^R(z) \partial_z \phi^R(z) : \;= \sum_{n=-\infty}^\infty \frac{L_n}{z^{n+2}},
\end{align}
with $L_n$ for $n\in \mathbb{Z}$ being the Virasoro generators.
Applying the residue theorem yields 
\begin{align}
H_0 = \frac{\pi v_F}{\LL} \oint_{|z|=1}\frac{\mathrm{d}z}{i2\pi} \sum_{n=-\infty}^\infty \frac{L_n}{z^{n+1}}=\frac{\pi v_F}{\LL} L_0,
\end{align}
which just means that $H_0$ implements radial dilations, i.e., time translations. Implementing the same transformation steps to $H_1$ gives 
\begin{align}
H_1 =\frac{i\Delta}{2}\oint_{|z|=1} \frac{\mathrm{d}z}{i2\pi} \, \frac{s(z)}{z}&\left[
:\e^{i\phi^R(z^{-1})}:\,:\e^{i\phi^R(z)}: \right. \nonumber \\  
&\left.- :\e^{-i\phi^R(z)}:\,:\e^{-i\phi^R(z^{-1})}:
\right],
\end{align}
with the square wave form in terms of $z$ given by
\begin{align}
s(z) := 2\sum_{n \text{ odd}} \frac{z^{-n}}{in\pi}.
\end{align}
Note that on the equal-time circle $|z|=1$, $s(z)$ is discontinuous and so the expression above has to be interpreted as a formal distribution.\cite{kac1998vertex} Finally, the fermionic quasiparticle operator Eqn.~(\ref{eqn:vertex_fermion}) becomes
\begin{align}
\widetilde{\psi}_E = \oint_{|z|=1} \frac{\mathrm{d}z}{i 2\pi z} \; \,\widetilde{\psi}(z) \, \sqrt{z}[:e^{i\phi^R(z)}:+ :e^{-i\phi^R(z)}:].
\end{align}
Also, $\sqrt{z}$ is not single-valued on $\mathbb{C}$ because of a branch cut which is taken here to lie on the negative real line. Hence, to maintain single-valuedness of the integrand, $\widetilde{\psi}(z)$ must also contain the same branch cut with $\widetilde{\psi}(\e^{i2\pi }z ) = -\widetilde{\psi}(z)$. This is nothing more than a restatement of antiperiodic NS boundary conditions.  

Now we need to define the operation of radial ordering, which is the equivalent of time ordering in radial quantization. We denote this ordering operation by $\mathcal{R}$ and for any two field operators $A(z)$ and $B(w)$ depending holomorphically on the complex variables $z,w$, we define
\begin{align}
\mathcal{R}\{A(z)B(w)\} = 
\begin{cases}
A(z)B(w) & |z|> |w|, \\
(-1)^{\uppi(A,B)} \, B(w)A(z) & |w|>|z|.
\end{cases} 
\end{align}
Here, the exchange sign $(-1)^{\uppi(A,B)}$ is set by the mutual braiding relation of $A$ and $B$. 
Since we are only dealing with fermions in the present situation, this phase is determined by the $\mathbb{Z}_2$ grading of operators into bosonic (even) and fermionic (odd) grades. Thus $(-1)^{\uppi(A,B)} = (-1)^{|A||B|}$, such that we have a negative sign if and only if both $A$ and $B$ are fermionic. The $\mathbb{Z}_2$ grade $|A|\in\{0,1\}$ of any graded operator $A$ is determined by its commutation relation with the fermion parity operator as
\begin{align}
(-1)^Q A (-1)^Q =(-1)^{|A|} A.   
\end{align}
More generally, for any two vertex operators $A(z) \;=\; {:\e^{i\alpha \phi^R(z)}:}$ and $B(w) \;=\; :\e^{i\beta \phi^R(w)}:$ we have
\begin{align}
(-1)^{\uppi(A,B)} = (-1)^{\alpha \beta}
\end{align}
due to the fundamental braiding relation in Eqn.~(\ref{eqn:vertex_alg}). 

\begin{figure}\begin{center}
\includegraphics[width=\columnwidth]{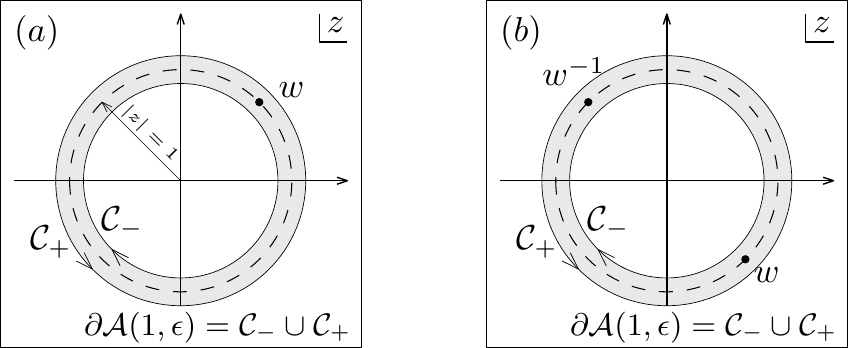}\end{center}
\caption{Contour plots used to compute commutators via OPEs in the radial quantization scheme. The shaded region denotes the annulus $\mathcal{A}(1,\epsilon)$ centered at $z=0$ with radii $(1-\epsilon,1+\epsilon)$. Its boundary $\partial\mathcal{A}(1,\epsilon)$ is composed of the two counterwinding contours $\mathcal{C}_-$ and $\mathcal{C}_+$. As $\epsilon$ tends to zero, the Cauchy residue theorem produces contributions only from singularities on the zero-time circle $|z|=1$. (a) The contour used to compute general commutators when there is only one pole at $z=w$. (b) The contour used to compute $[H_1,:\e^{\pm i \phi^R(w)}:]$, which now includes an extra singularity at $z=w^{-1}$.}
\label{fig:contour_pic}
\end{figure}

Consider next the following equal-time operator
\begin{align}
\mathcal{O}_A := \oint_{|z|=1}\frac{\mathrm{d}z}{i2\pi} \, A(z). 
\end{align}
Computing its \emph{generalized} commutator with another local operator $B(w)$ at the same zero time ($|w|=1$) yields
\begin{align}
[\mathcal{O}_A, B(w)]_\uppi 
&= \oint_{|z|=1}\frac{\mathrm{d}z}{i2\pi} \, [A(z),B(w)]_\uppi \nonumber \\ 
&= \oint_{\partial\mathcal{A}(1,\epsilon)}\frac{\mathrm{d}z}{i2\pi} \mathcal{R}\{A(z)B(w)\}.
\label{eqn:OPE_comm}
\end{align}
Here, $\epsilon$ is a positive infinitesimal number to be taken to zero, and $\mathcal{A}(1,\epsilon)$ is an annulus centered at 0 with radius 1 and thickness $2\epsilon$. The contour $\partial\mathcal{A}(1,\epsilon)$ denotes its oriented boundary with the outer boundary running anti-clockwise [see Fig.~\ref{fig:contour_pic}(a)]. Hence, to compute a commutator we need to evaluate a contour integral of a radially ordered product in the limit $\epsilon \rightarrow 0$. This then involves the use of the OPE between $A(z)$ and $B(w)$, the poles of which will yield the desired result. 

To see this method in action we consider the commutator $[H_0,:\e^{\pm i \phi^R(z)}:]$. Recall the well celebrated OPE between $T(z)$ and $:\e^{ i \alpha \phi^R(z)}:$,
\begin{align}
T(z):\e^{ i \alpha \phi^R(z)}:\; &=\; \frac{\alpha^2}{2(z-w)^2}:\e^{ i \alpha \phi^R(w)}: \nonumber \\
&+ \frac{1}{z-w} \partial_w \left(:\e^{ i \alpha \phi^R(w)}:\right)+ \text{regular},
\end{align}
which remains valid irrespective of whether $|w|<|z|$ or $|w|>|z|$. Plugging this expansion into (\ref{eqn:OPE_comm}) gives
\begin{align}
&[H_0, :\e^{\pm i\phi^R(w)}:] \nonumber \\
&= \frac{\pi v_F}{\LL} \oint_{\partial \mathcal{A}} \frac{\mathrm{d}z}{i2\pi} \, z \, \mathcal{R} \{ T(z) :\e^{\pm i\phi^R(w)}:\} \nonumber \\
&= \frac{\pi v_F}{\LL} \left[
w \partial_w (:\e^{\pm i\phi^R(w)}:)+ \tfrac{1}{2}:\e^{\pm i\phi^R(w)}:
\right].
\label{eqn:H_0_e}
\end{align}
Notice that the final expression on the RHS is really just a special case of the following general commutation relation\cite{belavin1984infinite} between the Virasoro generators and a primary field $\phi$ with scaling dimension $\Delta_\phi$
\begin{align}
[L_m, \phi(z)] = z^{m+1} \partial_z \phi(z) + \Delta_\phi (m+1) z^m \phi(z),
\end{align}
where the scaling dimension of $:\e^{i \alpha \phi^R(w)}:$ is $\alpha^2/2$. 
Moreover, we can express Eqn.~(\ref{eqn:H_0_e}) in a more compact form by multiplying with $\sqrt{w}$ such that
\begin{align}
[H_0, (\sqrt{w}:\e^{\pm i\phi^R(w)}:)]
= \frac{\pi v_F}{\LL} w \partial_w (\sqrt{w} :\e^{\pm i\phi^R(w)}:). 
\label{eqn:H_0_e_2}
\end{align}
The factor $\sqrt{w}$ essentially neutralizes the scaling dimension of the vertex operator.
From the bosonization identity
\begin{align}
R^{(\dagger)}(x) = \sqrt{\frac{w}{2\LL}} :\e^{\pm i\phi^R(w)}: 
\end{align}
with $w=\e^{-i\frac{\pi x}{\LL}}$, we then recover the Heisenberg equation of motion for the chiral fermionic field under time evolution by $H_0$,
\begin{align}
[H_0,R^{(\dagger)}(x) ] = iv_F \partial_x R^{(\dagger)}(x). 
\end{align}
Finally, using the result Eqn.~(\ref{eqn:H_0_e_2}) leads to the commutator
\begin{align}
&[H_0, \widetilde{\psi}_E] \\\nonumber
&= -\frac{\pi v_F}{\LL} \oint_{|z|=1}\frac{\mathrm{d}z}{i2\pi z} \;  z\partial_z \widetilde{\psi}(z) \left[\sqrt{z}(:\e^{i\phi^R(z)}:+:\e^{- i\phi^R(z)}:)\right]
\end{align}
after an integration by parts.

For the $H_1$ term, things are more complicated because there is now more than one singularity and we need to apply Wick's theorem when using the OPEs. First, the following OPEs can be derived from the fundamental product Eqn.~(\ref{eqn:vertex_alg}):
\begin{align}
:\e^{\pm i \phi^R(z)}:\,:\e^{\mp i \phi^R(w)}: \;& = 
\begin{cases}
+\frac{1}{z-w} + \text{reg.} & |z|>|w|, \\ \\
-\frac{1}{z-w} + \text{reg.} & |w|>|z|,
\end{cases} \\ \nonumber \\ 
:\e^{\pm i \phi^R(z^{-1})}:\,:\e^{\mp i \phi^R(w)}: \; &=
\begin{cases}
+\frac{\left(z/w\right)}{z-w^{-1}} + \text{reg.} & |z|>|w|,  \\ \\
- \frac{\left(z/w\right)}{z-w^{-1}} + \text{reg.} & |w|>|z|.
\end{cases}
\end{align} 
Applying these OPEs to the calculation of the commutator then yields\footnote{We only keep OPEs that yield singularities, whereas we can neglect products like $:\e^{i\phi^R(z)}::\e^{i\phi^R(w)}:$ that are regular.}
\begin{align}
[H_1, (\sqrt{w}:\e^{i \phi^R(w)}:)] 
&= i\Delta s(w) \sqrt{w^{-1}}:\e^{-i\phi^R(w^{-1})}:
\end{align}
for $w=\e^{-i\frac{\pi x}{\LL}}$. Firstly, when applying Wick's theorem, we have to include an exchange sign when permuting vertex operators. Since these vertex operators have mutual fermionic statistics, this introduces a negative sign. Secondly, there are now singularities at $z=w$ and $z=w^{-1}$ and so we have the situation depicted in Fig.~\ref{fig:contour_pic}(b). The same type of calculation then gives 
\begin{align}
[H_1, (\sqrt{w}:\e^{-i \phi^R(w)}:)]  = i\Delta s(w) \sqrt{w^{-1}} :\e^{i\phi^R(w^{-1})}:.
\end{align}
Together these commutators produce
\begin{align}
&[H_1,\widetilde{\psi}_E] = \nonumber \\
& i\Delta \oint_{|z|=1}\frac{\mathrm{d}z}{i2\pi z}\, \widetilde{\psi}(z) \, s(z) \sqrt{z^{-1}}\left[
:\e^{-i\phi^R(z^{-1})}:+ :\e^{i\phi^R(z^{-1})}:
\right] \nonumber \\
&= -i\Delta \oint_{|z|=1}\frac{\mathrm{d}z}{i2\pi z}\, s(z) \widetilde{\psi}(z^{-1})\sqrt{z}\left[
:\e^{i\phi^R(z)}:+ :\e^{-i\phi^R(z)}: \right]
\end{align}
after making the change of dummy variable $z \rightarrow 1/z$.

Putting the pieces together and comparison with Eqn.~(\ref{eqn:H_tilde_psi_com}) then requires the eigenvalue equation 
\begin{align}
\left(\frac{\pi v_F}{\LL}\right) z\frac{\partial \widetilde{\psi}(z)}{\partial z} + i \Delta s(z)\widetilde{\psi}(z^{-1})  = E \widetilde{\psi}(z)
\end{align}
to hold on the zero-time line $|z|=1$. Transforming back to the spatial coordinate $x$ and taking the real and imaginary parts of $\widetilde{\psi}(x) \equiv \widetilde{\psi}(-x)^*$ then gives the BdG eigenvalue equations Eqn.~(\ref{eqn:BdG_a_b}). Hence, although we have managed to avoid Nambu space, ultimately we still need to solve a single-particle BdG equation which describes the spatial profile of the Majorana modes. Fortunately, this has already been done in great detail in Sec.~\ref{sec:exact_MZM}. Nevertheless, this entire exercise was a good check to see if one can derive the same eigenvalue equation from purely bosonic methods. Finally, for the sake of completeness, the desired wavefunction $\widetilde{\psi}(x)$ from Sec.~\ref{sec:exact_MZM} is given by
\begin{align}
\widetilde{\psi}(x) &= \sqrt{\frac{\LL}{2}}[a(x) + i b(x)], \\ \nonumber \\
a(x) &=\mathcal{N}_\kappa \sinh(\kappa (L-|x|)),\quad 
b(x) =\mathcal{N}_\kappa \sinh(\kappa x),
\end{align}
where the quantities $\mathcal{N}_\kappa$ and $\kappa$ are as defined in Sec.~\ref{sec:exact_MZM}. The Majorana operators are themselves given by  
\begin{align}
\widetilde{\gamma}_0 &:= \int_{-\LL}^\LL \mathrm{d}x \; \frac{a(x)}{\sqrt{2\LL}} [:\e^{i \phi^R(x)}: + :\e^{-i\phi^R(x)}:]\e^{-i\frac{\pi x}{2\LL}}, \\
\widetilde{\gamma}_\LL &:= \int_{-\LL}^\LL \mathrm{d}x \; \frac{b(x)}{\sqrt{2\LL}} [:\e^{i \phi^R(x)}: + :\e^{-i\phi^R(x)}:]\e^{-i\frac{\pi x}{2\LL}}, 
\end{align}
which is nothing more than the bosonization of Eqns.~(\ref{eqn:gamma_0L}).


\section{Summary and discussion}\label{sec:summary_discussion}

In this work, a very simple continuum model of a one-dimensional fermionic SPT phase\cite{senthil2015symmetry} was analyzed in great detail using analytic methods. The topological non-trivialness of the phase is evidenced by the appearance of isolated Majorana zero modes (MZMs) localized on open boundaries. These MZMs are protected from weak disorder and interactions by fermion parity and spinless time-reversal symmetries. As was first established by Fidkowski and Kitaev,\cite{fidkowski2010effects} this topological class of fermionic SPT phases (BDI-class) possesses a $\mathbb{Z}_8$ classification. Consequently, we can regard the model studied here as a generator of this $\mathbb{Z}_8$ ``group of SPT phases'' by stacking multiple copies of it. In addition, the continuum model is realized effectively in the low-energy limit of Kitaev's Majorana chain model. 

What distinguishes this work from the previous studies is the exact analytic diagonalization of the Bogoliubov-de Gennes (BdG) mean-field Hamiltonian at finite lengths and open boundaries. The resulting free-fermion eigenmodes and their energies were then used to write down the exact quasi-degenerate ground states and their Bogoliubon quasiparticle excitations. In particular, at finite lengths, we uncover the exact MZM operators with boundary-localized wavefunctions and the associated exact Bogoliubon quasiparticle excitation energy. This energy eigenvalue becomes suppressed by large bulk lengths and is a symptom of the many-body quasi-degenerate energy splitting between opposite fermion number parity ground state sectors. Additionally, the symmetry fractionalization of the SPT phase can be confirmed from the transformation of the exact MZM operators under spinless time-reversal. 

With an in-depth characterization of the model in the language of free fermions, we then proceeded to perform finite-length bosonization analyses using normal-ordered vertex operators. We have also been careful to include zero modes, their associated conjugate momenta (topological windings), and Klein factors in the form of anticommuting Majorana operators (Clifford algebra generators). Bosonization was carried out in both open and closed wire geometries at finite lengths. This allowed for comparisons between the two geometry types, which we found to be qualitatively and quantitatively very different. Also, for both geometries we employed a novel rewriting of the bosonized Hamiltonian and classical Lagrangian that is explicit in zero modes, currents and charges. In $1+1$ dimensions, the quantum current and charge density fields satisfy an $U(1)$ current algebra which is reflected in our finite-length action. 

In a closed wire geometry, our bosonization approach was able to demonstrate, at a semi-classical level, fermion parity switching from twisting boundary conditions by a threaded magnetic flux. This effect, which may be taken to be another operational definition of an SPT phase, is limited to only the topological phase. We also clarified a subtle mathematical point which relates to the $\mathbb{Z}_2$ gauge fixing of products of Klein factors $i\eta^R \eta^L$ and total fermion number parity. Technically speaking, there is an often overlooked internal $\mathbb{Z}_2$ gauge symmetry which appears whenever the fermionic charge-raising/lowering operator (sometimes also called the Klein factor) is factored into a zero-mode exponential and a Majorana operator. Fixing the gauge of this internal symmetry then leads to a direct relationship between fermion parity and the products of Klein factors $i\eta^R \eta^L$.

With open boundary conditions, the situation is drastically different and much of it relates to the presence of topological MZMs. For one, unfolding consolidates left- and right-moving bosonic fields into a single chiral field in the extended domain. As a consequence, there is now only one zero mode $\vartheta_0$, and the compactification radii of the conjugate fields $\varphi(x),\vartheta(x)$ are doubled. This latter fact produces (quasi-)degenerate ground states in the relevant pinned phases of the sine-Gordon model. Such a spontaneous symmetry breaking by ordering in bosonic fields actually corresponds to a global symmetry breaking of fermion parity, which is forbidden by the boson-fermion superselection rule.\cite{PhysRev.88.101} Restoration of this symmetry then requires a finite amount of virtual tunneling between bosonic ground state sectors that can produce an exponentially suppressed energy splitting due to gapped bulk excitations.  

Secondly, consistent use of normal-ordered vertex operators leads to bosonic sine-Gordon potentials that are spatially modulated by the envelope function $\sin(\pi x/\LL)$. This is quite surprising and is another significant difference from the closed geometry case where such potentials are translationally invariant. We note that past studies, as far as we are aware, have overlooked this feature by misapplying the bosonization dictionary, and in particular often using unnormal-ordered vertex operators.
In particular, bosonizing the superconducting pairing potential results in a modulated cosine potential that vanishes at the boundaries. Considered in isolation, this cosine mass term will produce the topological phase with exact fermionic zero modes singularly localized at the boundary points. We then showed that kinetic energy fluctuations lift this exact degeneracy at finite lengths. We first demonstrate this at an effective semi-classical level. Our semi-classical analysis utilizes exact solutions to the Mathieu equation, which yield a finite-size quasi-degenerate energy splitting that scales as $\sim \e^{-a\sqrt{L}}$. This is in contrast to the $\sim \e^{-b {L}}$ behavior seen in the exact fermionic treatment and indicates that zero mode fluctuations, which are non-local in nature, are insufficient to qualitatively capture the restoration of fermion parity symmetry. Nevertheless, one can view this semi-classical result as an upper bound. 

Thirdly, the non-local semi-classical treatment points to the necessity of local bosonic fluctuations in order to properly capture the low-energy physics. Guided by our exact fermionic solutions, we then presented a full understanding of how this occurs entirely in the bosonic language. Our method uses vertex algebras and their operator product expansions in order to determine the bosonic analog of the BdG eigenvalue equation satisfied by the exact Majorana zero mode operators. We are then able to verify that the MZM operators in terms of bosons are identical to their fermionic counterparts. In particular, in the extreme SC limit, where kinetic fluctuations are quenched ($v_F = 0$), we recover local but singular forms of the MZM operators. Our analysis therefore provides a rigorous derivation of the results previously suggested by Refs.~\onlinecite{cheng2012superconducting,mazza2018poor}.


The totality of results -- especially the exact ones -- in this work reinforces our understanding of MZMs in one-dimensional SPT phases, whilst also demonstrating the many subtleties that can arise when applying bosonization. The fermionic and bosonic formulations are equivalent but complementary formulations of the same physics. As such, they naturally emphasize different aspects. The bosonic point of view of this SPT phase with open boundaries is essentially that of a spontaneously broken discrete symmetry, in this case Ising symmetry. Hence it more naturally describes the quasi-degenerate ground states of opposite fermion parity which are already accessible at the semi-classical level. However, it is at the expense of understanding the boundary MZMs which are truly quantum-mechanical objects. This aspect becomes more intuitive and natural in the fermionic picture. That complete equivalency is ever present between these dual points of view and the manner in which it occurs is probably one of the main takeaways of this work.


There are two extensions of this work which naturally suggest themselves. First, it will be interesting to study the effects of local density-density interactions to the exact free MZM operators. Previous studies including interactions have either been limited to closed wires\cite{gangadharaiah2011majorana} using renormalization group arguments or operate at the level of the lattice with the Jordan-Wigner transform.\cite{kells2015multiparticle} As is well known, bosonization is an indispensable tool to expose the integrability of the interacting Tomonaga-Luttinger liquid, which is obscured in the fermionic formulation. Therefore, a more detailed understanding of how the free fermion picture of boundary MZMs is changed by interactions at the level of quantum operators using bosonization is an attractive proposition. An approach based on the Schrieffer-Wolff transformation\cite{schrieffer1966relation,bravyi2011schrieffer} could serve this purpose.

The second extension would be to parafermionic theories, whereby the $\mathbb{Z}_2$ Ising symmetry as expressed in the bosonized theory is generalized to a $\mathbb{Z}_N$ group. The simplest theoretical setup which purportedly hosts topologically protected parafermionic boundary operators and the associated quasi-degenerate ground states involves proximity-induced SC on fractional quantum Hall edge states \cite{clarke2013exotic,lindner2012fractionalizing,vaezi2013fractional} and fractional topological insulators.\cite{klinovaja2014kramers} Thus far these models are best understood within a bosonized language, but a quantitatively detailed and rigorous description of the quasi-degenerate energy splitting between ground states and the topological parafermionic zero mode (PZM) operators remains lacking. The most common realization of PZMs in this context is that of interface quasiparticle operators that straddle domains of oppositely pinned bosonic order.

\begin{acknowledgments}
%
This work was supported by the Swiss National Science Foundation and NCCR QSIT.
We also would like to thank C. Reeg, M. Thakurathi, P. Aseev, S. Hoffman, D. Chevallier and S. D\'{i}az for useful discussions. VC would like to thank especially E. Fradkin for teaching him about (higher-dimensional) bosonization, Schwinger terms and how to read off commutators from classical actions. 

\end{acknowledgments}



\appendix


\section{The topological contribution to correlation functions}\label{app:corr}

\begin{figure}
	\begin{center}
		\includegraphics[width=0.45\textwidth]{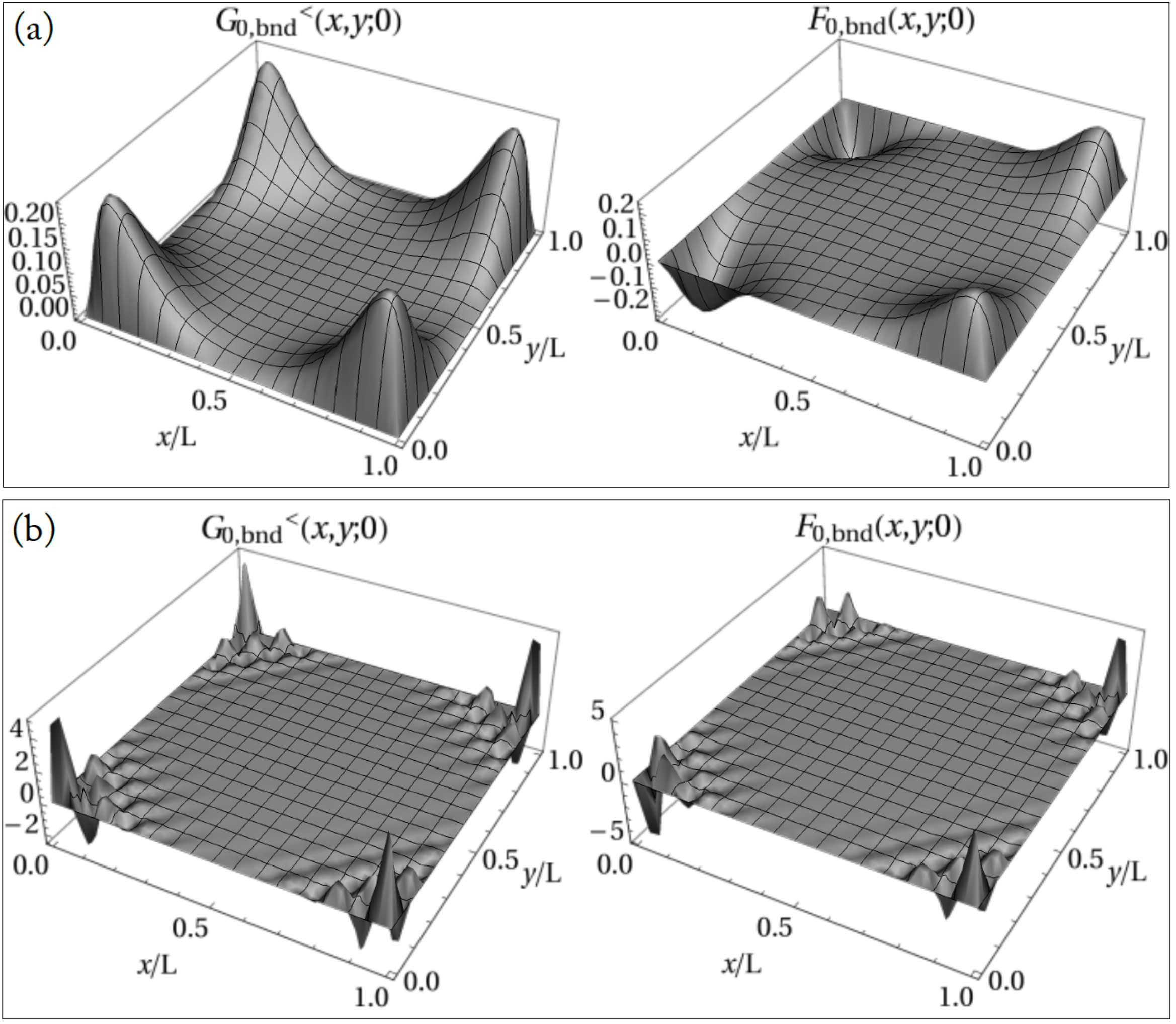}
	\end{center}
	\caption{Plots of the ground state static correlation functions $G_{0,\text{bnd}}^{<}(x,y;0)$ and $F_{0,\text{bnd}}(x,y;0)$ as defined in the main text. These are the topological boundary mode contributions to the $\langle \Psi^\dagger(x)\Psi(y)\rangle$ and $\langle \Psi(x)\Psi(y)\rangle$ correlation functions, respectively. The data shown is with $\kappa=12\LL^{-1}$ and Fermi momentum (a) $k_F = 1.5\pi/\LL$, (b) $k_F = 15.5\pi/\LL$.}
	\label{fig:GF0_bnd}
\end{figure}

In this appendix, we calculate the contribution to all single-particle correlation functions due to either the occupation or vacancy of the quasi-zero mode $\psi_{E_{i\kappa}}$ relative to the ground state $|\Omega_0\rangle$. The non time-ordered single particle Green's functions are defined by
\begin{subequations}
	\begin{align}
	G^<(x,y;t)&:= \langle \e^{iHt}\Psi^\dagger(x)\e^{-iHt} \Psi(y) \rangle, \\
	G^>(x,y;t)&:= \langle \e^{iHt}\Psi(x)\e^{-iHt} \Psi^\dagger(y) \rangle, \\
	F(x,y;t)&:= \langle \e^{iHt}\Psi(x)\e^{-iHt}\Psi(y)\rangle,  
	\end{align}
\end{subequations}
where $H$ is the many-body mean-field Hamiltonian. Retarded, advanced and time-ordered Green's functions may then be obtained from those above. Expansion of $\Psi^{(\dagger)}$ in terms of exact BdG quasi-particle creation and annihilation operators gives
\begin{subequations}
	\begin{align}
	&G^<(x,y;t)= \sum_E \left[
	\{\Psi^\dagger(x),\psi^\dagger_E\}\{\psi_E,\Psi(y)\}\langle \psi_E \psi_E^\dagger \rangle \e^{-iEt}\right. \nonumber \\ 
	&\hspace{0.25cm}\left.+\{\Psi^\dagger(x),\psi_E\}\{\psi^\dagger_E,\Psi(y)\}\langle \psi^\dagger_E\psi_E\rangle \e^{+iEt} \right],\\
	\nonumber \\
	&F(x,y;t)= \sum_E \left[
	\{\Psi(x),\psi^\dagger_E\}\{\psi_E,\Psi(y)\}\langle \psi_E \psi_E^\dagger \rangle \e^{-iEt}\right. \nonumber \\ 
	&\hspace{0.3cm}\left.+\{\Psi(x),\psi_E\}\{\psi^\dagger_E,\Psi(y)\}\langle \psi^\dagger_E\psi_E\rangle \e^{+iEt} \right],
\end{align}
\label{eqn:Correlators_fermions}%
\end{subequations}
with $G^>(x,y;t)$ obtained by the replacement $\Psi^\dagger \leftrightarrow \Psi$ in $G^<(x,y;t)$. 

As was stated, the difference between correlation functions in the quasi-degenerate ground states $|\Omega_0\rangle$ and $|\Omega_1\rangle$ is the occupation configuration of the boundary-localized quasi-zero mode $\psi_{E_{i\kappa}}= (\gamma_0 -i \gamma_\LL)/{2}$ composed of the localized Majoranas. Hence, we define the topological contributions to the Green's functions by 
\begin{align}
&G^<_{0,\text{bnd}}(x,y;t) = \nonumber \\
&\quad\{\Psi^\dagger(x),\psi_{E_{i\kappa}}^\dagger\}
\{\psi_{E_{i\kappa}},\Psi(y)\}\langle \Omega_0 |\psi_{E_{i\kappa}}\psi_{E_{i\kappa}}^\dagger|\Omega_0\rangle \e^{-i E_{i\kappa}t}, \\
&G^<_{1,\text{bnd}}(x,y;t) = \nonumber \\
&\quad\{\Psi^\dagger(x),\psi_{E_{i\kappa}}\}
\{\psi_{E_{i\kappa}}^\dagger,\Psi(y)\}\langle \Omega_1 |\psi_{E_{i\kappa}}^\dagger\psi_{E_{i\kappa}}|\Omega_1\rangle \e^{+i E_{i\kappa}t},
\end{align}
with $G^>_{(0,1),\text{bnd}}(x,y;t)$ defined analogously with $\Psi^\dagger \leftrightarrow \Psi$, and
\begin{align}
&F_{0,\text{bnd}}(x,y;t) = \nonumber \\ 
&\quad\{\Psi(x),\psi_{E_{i\kappa}}^\dagger\}
\{\psi_{E_{i\kappa}},\Psi(y)\}\langle \Omega_0 |\psi_{E_{i\kappa}}\psi_{E_{i\kappa}}^\dagger|\Omega_0\rangle \e^{-i E_{i\kappa}t}, \\
&F_{1,\text{bnd}}(x,y;t) = \nonumber \\ 
&\quad\{\Psi(x),\psi_{E_{i\kappa}}\}
\{\psi_{E_{i\kappa}}^\dagger,\Psi(y)\}\langle \Omega_1 |\psi_{E_{i\kappa}}^\dagger\psi_{E_{i\kappa}}|\Omega_1\rangle \e^{+i E_{i\kappa}t}.
\end{align}
Next, we define the auxiliary functions 
\begin{align}
&A_\kappa^{(\pm)}(x,y) := \nonumber \\
&\quad\mathcal{N}_\kappa^2 \left[
\sin(k_F x)\sin(k_F y) \sinh(\kappa[\LL-|x|]) \sinh(\kappa[\LL-|y|]) \right. \nonumber \\
&\quad\left.\hspace{0.2cm} \pm \cos(k_F x)\cos(k_F y)\sinh(\kappa x)\sinh(\kappa y)\right],\\
\nonumber \\
&B_\kappa^{(\pm)}(x,y) := \nonumber \\
&\quad\mathcal{N}_\kappa^2 \left[
\sin(k_F x)\cos(k_F y) \sinh(\kappa[\LL-|x|]) \sinh(\kappa y) \right. \nonumber \\
&\quad\left.\hspace{0.2cm} \pm \cos(k_F x)\sin(k_F y)\sinh(\kappa x)\sinh(\kappa [\LL-|y|])\right].
\end{align}   
The topological contributions then take the form
\begin{align}
G^{<(>)}_{0,\text{bnd}}(x,y;t) &= \left[A_\kappa^{(+)}(x,y) \mp B_\kappa^{(+)}(x,y) \right] \e^{-i E_{i\kappa}t}, \\
G^{<(>)}_{1,\text{bnd}}(x,y;t) &= \left[A_\kappa^{(+)}(x,y) \pm B_\kappa^{(+)}(x,y) \right] \e^{+i E_{i\kappa}t},\\
F_{0,\text{bnd}}(x,y;t) &= -\left[A_\kappa^{(-)}(x,y) - B_\kappa^{(-)}(x,y) \right] \e^{-i E_{i\kappa}t}, \\
F_{1,\text{bnd}}(x,y;t) &= -\left[A_\kappa^{(-)}(x,y) + B_\kappa^{(-)}(x,y) \right] \e^{+i E_{i\kappa}t}. 
\end{align}
Notice that the $B^{(\pm)}_\kappa(x,y)$ pieces represent non-local correlations involving bulk-separated boundaries. Nevertheless, the relative signs between $A^{(\pm)}_\kappa$ and $B^{(\pm)}_\kappa$ in the superpositions ensure that locality is restored in the limit that $E_{i\kappa}=0$ as $\LL \rightarrow \infty$. 

Shown in Fig.~\ref{fig:GF0_bnd} are example plots of the functions $G_{0,\text{bnd}}^{<}(x,y;t)$ and $F_{0,\text{bnd}}(x,y;t)$ in the equal-time limit $t=0$. They decay into the central bulk region (at a rate $\kappa$) and exhibit $k_F$ oscillations. Analogous plots for the first excited state $|\Omega_1\rangle$ show similar behavior but with opposite signs for correlation functions between boundary points. In principle, these non-local topological contributions to the correlation functions could be measured by scanning tunneling microscopy in short topological wires.


\begin{widetext}

\section{Current algebra and the bosonic action}\label{app:bosonic_action}

In this lengthy appendix, we present a careful derivation of the classical action for the bosonized continuum fermionic model in Eqn.~(\ref{eqn:L_setup_kF}). Specifically, the following equivalent forms of the action $S_\text{tot}$ will be derived, 
\begin{subequations}
\begin{align}
S_\text{tot} 
&\equiv -\pi \int \mathrm{d}t \int_{-\LL/2}^{\LL/2} \mathrm{d}x \int_{-\LL/2}^{\LL/2} \mathrm{d}y \; \zeta_1(x-y) \;j(x) \partial_t \rho(y) -  \int \mathrm{d}t \int_{-\LL/2}^{\LL/2} \mathrm{d}x\, \frac{\pi v_F}{2}\left( j(x)^2 + \rho(x)^2\right)\nonumber \\  
&\quad\,+ \int \mathrm{d}t  \left(
J\, \partial_t \varphi_0 - Q \, \partial_t \vartheta_0 - \frac{\pi v_F }{2\LL} (J^2 + Q^2) - \frac{\pi v_F }{\LL} ( Q  [\delta_Q-1]  +  J  \delta_J)
\right)  \\
&\equiv \int \mathrm{d}t\int_{-\LL/2}^{\LL/2} \mathrm{d}x \left\{
\frac{1}{\pi}\partial_x \vartheta \, \partial_t \varphi - \frac{v_F}{2\pi}(\partial_x\vartheta^2 + \partial_x \varphi^2)
\right\} - \int \mathrm{d}t \left[ Q  \partial_t \vartheta_0  + \frac{\pi v_F }{\LL} ( Q [\delta_Q-1]  +  J  \delta_J) \right] \\
&\equiv \int \mathrm{d}t\int_{-\LL/2}^{\LL/2} \mathrm{d}x \left\{
\frac{1}{\pi}\partial_x \varphi \, \partial_t \vartheta - \frac{v_F}{2\pi}(\partial_x\vartheta^2 + \partial_x \varphi^2)
\right\} + \int \mathrm{d}t \left[ J \partial_t\varphi_0 - \frac{\pi v_F }{\LL} ( Q  [\delta_Q-1]  +  J  \delta_J) \right],
\end{align}
\label{eqn:S_usual}
\end{subequations}
\end{widetext}
valid in the massless limit ($\mathrm{M}=\Delta=0$) with twisted boundary conditions (parametrized by $\delta_Q$, $\delta_J$) and finite length $\LL$. This derivation will involve the discussion of Schwinger terms in current algebras ($\rho,j$), zero momentum modes ($\vartheta_0,\varphi_0$), topological windings ($Q,J$), and the strict adherence to a finite length. 

There are many reasons for desiring a classical action of the bosonized theory, some of the most important ones being the application of perturbative renormalization group techniques and the development of semi-classical approximations at strong pinning. We will be mainly interested in the latter as we discuss fermion parity switching and the bosonized MZMs in the infinitely pinned limit. However, in this appendix we shall consider only the case of closed boundaries in the massless limit because the case of open boundaries with cosine mass terms is specially treated in the main text. 

The bosonic action is often taken for granted and is by now established textbook material. In spite of that, most standard references\cite{gogolin1999bosonization,giamarchi2004quantum,altland2010condensed,fradkin2013field} on the subject remain unclear about the fate of zero modes and twisted boundary conditions. In fact, the action is often quoted in its sine-Gordon form which follows from the integration of either one of the fields $\vartheta,\varphi$ at the expense of the other. However, the actions quoted above represent `phase-space' actions where the quantum commutation relations between conjugate pairs $(\vartheta,\varphi)$ and $(\rho,j)$ can be easily read off. This form of the action is particularly suited for the discussion of competing pinning potentials in $\vartheta$ and $\varphi$, which is awkward to do with the sine-Gordon formulation. Furthermore, there are additional terms on the right-hand sides of $S_\text{tot}$ above that are typically absent in most standard treatments. Yet they explicitly involve the zero modes, topological windings and twisted boundary conditions.

\subsection{Bosonization convention}

The existence of several competing bosonization conventions in the literature often leads to confusion and misunderstandings, %
	\footnote{This is acutely frustrating for graduate students learning bosonization for the first time. Ordinarily, specialists adhere to a single convention over time and lose the ability to translate between conventions.} 
see for example the comments by von Delft and Schoeller\cite{von_Delft_Schoeller} and the tower of babel appendix by Giamarchi.\cite{giamarchi2004quantum} To make matters worse, the form of the action $S_\text{tot}$ can depend on these details. For example, if one chooses to use non-commuting zero modes instead of Klein factors, then there should be an additional $ i\vartheta_0 \partial_t \varphi_0$ term in the Lagrangian. In another case, if one decides not to normal-order then the final bosonized action can acquire an explicit dependence on a UV cutoff scale. It is for reasons like these that we are compelled to restate our bosonization convention here. For the experts, we will just mention that our convention is closer to the one used by string theorists\cite{ginsparg1989applied,goddard1986kac,blumenhagen2009basics} and is a form of constructive or operator bosonization.\cite{haldane1981luttinger,von_Delft_Schoeller}

First, we recall the following bosonization identities [Eqn.~(\ref{eqn:bosonized_RL_normal})] for the right- and left-moving Fermi fields in a system of finite length $\LL$, 
\begin{subequations}
\begin{align}
R(x) &\equiv \frac{\eta^R}{\sqrt{\LL}}:\e^{i\phi^R(x)}:\,\e^{i\frac{\pi x}{\LL}(\delta_J + \delta_Q-1)},  \\
L(x) & \equiv \frac{\eta^L}{\sqrt{\LL}}:\e^{i\phi^L(x)}:\,\e^{i\frac{\pi x}{\LL}(\delta_J - \delta_Q+1)},
\end{align}
\end{subequations}
expressed in normal-ordered form. The chiral bosonic fields are compact bosons with the compactification radii
\begin{align}
\phi^R \sim \phi^R + 2\pi, \qquad \phi^L \sim \phi^L + 2\pi.
\end{align}
They have the mode expansions
\begin{align}
\phi^R(x) &=\phi^R_0 + \frac{2\pi Q^R x}{\LL} \nonumber \\ 
&+ \sum_{q\neq 0} \left({\frac{2\pi}{|q|\LL}}\right)^{\frac{1}{2}}\left(
\Theta(q)a_q + \Theta(-q)a^\dagger_{-q}
\right) \e^{i q x},\\
\phi^L(x) &=\phi^L_0 - \frac{2\pi Q^L x}{\LL} \nonumber \\
&- \sum_{q\neq 0} \left({\frac{2\pi}{|q|\LL}}\right)^{\frac{1}{2}}\left(
\Theta(-q)a_q + \Theta(q)a^\dagger_{-q}
\right) \e^{i q x},
\end{align}
where $q \in \frac{2\pi}{\LL}\mathbb{Z}$ and $\Theta(x)$ is the Heaviside step function. The only non-zero commutation relations amongst the operators are
\begin{align}
[\phi_0^R,Q^R] = -i, \quad [\phi_0^L,Q^L] = -i , \quad [a_q,a_{q'}^\dagger] = \delta_{qq'}.
\end{align}
Due to the compactification radii of $\phi^{R,L}$, the operators $Q^R,Q^L$ are integer-quantized and represent occupation numbers of the $R$ and $L$ fermions. The operators $\phi^{R,L}_0$ are known as the zero-momentum $(q=0)$ modes. The bosonic normal ordering $:\;:$ is defined such that all annihilation operators $a_q$ are ordered to the right of the creation operators $a_q^\dagger$ and the charges $Q^{R,L}$ are ordered to the right of the zero modes $\phi^{R,L}_0$. This entails Wick-ordering for the annihilation and creation operators and $qp$-ordering for the zero mode and number operators. For example, this means that 
\begin{align}
:\e^{i\phi^R(x)}: \;\equiv \e^{i\phi^R_0}\e^{i\frac{2\pi }{\LL}Q^R x}\,\e^{i\phi^R_+(x)} \e^{i\phi^R_-(x)},
\end{align}
where
\begin{align}
\phi_+^R(x) =  \sum_{q>0} \left({\frac{2\pi}{q\LL}}\right)^{\frac{1}{2}}a^\dagger_{q}
\e^{-i q x},\qquad 
\phi_-^R(x) = [\phi_+^R(x)]^\dagger,
\end{align}
and analogously for $\phi_\pm^L(x)$. With these relations, one can explicitly verify the commutation relations  
\begin{align}
[\phi^R(x),\phi^R(y)] &= +i 2\pi\,  \varepsilon_1(x-y), \\
[\phi^L(x),\phi^L(y)] &= -i 2\pi\,  \varepsilon_1(x-y),\\
[\phi^L(x),\phi^R(y)] &= 0,
\end{align}
and 
\begin{align}
[\phi_-^R(x+i0^+),\phi_+^R(y)] &= -\ln \left(1-\e^{i\frac{2\pi}{\LL}(x-y+i0^+)}\right),\\
[\phi_-^L(x+i0^+),\phi_+^L(y)] &= -\ln \left(1-\e^{-i\frac{2\pi}{\LL}(x-y-i0^+)}\right),
\end{align}
where the $i0^+$ convergence factors are remnants of the Wick normal-ordering prescription. The function $\varepsilon_1$ is defined as
\begin{align}
\varepsilon_1(x-y) &:= \frac{1}{\LL}(x-y) + \frac{i}{2\pi}
\ln \left(\frac{1-\e^{+\frac{i2\pi}{\LL}(x-y+ i 0^+)}}{1-\e^{-\frac{i2\pi}{\LL}(x-y-i0^-)}}\right) \nonumber \\
&= \Bigl \lceil \frac{x-y}{\LL} \Bigr\rceil -\frac{1}{2},
\end{align}
where $\lceil * \rceil$ denotes the integer ceiling. Because $\varepsilon_1$ is a function that depends on the difference of two logarithmic functions which are holomorphic on mutually exclusive regions of the complex plane, $\varepsilon_1$ should be understood to be a hyperfunction.\cite{graf2010introduction} It is also implied here that principal branches are used when evaluating the $\arg$ and $\ln$ multi-functions. More importantly, $\varepsilon_1$ is a Green's function\cite{stone2009mathematics} in the sense that 
\begin{align}
\varepsilon_1'(x-y)= \sum_{m \in \mathbb{Z}}\delta(x-y+m\LL),
\label{eqn:del_epsilon} 
\end{align}
where the RHS is the $\LL$-periodic Dirac delta function. One can also see from plotting $\varepsilon_1(x)$ that it has discontinuities at intervals of $\LL$, in accordance to its derivative being a Dirac delta comb. Finally, for $|x-y|<\LL$, we have 
\begin{align}
\varepsilon_1(x-y) = \frac{1}{2} \text{sgn}(x-y),
\end{align}
which gives the usual non-local commutator for $\phi^{R,L}$  in the infinite $\LL$ limit. 

Using these bosonic commutators, all of the canonical anticommutation relations for $R(x),L(x)$ can be reproduced; with the effects of twisted boundary conditions included. Verifying this may be done using the vertex operator product identities\cite{stone1994bosonization,blumenhagen2009basics}
\begin{subequations}
\begin{align}
&:\e^{i \alpha \phi^R(x)}:\;:\e^{ i\beta \phi^R(y)}:  \nonumber \\
&= 
\left[ \e^{-i\frac{2\pi x}{\LL}} - \e^{-i\frac{2\pi y}{\LL}}
\right]^{\alpha \beta} :\e^{i [\alpha \phi^R(x) + \beta \phi^R(y)]}:\,, \\ \nonumber \\
&:\e^{i \alpha \phi^L(x)}:\;:\e^{ i\beta \phi^L(y)}:  \nonumber \\
&= 
\left[ \e^{i\frac{2\pi x}{\LL}} - \e^{i\frac{2\pi y}{\LL}}
\right]^{\alpha \beta} :\e^{i [\alpha \phi^L(x) + \beta \phi^L(y)]}:
\end{align}
\end{subequations}
for $\alpha,\beta \in \mathbb{Z}$. These are derived by adhering to the normal-ordered product convention and using the Baker-Campbell-Hausdorff identity.

The bosonization and operator product identities may then be used to bosonize the fermion densities giving 
\begin{align}
&:R^\dagger(x)R(x):\;=\;:\rho^R(x): = +\frac{1}{2\pi} \partial_x \phi^R(x), \\
&:L^\dagger(x)L(x):\;=\;:\rho^L(x): = -\frac{1}{2\pi} \partial_x \phi^L(x),
\end{align}
where fermionic normal ordering is defined using point-splitting
\begin{align}
:\mathcal{A}(x)\mathcal{B}(x):\;=\; \lim_{\epsilon\rightarrow0} \left[\mathcal{A}(x+\epsilon)\mathcal{B}(x) - \langle 0|\mathcal{A}(x+\epsilon)\mathcal{B}(x)|0\rangle\right]
\label{eqn:point_split}
\end{align}
with $|0\rangle$ being the filled Fermi sea. Next, the local conjugate fields are defined by
\begin{align}
\vartheta(x) =\frac{\phi^L (x)+\phi^R(x)}{2}, \quad 
\varphi(x) = \frac{\phi^L(x)-\phi^R(x)}{2},
\end{align}
and have the commutators
\begin{subequations}
\begin{align}
[\varphi(x),\vartheta(y)] &= -i \pi \varepsilon_1(x-y), \\
[\varphi(x),\varphi(y)] &= [\vartheta(x),\vartheta(y)] =0.
\end{align}
\label{eqn:commute}%
\end{subequations}%
These fields are related to the charge density and current by 
\begin{align}
&\rho_\text{tot}(x)=\;:\rho^R(x) + \rho^L(x):\; = -\frac{1}{\pi}\partial_x \varphi(x), \\
&j_\text{tot}(x)= \;:\rho^R(x) - \rho^L(x):\; = +\frac{1}{\pi}\partial_x \vartheta(x). \label{eqn:j_theta} 
\end{align}
They have the mode expansions
\begin{align}
\varphi(x) &= \varphi_0 - \frac{\pi Q x}{\LL} - \sum_{q \neq 0}\left(\frac{\pi}{2|q|\LL}\right)^{\frac{1}{2}} (a_q + a_{-q}^\dagger)\e^{i q x}, \\
\vartheta(x) &= \vartheta_0 + \frac{\pi J x}{\LL} + \sum_{q \neq 0}\left(\frac{\pi }{2|q|\LL}\right)^{\frac{1}{2}}\, \text{sgn}(q) \,(a_q -a_{-q}^\dagger)\e^{iqx},
\end{align}
accompanied by the non-zero commutation relations
\begin{align}
[\vartheta_0,Q] = -i, \qquad [\varphi_0,J] = i.
\end{align}

The bosonic fields $\varphi,\vartheta$ are also compact bosons with compactification radii  
\begin{align}
\varphi \sim \varphi + \pi,\qquad \vartheta \sim \vartheta + \pi.
\end{align}
Their respective topological windings $Q,J$ are integer-quantized and subject to the parity condition
\begin{align}
(-1)^Q = (-1)^J = 1 \quad \Leftrightarrow \quad Q = J \text{ mod 2}.  
\end{align}
Employing the presented fermion-boson dictionary then translates the free-fermion Hamiltonian 
\begin{align}
H_0 = \int_{0}^{\LL} \left\{
-i v_F R^\dagger \partial_x R + i v_F L^\dagger \partial_x L 
\right\}\mathrm{d}x
\end{align}
into the bosonized Hamiltonian
\begin{align}
H_0 =& \int_{0}^{\LL} 
\frac{v_F}{2\pi}:\left(
[\partial_x \vartheta]^2 + [\partial_x \varphi]^2
\right):\mathrm{d}x \nonumber \\
&\hspace{1cm}+ \frac{\pi v_F }{\LL} \left(Q[\delta_Q-1]  +  J\delta_J\right).
\label{eqn:H_boson}
\end{align}
The additional $J\delta_J$ term arises from the diamagnetic response to the applied magnetic flux, while the $Q(\delta_Q-1)$ term may be interpreted as a Fermi pressure term.

The next goal is now to determine a classical action which, when quantized, does not only give the correct bosonized Hamiltonian but also the correct commutators between the bosonic fields $\varphi$ and $\vartheta$ as described above. This will be extremely useful when performing semi-classical approximations based on saddle points.

\subsection{Current algebra and Schwinger terms}

An unconventional way to dequantize the bosonic theory is to work directly with the non-zero-mode $(q \neq 0)$ contributions to the current and density. These are defined to be
\begin{align}
\rho(x) := \rho_\text{tot}(x) - \frac{Q}{\LL}, \quad 
j(x) := j_\text{tot}(x) - \frac{J}{\LL},
\end{align}
and constitute non-zero contributions to the current and density obeying
\begin{align}
&\int_{0}^{\LL} \rho(x) \mathrm{d}x =
\int_{0}^{\LL} j(x) \mathrm{d}x = 0.
\label{eqn:rho_j_constraints}
\end{align}
Their mode expansions are
\begin{subequations}
\begin{align}
\rho(x) &= \frac{i}{\LL}\sum_{q\neq 0}\left(\frac{|q|\LL}{2\pi}\right)^{\frac{1}{2}}\text{sgn}(q)\, [a_q+a_{-q}^\dagger]\,\e^{iqx}, \\
j(x) &= \frac{i}{\LL}\sum_{q \neq 0} \left(\frac{|q|\LL}{2\pi}\right)^{\frac{1}{2}} [a_q-a_{-q}^\dagger]\,\e^{iqx}.
\end{align}
\label{eqn:rho_j_mode}%
\end{subequations}
These expressions are independent of $Q$ and $J$ since  
\begin{align}
&[\rho,Q]=[\rho,J]=0, \quad [j,Q]=[j,J]=0.
\end{align}
Thus, $Q$ and $J$ are the constant background charge and current. However, although $Q$ and $J$ commute with each other, the non-zero-mode counterparts $\rho$ and $j$ do not. Instead they satisfy a special commutation relation known as \emph{current algebra} 
\begin{align}
[\rho(x),j(y)] = -\frac{i}{\pi} \sum_{m \in \mathbb{Z}}\delta'(x-y-m\LL) = -\frac{i}{\pi}\varepsilon_1''(x-y).
\label{eqn:current_alg}
\end{align}
The right-hand side of the above equation is understood in the sense of distributions, such that
\begin{align}
\int_{0}^{\LL} \mathrm{d}x \int_{0}^{\LL} \mathrm{d}y\; g(x)\, \delta'(x-y)\, f(y) 
= \int_{0}^{\LL} \mathrm{d}x \; g(x) \,\partial_x f(x).
\end{align}
The $\delta'$ term in Eqn.~(\ref{eqn:current_alg}) is known as a \emph{Schwinger term} and is an ultra-local term because it depends on the derivative of the Dirac delta function.

Now, reversing the Dirac quantization scheme yields the classical bracket
\begin{align}
\{\rho(x),j(y)\} = -\frac{1}{\pi} \sum_{m\in \mathbb{Z}}\delta'(x-y+m\LL). 
\label{eqn:classical_current_alg}
\end{align}
More generally, the Poisson bracket for functionals $F[\rho,j],G[\rho,j]$ is implied to be 
\begin{subequations}
\begin{align}
\{F,G\} &:= \int \mathrm{d}x\int \mathrm{d}y\; \kappa(x,y)
\left(
\tfrac{\delta F}{\delta \rho(x)}\tfrac{ \delta G}{\delta j(y)}
-\tfrac{\delta G}{\delta \rho(x)}\tfrac{ \delta F}{\delta j(y)}
\right) \nonumber \\
&= -\tfrac{1}{\pi} \int \mathrm{d}x 
\left[
\tfrac{\delta F}{\delta \rho(x)} \partial_x \left(\tfrac{ \delta G}{\delta j(x)}\right)
-\tfrac{\delta G}{\delta \rho(x)}\partial_x \left(\tfrac{ \delta F}{\delta j(x)}\right)
\right],
\end{align}
where \begin{align}
\kappa(x,y) &:= -\tfrac{1}{\pi}\sum_{m\in \mathbb{Z}}\delta'(x-y+m\LL). 
\end{align}
\end{subequations}
One can check that all the properties of a Poisson bracket are satisfied: skew symmetry, associativity (Leibniz rule) and Jacobi identity. 

Generally, the classical Poisson bracket and the quantum (Dirac) commutator do not necessarily agree. The situations when they do not agree require the methods of Deformation Quantization.\cite{sternheimer1998deformation} Fortunately, within the context of one-dimensional bosonization, the underlying commutator algebra is that of the simple harmonic oscillator -- i.e., the Heisenberg algebra -- such that the Dirac prescription remains exact.

Since the Schwinger term $\delta'$ is ultra-local, the (functional) symplectic form that is taken from the inverse of the Poisson bracket kernel function is expected to be non-local. Let $X,Y$ be functional vector fields
\begin{align*}
X &= \int \mathrm{d}x \left(
X_\rho (x)\frac{\delta}{\delta \rho(x)} + X_j(x) \frac{\delta }{\delta j(x)}
\right), \\
Y &= \int \mathrm{d}x \left(
Y_\rho (x)\frac{\delta}{\delta \rho(x)} + Y_j(x) \frac{\delta }{\delta j(x)}
\right). 
\end{align*}
Expressing the symplectic 2-form $\Omega$ as
\begin{align}
\Omega[X,Y] := \int \mathrm{d}x\int \mathrm{d}y \; \uplambda(x,y) \left[
X_j(x)Y_\rho(y) - Y_j(x)X_\rho(y)
\right]
\end{align}
then requires that 
\begin{align}
\int \mathrm{d}z \, \kappa(x,z)\uplambda(z,y) = -\tfrac{1}{\pi}\partial_x\uplambda(x,y) =\sum_{m\in \mathbb{Z}}\delta(x-y+m\LL).
\end{align}
Thus, $\uplambda$ is a Green's function to the derivative operator $\partial_x$, and using Eqn.~(\ref{eqn:del_epsilon}) it must have the form
\begin{align}
\uplambda(x,y) = -\pi \left[ \varepsilon_1(x-y) + C(y) \right], 
\end{align}
where $C(y)$ is only a function of $y$. Typically this is resolved by selecting boundary conditions. The most natural choice happens to be 
\begin{align}
C(y) \equiv  0,
\end{align}
which ensures that $\uplambda(x,y) = -\uplambda(y,x)$. Taking this on faith for the moment, we then have the symplectic 2-form 
\begin{align}
&\Omega[X,Y] \nonumber\\
&=  -\pi \int_{0}^{\LL} \mathrm{d}x \int_{0}^{\LL} \mathrm{d}y \; \varepsilon_1(x-y) \left[
X_j(x)Y_\rho(y) - Y_j(x)X_\rho(y) 
\right].
\label{eqn:symplectic}
\end{align}
This leads to the real-time finite $\LL$ phase-space classical Lagrangian for $\rho$ and $j$
\begin{subequations}
\begin{align}
\mathcal{L}[\rho,j] &= -\pi \int_{0}^{\LL} \mathrm{d}x \int_{0}^{\LL} \mathrm{d}y \; \varepsilon_1(x-y) \;j(x) \partial_t \rho(y)\nonumber \\ 
&\quad - H[\rho,j], \\
H[\rho,j] &:=  \int_0^\LL \mathrm{d}x\, \frac{\pi v_F}{2}\left( j(x)^2 + \rho(x)^2\right),
\end{align}
\end{subequations}
where the Hamiltonian is inferred from Eqn.~(\ref{eqn:H_boson}). Here $\rho$ is taken to be the ``position'' coordinate and $j$ the ``momentum'', although they are clearly interchangeable by an integration by parts in time. Variation of the action leads to the equations of motion
\begin{subequations}
 \begin{align}
\rho(x,t) &= -\frac{1}{v_F} \int  \mathrm{d}y \, \varepsilon_1(x-y) \, \partial_t j(y,t),   \\
j(x,t) &= -\frac{1}{v_F} \int  \mathrm{d}y \, \varepsilon_1(x-y) \, \partial_t \rho(y,t).
\end{align}
\end{subequations}
Comparing these equations of motion with the mode expansions Eqn.~(\ref{eqn:rho_j_mode}) and the known time-dependencies $a_q \e^{-i |q|t}$, $a_{-q}^\dagger \e^{i|q|t}$ demonstrates that they are \emph{incorrect}. However, this gives us hints towards the correct form of the RHS. We note that we require a function $\zeta_1(x-y)$ with the properties that
\begin{align}
&\zeta_1(x-y) = - \zeta_1(y-x) 
\end{align} and \begin{align}
&\int_0^\LL \mathrm{d}y \, \zeta_1(x-y) \, \partial_y(\e^{i q y}) = \e^{iq x}
\end{align}
whenever $q \in \frac{2\pi}{\LL} \mathbb{Z}$ and $q \neq 0$. This then leads to the unique solution
\begin{align}
\zeta_1(x-y) &:= \varepsilon_1(x-y) -\frac{x-y}{\LL} \nonumber \\
&= -\frac{1}{\pi}\text{arg}(1-\e^{+\frac{i2\pi}{\LL}(x-y+i0^+)}) \nonumber \\
&= -\frac{i}{ 2\pi}\sum_{n \neq 0} \frac{\left(\e^{\frac{i2\pi }{\LL}(x-y)}\right)^n}{n} \nonumber \\
&= \sum_{n> 0}\frac{\sin\left[\frac{2\pi n}{\LL}(x-y)\right]}{\pi n},
\end{align}
which is just the descending saw-tooth function of period $\LL$. This function is not quite a Green's function because 
\begin{align}
\partial_x \zeta_1(x-y) = \sum_{m\in\mathbb{Z}}\delta(x-y+ m\LL) - \frac{1}{\LL}.
\end{align}
However, when limiting ourselves to functions which integrate to zero over the interval $[0,\LL]$, the above expression still shows the behavior expected of a Green's function. Incorporating this subtle issue, we arrive at the correct final $\LL$-periodic Lagrangian  
\begin{subequations}
\begin{align}
\mathcal{L}[\rho,j] &= -\pi \int_{-\LL/2}^{\LL/2} \mathrm{d}x \int_{-\LL/2}^{\LL/2} \mathrm{d}y \; \zeta_1(x-y) \;j(x) \partial_t \rho(y)\nonumber \\ 
&\quad - H[\rho,j], \\
H[\rho,j] &:=  \int_0^\LL \mathrm{d}x\, \frac{\pi v_F}{2}\left( j(x)^2 + \rho(x)^2\right),
\end{align}
\label{eqn:action_rho_j}
\end{subequations}
which yields the equations of motion 
\begin{subequations}
 \begin{align}
\rho(x,t) &= -\frac{1}{v_F} \int  \mathrm{d}y \, \zeta_1(x-y) \, \partial_t j(y,t),   \\
j(x,t) &= -\frac{1}{v_F} \int  \mathrm{d}y \, \zeta_1(x-y) \, \partial_t \rho(y,t).
\end{align}
\end{subequations}
The correctness of these equations can be verified from the mode expansions of $\rho$ and $j$. These give the conservation laws
\begin{align}
\partial_t \rho + v_F \partial_x j =0,\qquad
\partial_t j + v_F \partial_x \rho =0,
\end{align}
where the first equation is the continuity equation and the second equation comes from the duality $\rho \leftrightarrow j$ that is only valid for (1+1)-dimensional massless relativistic fluids. Both lead to the more conventional massless Klein-Gordon equations of motion 
\begin{align}
(\partial_t^2 - v_F^2 \partial_x^2) \rho = 0 ,\qquad
(\partial_t^2 - v_F^2 \partial_x^2) j= 0.
\end{align}
As a final technical remark, we note that in determining the classical action, we have made use of the fact that the quantum Hamiltonian Eqn.~(\ref{eqn:H_boson}) is boson normal-ordered according to our earlier described prescription. Implicitly, this means that the construction of the partition function/time-evolution operator in terms of a functional integral unambiguously employs the Wick-ordering of the $a_q,a_q^\dagger$ operators. It should be stressed that different operator ordering conventions will in general yield different functional integral actions.\cite{zinn1996quantum,berezin1980feynman}  

\subsection{Full action for the bosonic fields}

The non-local action in Eqn.~(\ref{eqn:action_rho_j}) was originally written down in a related form for the chiral boson by Floreanini and Jackiw\cite{floreanini1987self} in the case of infinite $\LL$. However, this formulation is rarely encountered in the literature. To connect it to Eqn.~(\ref{eqn:S_usual}) we recall that 
\begin{align}
\rho(x) = -\frac{1}{\pi} \partial_x \varphi(x) - \frac{Q}{\LL}, \quad
j(x) = \frac{1}{\pi}\partial_x \vartheta(x) - \frac{J}{\LL}.
\end{align}
This leads to the following relationship between the angular fields $\vartheta,\varphi$ and $\rho$,$j$:
\begin{align}
\varphi(x) &= \varphi_0 - \frac{\pi Q}{\LL}x - \pi \int_{-\LL/2}^{\LL/2} \zeta(x-y) \rho(y)\, \mathrm{d}y, \\
\vartheta(x) &= \vartheta_0 + \frac{\pi J}{\LL}x + \pi \int_{-\LL/2}^{\LL/2} \zeta(x-y) j(y)\, \mathrm{d}y.
\end{align}
In fact, these expressions can be verified by substituting the mode expansions Eqns.~(\ref{eqn:rho_j_mode}) and performing a spatial integration by parts. Thus we have succeeded in simply expressing the angular fields $\vartheta,\varphi$ as a \emph{functional} of the zero modes $\vartheta_0,\varphi_0$, the constant background charges $Q,J$ and the current/density fields $j,\rho$. The advantage of these expressions is that they allow us to appreciate the non-local dependence between the $\vartheta,\varphi$ fields and the normally measurable observables $Q,J$ and $\rho,j$. 

An important point to notice is that the zero modes $\varphi_0,\vartheta_0$ represent spatial averages of $\varphi(x),\vartheta(x)$ and may be extracted according to 
\begin{align}
\varphi_0 = \frac{1}{\LL}\int_{-\LL/2}^{\LL/2} \varphi(x) \, \mathrm{d}x, \quad
\vartheta_0 =  \frac{1}{\LL}\int_{-\LL/2}^{\LL/2} \vartheta(x) \, \mathrm{d}x.
\end{align}
However, these relations are very much dependent on the limits of integration despite the $\LL$-periodicity. Moreover, it may be sometimes desirable to use different limits of integration such as   
\begin{align}
\varphi_0 = \frac{1}{\LL}\int_{0}^{\LL} \tilde{\varphi}(x) \, \mathrm{d}x, \quad
\vartheta_0 =  \frac{1}{\LL}\int_{0}^{\LL} \tilde{\vartheta}(x) \, \mathrm{d}x,
\end{align}
which requires using a different set of fields defined by 
\begin{align}
\tilde{\varphi}(x) &= \varphi_0 - \frac{\pi Q}{\LL}\left(x-\frac{\LL}{2}\right) - \pi \int_{0}^{\LL} \zeta(x-y) \rho(y)\, \mathrm{d}y, \\
\tilde{\vartheta}(x) &= \vartheta_0 + \frac{\pi J}{\LL}\left(x-\frac{\LL}{2}\right) + \pi \int_{0}^{\LL} \zeta(x-y) j(y)\, \mathrm{d}y.
\end{align}
These fields  differ from $\varphi(x),\vartheta(x)$ by multiples of $\pi Q,\pi J$. Equivalently, the zero modes of $\tilde{\varphi}(x),\tilde{\vartheta}(x)$ and $\varphi(x),\vartheta(x)$ differ by multiples of $\pi Q,\pi J$. In fact, the commutation relations 
\begin{align}
[\vartheta_0,Q] = -i, \quad [\varphi_0,J] = i, \quad [\vartheta_0,\varphi_0]=0 
\end{align}
remain unchanged under the `gauge' transformation 
\begin{align}
\vartheta_0 \rightarrow \vartheta_0 +  c J, \quad \varphi_0 \rightarrow \varphi_0 -  c Q
\end{align}
for any $c \in \mathbb{R}$. Thus, fixing a spatial averaging convention implicitly fixes this gauge choice in the zero modes. Counterintuitively, this also means that the zero modes are not coordinate invariant.

Finally, we can add the action capturing the physics of $Q,J$ and $\varphi_0,\vartheta_0$ to the current-algebra Lagrangian $\mathcal{L}[\rho,j]$ in Eqn.~(\ref{eqn:action_rho_j}), which yields the total action
\begin{widetext}
\begin{align}
S_\text{tot}[\varphi,\vartheta] &= -\pi \int \mathrm{d}t \int_{-\LL/2}^{\LL/2} \mathrm{d}x \int_{-\LL/2}^{\LL/2} \mathrm{d}y \; \zeta_1(x-y) \;j(x) \partial_t \rho(y) -  \int \mathrm{d}t \int_{-\LL/2}^{\LL/2} \mathrm{d}x\, \frac{\pi v_F}{2}\left( j(x)^2 + \rho(x)^2\right)\nonumber \\  
&\quad\,+ \int \mathrm{d}t  \left(
J\, \partial_t \varphi_0 - Q \, \partial_t \vartheta_0 - \frac{\pi v_F }{2\LL} (J^2 + Q^2) - \frac{\pi v_F }{\LL} ( Q  [\delta_Q-1]  +  J  \delta_J)
\right).
\label{eqn:S_best} 
\end{align}
After some manipulation, this gives us the other two equivalent forms in Eqns.~(\ref{eqn:S_usual}). However, the form given in Eqn.~(\ref{eqn:S_best}) reveals the separation between the zero and non-zero modes. In fact, one can already read off the correct commutation relations amongst the zero modes and the topological charge/current from the action above. The other more common forms of the action in Eqns.~(\ref{eqn:S_usual}) obscure this separation between zero and non-zero modes, but highlight the need to include the zero-mode contributions $Q \partial_t \varphi_0$ and $ J \partial_t \vartheta_0$, which are often wrongly neglected in the infinite $\LL$ limit. 
It is furthermore worth emphasizing that the limits of integration play an important role here. To illustrate this, let us outline one of the necessary derivations connecting Eqn.~(\ref{eqn:S_best}) and the rest of Eqns.~(\ref{eqn:S_usual}). The temporal term for $\rho,j$ can be seen to be  
\begin{align*}
&-\pi \int_{-\LL/2}^{\LL/2}\mathrm{d}x \int_{-\LL/2}^{\LL/2}\mathrm{d}y \; j(x) \partial_t [\zeta(x-y)\rho(y)] \\
&= \frac{1}{\pi}\int_{-\LL/2}^{\LL/2} \mathrm{d}x
\left[
\partial_x \vartheta(x) + \frac{\pi J}{\LL} 
\right] \partial_t
\left[
\varphi(x) - \varphi_0 + \frac{\pi Q}{\LL}x 
\right] \\
&= \frac{1}{\pi}\int_{-\LL/2}^{\LL/2} \mathrm{d}x \, \partial_x \vartheta(x) \partial_t \varphi(x) - J \partial_t \varphi_0 
+ \partial_t Q \int_{-\LL/2}^{\LL/2}\mathrm{d}x\left(\frac{x}{\LL}\right)\partial_x \vartheta(x). 
\end{align*}
The last term on the RHS can be integrated by parts to give zero due to boundary conditions. Thus we have that 
\begin{align*}
-\pi \int_{-\LL/2}^{\LL/2}\mathrm{d}x \int_{-\LL/2}^{\LL/2}\mathrm{d}y \; j(x) \partial_t [\zeta(x-y)\rho(y)] + J \partial_t \varphi_0  = \frac{1}{\pi}\int_{-\LL/2}^{\LL/2} \mathrm{d}x \, \partial_x \vartheta(x) \partial_t \varphi(x)
\end{align*}
\end{widetext}
as desired. If we had instead chosen the less symmetric limits $\int_0^\LL\mathrm{d}x$, then  the action for $\tilde{\varphi}$ and $\tilde{\vartheta}$ would have been obtained.

Finally, we may also express the action in terms of the chiral fields $\phi^R(x),\phi^L(x)$. By the change of variables 
\begin{align}
\phi^R(x) &= \vartheta(x)-\varphi(x), \\ 
\phi^L(x) &= \vartheta(x)+\varphi(x),
\end{align}
one obtains
\begin{align}
S_\text{tot} [\phi^R,\phi^L] &= \int \mathrm{d}t \left( L_R[\phi^R] + L_L[\phi^L] \right) 
\end{align} with the Lagrangians
\begin{align}
&L_R[\phi^R]  =-\frac{1}{4\pi} \int_{-\LL/2}^{\LL/2}  \left(
\partial_t \phi^R + v_F \partial_x \phi^R
\right) \partial_x \phi^R\; \mathrm{d}x \nonumber \\
&= -Q^R \partial_t \phi^R_0 -\pi \int_{-\LL/2}^{\LL/2}\mathrm{d}x \int_{-\LL/2}^{\LL/2}\mathrm{d}y \; \rho^R(x)  \zeta(x-y) \partial_t\rho^R(y) \nonumber \\
&\quad\,-\frac{\pi v_F}{\LL}(Q^R)^2 - \pi{v_F}\int_{-\LL/2}^{\LL/2} [\rho^R(x)]^2 \mathrm{d}x
\end{align} and
\begin{align}
&L_L[\phi^L]  =+\frac{1}{4\pi} \int_{-\LL/2}^{\LL/2}  \left(
\partial_t \phi^L - v_F \partial_x \phi^L
\right) \partial_x \phi^L\; \mathrm{d}x \nonumber \\
&= -Q^L \partial_t \phi^L_0 + \pi \int_{-\LL/2}^{\LL/2}\mathrm{d}x \int_{-\LL/2}^{\LL/2}\mathrm{d}y \; \rho^L(x)  \zeta(x-y) \partial_t\rho^L(y) \nonumber \\
&\quad\,-\frac{\pi v_F}{\LL}(Q^L)^2 - \pi {v_F}\int_{-\LL/2}^{\LL/2} [\rho^L(x)]^2 \mathrm{d}x,
\end{align}
where we also have the mode expansion
\begin{align}
\phi^R(x) &= \phi^R_0 + \frac{2\pi Q^R}{\LL} x + 2\pi \int_{-\LL/2}^{\LL/2} \zeta(x-y) \rho^R(y)\, \mathrm{d}y, \\
\phi^L(x) &= \phi^L_0 + \frac{2\pi Q^L}{\LL} x + 2\pi \int_{-\LL/2}^{\LL/2} \zeta(x-y) \rho^L(y)\, \mathrm{d}y.
\end{align}



\section{Direct verification of the diagonalized Hamiltonian}\label{app:direct_check}

This appendix presents an explicit verification of the Hamiltonian diagonalization in Eqn.~(\ref{eqn:Ham_final_diag}). Recall first that the Hamiltonian is
\begin{align}
H &= \int_0^\LL \mathrm{d}x \left[-iv_F (R^\dagger \partial_x R - L^\dagger \partial_x L) + i \Delta(R^\dagger L^\dagger -LR)  \right], 
\end{align}
which upon unfolding into the extended domain $[-\LL,\LL]$ via
\[
L(x) \equiv -R(-x),\qquad R(x+2\LL)= -R(x)	
\]
leads to
\begin{subequations}
\begin{align}
H &:= H_0 + H_1, \\
H_0 &:= \int_{-\LL}^\LL\mathrm{d}x\left(
-iv_F R^\dagger \partial_x R \right), \\
H_1 &:= \int_{-\LL}^\LL\mathrm{d}x\left(- \tfrac{i \Delta}{2}s(x) [R^\dagger(x)R^\dagger(-x) - R(-x)R(x)] \right),
\end{align}
\end{subequations}
where $s(x) := \text{sgn}(\sin(\tfrac{\pi x}{\LL}))$ is the periodic square wave form. The presence of $s(x)$ introduces two domain walls in the SC pairing potential at $x=0,\pm \LL$, where we expect to find localized Majorana modes. 

\subsection{Majorana zero mode operators}

The Majorana operator localized around $x=0$ is claimed to be 
\begin{align}
\gamma_0 &= \int_{-\LL}^\LL\mathrm{d}x\; a(x) [R(x)+R^\dagger(x)] = \gamma_0^\dagger ,
\end{align}
where 
\begin{align}
a(x) &= \mathcal{N}_\kappa \sinh(\kappa[\LL-|x|])\qquad \text{for}\; x \in [-\LL,\LL], \\
\mathcal{N}_\kappa &= \frac{1}{\sqrt{\LL}}\left(\frac{\sinh(2\kappa \LL)}{2\kappa \LL}-1\right)^{-1/2}.
\end{align}
The normalization $\int_{-\LL}^\LL a(x)^2 \mathrm{d}x = 1 $ implies $\gamma_0^2 = 1$. The inverse decay length $\kappa >0 $ satisfies the gap equation 
\begin{align}
\frac{v_F \kappa }{\Delta} = \tanh(\kappa \LL),
\end{align}
which implies the relations
\begin{subequations}
\begin{align}
v_F\kappa  &= E_{i\kappa} \sinh(\kappa \LL), \\
\Delta &= E_{i\kappa} \cosh(\kappa \LL), 
\end{align}
\label{eqn:relations}
\end{subequations}
where 
\begin{align}
E_{i\kappa} = \sqrt{\Delta^2+v_F^2 (i\kappa)^2} = \sqrt{\Delta^2-v_F^2 \kappa^2}
\end{align}
is the quasi-degenerate energy splitting. Note that $a(x)$ is continuous throughout $[-\LL,\LL]$ and is an even function, $a(-x) = a(x)$. 

The goal is then to verify that 
\begin{align}
H = -\tfrac{i}{2}E_{i\kappa}\gamma_0\gamma_\LL + \ldots,
\end{align}
where $\gamma_\LL$ is the other Majorana operator localized around $x=\pm\LL$. This is achieved by checking that 
\begin{align}
[H,\gamma_0] = i E_{i\kappa}\gamma_\LL.
\end{align}
First, the equation of motion for $R(x),R^\dagger(x)$ under time evolution by $H_0$ 
\begin{align}
&R^{(\dagger)}(x,t) := \e^{iH_0 t}R^{(\dagger)}(x)\e^{-iH_0 t} = R^{(\dagger)}(x-v_F t) 
\end{align}
produces the commutation relation
\begin{align}
i[H_0, R^{(\dagger)}(x)] = -v_F \partial_x R^{(\dagger)}(x) 
\label{eqn:eqm_R_H0}
\end{align}
and hence 
\begin{align}
[H_0,R(x) + R^\dagger(x)] = iv_F [\partial_x R(x) + \partial_x R^\dagger(x)]. 
\end{align}
Thus, we find
\begin{align}
[H_0,\gamma_0] 
&=iv_F \int_{-\LL}^\LL \mathrm{d}x \; a(x) \, [\partial_x R(x) + \partial_x R^\dagger(x)] \nonumber\\
&= E_{i\kappa}\mathcal{N}_\kappa \int_{-\LL}^\LL\mathrm{d}x\; s(x)\sinh(\kappa\LL)\cosh(\kappa[\LL-|x|])\nonumber\\ &\quad\times[R(x)+R^\dagger(x)], 
\end{align}
where we have integrated by parts and used Eqns.~(\ref{eqn:relations}) on the final line. In performing the integration by parts, we have used the fact that there are no boundary term contributions because $a(\pm \LL)=0$. 

\begin{widetext}
Next, we also have
\begin{align}
[H_1,\gamma_0] 
&=  -\frac{i\Delta}{2}\int_{-\LL}^\LL\mathrm{d}x\int_{-\LL}^\LL\mathrm{d}y\, 
s(x)a(y)\nonumber \\
&\hspace{1cm}\times \left\{
R^\dagger(x)\delta(x+y) -R^\dagger(-x)\delta(x-y) -R(-x)\delta(x-y) + R(x)\delta(x+y)
\right\} \nonumber\\
&= -\frac{i\Delta}{2}\int_{-\LL}^\LL\mathrm{d}x \, s(x) a(x)
\left\{
R^\dagger(x) -R^\dagger(-x) -R(-x) + R(x)\right\}\quad \text{because $a(x)$ is even} \nonumber \\
&= -{i\Delta}\int_{-\LL}^\LL\mathrm{d}x \, s(x) a(x)
\left\{
R^\dagger(x)+R(x)\right\}\quad \text{because $s(x)$ is odd}.
\end{align}
Then, substituting the $\sinh$ expression for $a(x)$ and using Eqns.~(\ref{eqn:relations}) results in
\begin{align}
[H_1,\gamma_0]  &=-i\Delta \mathcal{N}_\kappa \int_{-\LL}^\LL \mathrm{d}x\, s(x) \sinh(\kappa[\LL-|x|])\, [R^\dagger(x)+R(x)] \nonumber\\
&= -iE_{i\kappa}\mathcal{N}_\kappa \int_{-\LL}^\LL\mathrm{d}x\, \cosh(\kappa \LL) \sinh(\kappa [\LL-|x|])[R(x)+R^\dagger(x)].
\end{align}
Assembling things together finally gives
\begin{align}
[H_0+H_1,\gamma_0 ] &= iE_{i\kappa}\mathcal{N}_\kappa \int_{-\LL}^\LL \mathrm{d}x\, s(x) \left\{
\sinh(\kappa\LL)\cosh(\kappa[\LL-|x|]) - \cosh(\kappa \LL)\sinh(\kappa[\LL-|x|]) \right\} \, [R(x)+R^\dagger(x)] \nonumber \\
&= iE_{i\kappa} \int_{-\LL}^\LL \mathrm{d}x\, \mathcal{N}_\kappa s(x) \sinh(\kappa|x|)\,[R(x)+R^\dagger(x)] \nonumber\\
&= iE_{i\kappa}\gamma_\LL
\end{align}
\end{widetext}
as claimed because the other Majorana operator is 
\begin{align}
\gamma_\LL = \int_{-\LL}^\LL \mathrm{d}x\, \mathcal{N}_\kappa \sinh(\kappa x )\,[R(x)+R^\dagger(x)].
\end{align}
By similar manipulations as above, one can easily show that 
\begin{align}
[H,\gamma_\LL] = -i E_{i\kappa}\gamma_0.
\end{align}
In this case, however, one needs to use the fact that $R(\LL)+R(-\LL)=0$ at the boundary. 

\subsection{Extended states operators}

We can follow the same strategy as above to verify that 
\begin{align}
[H,\phi_{n}^\dagger] = E_{k_{n,1}}\phi_n^\dagger, \qquad 
[H,\varphi_n^\dagger] = E_{k_{n,2}}\varphi_n^\dagger,
\end{align}
where we have the purported eigensolutions
\begin{subequations}
\begin{align}
\phi_{n} &= \int_{-\LL}^\LL \mathrm{d}x\; \frac{u_{k_{n,1}}(x)^*}{\sqrt{2}}\left[R(x) + R^\dagger(x)\right], \\
u_k(x) &= \frac{1}{\sqrt{\LL \mathrm{N}_k}}\left(
\cos(k|x| + 2\theta_k)+ i \sin(kx) 
\right), 
\end{align} 
\end{subequations}
\begin{subequations}
\begin{align}
\varphi_n &= \int_{-\LL}^\LL \mathrm{d}x\; \frac{v_{k_{n,2}}(x)^*}{\sqrt{2}}\left[R(x) - R^\dagger(x)\right], \\
v_k(x) &= \frac{1}{\sqrt{\LL \mathrm{N}_k}}\left(
\cos(k|x| - 2\theta_k)+ i \sin(kx)
\right).
\end{align}
\end{subequations}
The momenta $k_{n,1}$ and $k_{n,2}$ above have to satisfy the quantization conditions
\begin{align}
&k_{n,1} = \left(n +\frac{1}{2}\right)\frac{\pi}{\LL} - \frac{2\theta_{k_{n,1}}}{\LL} , \\
&k_{n,2} = \left(n +\frac{1}{2}\right)\frac{\pi}{\LL} + \frac { 2\theta_{k_{n,2}}} {\LL},
\end{align}
with $n \in \mathbb{N}$ and   
\begin{align}
2\theta_k = \tan^{-1}\left(\frac{\Delta}{v_F k}\right).
\end{align}
Their associated energies are 
\begin{align}
E_{k_{n,i}} := \sqrt{\Delta^2 + v_F^2 k_{n,i}^2}, \qquad i = 1,2.
\end{align}
We use the principle branch cut for the inverse tangent function. 
Firstly, using Eqn.~(\ref{eqn:eqm_R_H0}) again yields
\begin{align}
[H_0, \phi_n^\dagger] &= \int_{-\LL}^\LL \frac{-iv_F \partial_x u_{k_{n,1}}(x)}{\sqrt{2}}\left[R(x)+R^\dagger(x)\right], \\
[H_0, \varphi_n^\dagger] &= \int_{-\LL}^\LL \frac{-iv_F \partial_x v_{k_{n,2}}(x)}{\sqrt{2}}\left[R^\dagger(x)-R(x)\right]
\end{align}
after an integration by parts and by using the fact that 
\begin{align*}
u_{k_{n,1}}(-\LL)&= - u_{k_{n,1}}(\LL), \quad v_{k_{n,2}}(-\LL)&= - v_{k_{n,2}}(\LL) 
\end{align*}
and $R(\LL)+R(-\LL)=0$ to drop boundary terms. 

\begin{widetext}
However, in this case, one has 
\begin{align}
[H_1,\phi^\dagger_n] &= -\frac{i\Delta}{2}\int_{-\LL}^\LL\mathrm{d}x\int_{-\LL}^\LL\mathrm{d}y\, 
s(x)\frac{u_{k_{n,1}}(y)}{\sqrt{2}} \;[R^\dagger(x)R^\dagger(-x)-R(-x)R(x)\;, \;R(y)+R^\dagger(y)] \nonumber\\
&= -i\Delta\int_{-\LL}^\LL\mathrm{d}x\; \frac{s(x)}{\sqrt{2}} u_{k_{n,1}}^*(x)[R(x)+R^\dagger(x)],
\end{align}
\end{widetext}
where we have used the fact that $u_k(-x)=u^*_k(x)$ and that $s(x)$ is an odd function. Likewise, the same type of calculation yields 
\begin{align}
[H_1,\varphi^\dagger_n] &=
-i\Delta\int_{-\LL}^\LL\mathrm{d}x\; \frac{s(x)}{\sqrt{2}} v_{k_{n,2}}^*(x)[R(x)-R^\dagger(x)].  
\end{align}
Thus, we have the expressions
\begin{align}
[H,\phi^\dagger_n] = \int_{-\LL}^\LL \mathrm{d}x\frac{1}{\sqrt{2}}&\left\{
-iv_F \partial_x u_{k_{n,1}}(x) - i\Delta s(x) u_{k_{n,1}}^*(x)  
\right\} \nonumber \\ 
&\times [R(x)+R^\dagger(x)], \\
[H,\varphi^\dagger_n] = \int_{-\LL}^\LL \mathrm{d}x\frac{1}{\sqrt{2}}&\left\{
+iv_F \partial_x v_{k_{n,2}}(x) - i\Delta s(x) v_{k_{n,2}}^*(x)  
\right\}\nonumber \\
&\times[R(x)-R^\dagger(x)].
\end{align}
Then, by direct substitution and making use of the trigonometric identities with 
\begin{align}
v_F k = E_k \cos(2\theta_k), \qquad 
\Delta = E_k \sin(2\theta_k),
\end{align}
we have that
\begin{align}
&-iv_F \partial_x u_k(x) -i \Delta s(x) u_k^*(x) \nonumber \\
&=
\frac{E_k}{\sqrt{\LL}} \left[
\cos (k|x|+ 2\theta_k) +i \sin(kx) 
\right], \\ \nonumber \\
&iv_F \partial_x v_k(x) -i \Delta s(x) v_k^*(x) \nonumber \\
&= -
\frac{E_k}{\sqrt{\LL}} \left[
\cos (k|x|- 2\theta_k) +i \sin(kx) 
\right].
\end{align}
This then gives the desired relations
\begin{align*}
[H,\phi_n^\dagger] = E_{k_{n,1}}\phi_n^\dagger, \qquad 
[H,\varphi_n^\dagger] = E_{k_{n,2}}\varphi_n^\dagger.
\end{align*}
Finally, one can also confirm numerically that the set of fermionic modes $\phi_n,\varphi_{n'}, \psi_{E_{i\kappa}}$ mutually anticommute because their mode wavefunctions are orthogonal.


\section{Numerical checks with a lattice model}\label{app:Maj_chains}

In this appendix, we present a detailed derivation of a lattice model which yields the continuum Hamiltonian given by Eqn.~(\ref{eqn:Ham_fermion2}) at low energies and long wavelengths. By doing so, we can numerically confirm the correctness of the exact continuum solutions of Sec.~\ref{sec:exact_fermion}.

The task of finding a suitable lattice model is more subtle than it seems at first because the $R$ and $L$ fields are relativistic. Nevertheless, the most direct way to proceed is to express the model purely in terms of Majorana fields (cf. Sec.~\ref{sec:in_terms_of_majoranas})
\begin{align}
\lambda = \frac{1}{2}\begin{pmatrix}
R+R^\dagger \\ L+L^\dagger
\end{pmatrix}, \quad
\lambda' = \frac{1}{2i}\begin{pmatrix}
R-R^\dagger \\ L-L^\dagger
\end{pmatrix}.
\label{eqn:lambdas}
\end{align}
These obey the anticommutation relations
\begin{align}
\{\lambda_\alpha (x), \lambda_\beta (y) \} =
\{\lambda'_\alpha (x), \lambda'_\beta (y) \} = \frac{1}{2}\delta_{\alpha \beta} \delta(x-y).
\end{align}
 In this Majorana basis of field operators, the Hamiltonian given in Eqn.~(\ref{eqn:Ham_fermion2}) simply decomposes into 
\begin{subequations}
\begin{align}
H &= H_+ + H_-, \\
H_+ &= \int \mathrm{d}x \,\lambda^T (-iv_F \sigma^z \partial_x - \Delta \sigma^y) \lambda, \\
H_- &= \int \mathrm{d}x \,(\lambda')^T (-iv_F \sigma^z \partial_x + \Delta \sigma^y) \lambda'.
\end{align}
\label{eqn:H_lambdas}  
\end{subequations}
The mode Hamiltonians $\mathcal{H}_{\pm} = -iv_F \sigma^z \partial_x \mp \Delta \sigma^y$ bear striking resemblance to the low-energy effective Hamiltonian of the SSH model,\cite{unpublished} but with opposite mass signs. This connection suggests that we consider the single Kitaev chain Hamiltonian\cite{kitaev2001unpaired} 
\begin{align}
H_+ = \sum_{j=1}^N \left(
	\frac{it}{2}\,\gamma_{2j+2}\gamma_{2j+1}- \frac{it'}{2}\,\gamma_{2j}\gamma_{2j+1}
\right)
\label{eqn:H_Kit_wire}
\end{align}
with $t,t'>0$ and where the $\gamma_j$'s are local Majorana operators obeying $\{\gamma_j,\gamma_{j'} \} = 2\delta_{jj'}$. 

Next, by taking periodic boundary conditions for the moment, we proceed to transform to momentum space according to 
\begin{align}
\gamma_{k \alpha} :=\frac{1}{\sqrt{2N}}\sum_{j=1}^N \e^{-i k(j a + r_\alpha)}\, \gamma_{2j+\alpha},  
\end{align}
where $\alpha =0,1$ is an orbital (sublattice) index, $r_0 = -a/4$, $r_1 = +a /4$ are intra-cell site positions, and $a=2$ is the primitive lattice constant. The momentum $k$ is quantized in units of $2\pi/(Na)$ and we will take $N$ to be even such that $k=\pi/a$ is an allowed momentum. Moreover, these $k$-space Majorana operators can be shown to obey the Hermitian conjugate and anticommutation relations 
\begin{align}
\gamma_{k\alpha}^\dagger = \gamma_{-k \alpha}, \qquad 
\{\gamma_{k\alpha}, \gamma_{k'\beta} \} = \delta_{\alpha \beta} \,\delta_{-k,k'}. 
\end{align}
As usually, this implies that the operators $\gamma_{0\alpha}$ and $\e^{i\frac{\pi r_{\alpha}}{a}}\gamma_{\pi \alpha}$ are Hermitian and hence represent Majorana operators. Now the inverse Fourier transform
\begin{align}
\gamma_{2j+\alpha} = \sqrt{\frac{2}{N}} \sum_k \e^{+i k(ja + r_\alpha)}\gamma_{k\alpha}
\end{align}
yields the $k$-space BdG Hamiltonian
\begin{subequations}
\begin{align}
H_+ &= \frac{1}{2}\sum_k \begin{pmatrix}\gamma_{-k0}& \gamma_{-k1}\end{pmatrix}
\mathcal{H}_+(k) \begin{pmatrix} \gamma_{k0}\\ \gamma_{k1}\end{pmatrix}, \\
\mathcal{H}_+(k) &= (t+t')\sin\left(\tfrac{ka}{2}\right)\sigma^x - (t-t')\cos\left(\tfrac{ka}{2}\right)\sigma^y.
\end{align}
\end{subequations}
%
The (chiral) topological winding number
\begin{align}
\nu = -\frac{i}{4\pi a}\int_{-\pi/a}^{\pi/a}\mathrm{d}k \, \text{tr}\left[
\mathcal{H}_+(k)^{-1}\sigma^z \nabla_k \mathcal{H}_+(k) 
\right]
\end{align}
is $\mathbb{Z}$ quantized whenever $\det \mathcal{H}_+(k) \neq 0$, which is the case when $t \neq t'$. Here the $k$-space covariant derivative is defined by 
\begin{align}
\nabla_k := \partial_k + i [\hat{r},\cdot], \qquad \hat{r} := \frac{a}{4}\sigma^z,
\end{align}
where the operator $\hat{r}$ is the orbital position operator and accounts for the choices of unit cell and boundary surface termination. 

From this Bloch Hamiltonian, we specialize to the limit where $t\approx t'$. Since $t,t'>0$, the band minimum lies at $k=0$. Performing a gradient expansion about $k=0$ then yields
\begin{align}
\mathcal{H}_+(k)\approx \frac{(t+t')a}{2}\,k\, \sigma^x - (t-t')\sigma^y, 
\end{align}
which is a valid approximation whenever $|k|\ll \pi/a$ and $|t-t'| \ll (t+t')/2$. Going back to real space then produces 
\begin{align}
\mathcal{H}_+ = -i \frac{(t+t')a}{2} \sigma^x \partial_x - (t-t') \sigma^y,
\end{align}
from which we identify a Fermi velocity and SC gap by 
\begin{align}
v_F = \frac{(t+t')a}{2}, \qquad \Delta = t-t'. 
\end{align}
Note that this gradient expansion is somewhat more faithful to the relativistic continuum model than it might appear at first. This is because the critical theory (when $t=t'$) has a dispersion that possesses no band curvature.
In this way, band curvature effects are reduced to a minimum for $\Delta\neq0$.

Next, it helps to perform an internal $O(2)$ rotation
\begin{align}
\begin{pmatrix} \gamma_{k0} \\ \gamma_{k1} \end{pmatrix} =
\frac{1}{\sqrt{2}} \begin{pmatrix} 1 & -1 \\ 1 & 1\end{pmatrix}
\begin{pmatrix} \lambda_{k1} \\ \lambda_{k2} \end{pmatrix} 
\end{align}
that diagonalizes $\sigma^x$ and is essentially a rotation into an anti-bonding ($\lambda_{k1}$) and bonding ($\lambda_{k2}$) basis of $k$-space Majoranas. This leads to the following form of the Hamiltonian
\begin{align}
H_+ &= \frac{1}{2}\sum_k \begin{pmatrix} \lambda_{-k 1} & \lambda_{-k2} \end{pmatrix} \nonumber \\
&\quad\times\left[
(t+t')\sin\left(\tfrac{ka}{2}\right)\sigma^z - (t-t')\cos\left(\tfrac{ka}{2}\right)\sigma^y 
\right] \begin{pmatrix} \lambda_{k 1} \\ \lambda_{k2} \end{pmatrix} \nonumber \\
&\approx \frac{1}{2}\sum_{|k|<\Lambda} \begin{pmatrix} \lambda_{-k 1} & \lambda_{-k2} \end{pmatrix} 
\left[
v_F k \sigma^z - \Delta \sigma^y 
\right]  
\begin{pmatrix} \lambda_{k 1} \\ \lambda_{k2} \end{pmatrix} 
\end{align}
with $\Lambda \ll \pi/a$ being a short-distance cutoff defining the effective low-energy theory. Now we define real-space continuum Majorana fields by 
\begin{align}
\lambda_{\alpha}(x) = \frac{1}{\sqrt{2\LL}} \sum_{|k|<\Lambda} \e^{ik(x+r_\alpha) } \lambda_{k\alpha},
\end{align}
where $\LL= Na $ is the length of the system. These fields -- in the long wavelength limit -- also obey the desired anticommutation relations 
\begin{align}
\{\lambda_{\alpha}(x), \lambda_{\beta}(y)\} = \frac{1}{2}\delta_{\alpha \beta} \delta(x-y).
\end{align}
Finally, transforming back to real space produces the desired low-energy relativistic theory
\begin{align}
H_+ = \int \mathrm{d}x \, \lambda^T (-iv_F \sigma^z \partial_x - \Delta \sigma^y)\lambda
\end{align}
valid whenever $|\Delta| \ll v_F /a$ and for momenta smaller than $\Lambda \ll \pi/a.$ 

By the same arguments we can construct a lattice realization of $H_-$ by making the change $\Delta \rightarrow -\Delta$. In summary, the full form of our desired lattice Hamiltonians is 
\begin{subequations} 
\begin{align}
H_+ &=  \frac{i}{2}\sum_{j=1}^N \left(
	\left[\frac{v_F}{a}+\frac{\Delta}{2}\right] \gamma_{2j+2}\gamma_{2j+1} - \left[\frac{v_F}{a}-\frac{\Delta}{2}\right] \gamma_{2j}\gamma_{2j+1} 
\right), \\
H_- &=  \frac{i}{2}\sum_{j=1}^N \left(
	\left[\frac{v_F}{a}-\frac{\Delta}{2}\right] \gamma'_{2j+2}\gamma'_{2j+1} - \left[\frac{v_F}{a}+\frac{\Delta}{2}\right] \gamma'_{2j}\gamma'_{2j+1} 
\right), 
\end{align}
\label{eqn:H_Maj_lattice}%
\end{subequations}
with $\gamma'_{j}$ being another independent set of local lattice Majorana operators. 

Before we move on to the issue of boundary conditions in an open system, we have to consider first the symmetries of the lattice models. Although it is manifestly clear that in the limit $\Delta=0$ the continuum model possesses a global $U(1)$ symmetry, this is obscured in the lattice model. Nevertheless, it is indeed $U(1)$ symmetric. Specializing to the limit $\Delta=0$ leads to 
\begin{subequations}
\begin{align}
H_+ &= \frac{it}{2}\sum_j \left( \gamma_{2j+2} \gamma_{2j+1} - \gamma_{2j} \gamma_{2j+1}  \right), \\
H_- &= \frac{it}{2}\sum_j \left( \gamma'_{2j+2} \gamma'_{2j+1} - \gamma'_{2j} \gamma'_{2j+1}  \right).
\end{align}
\end{subequations}
We then relabel the local Majorana operators alternatingly according to even and odd sites
\begin{subequations}
\begin{align}
&a_{2j-1} := \gamma_{2j-1},\qquad   a_{2j} := \gamma'_{2j}, \\
&b_{2j-1} := -\gamma'_{2j-1},\quad\;\; b_{2j} := \gamma_{2j},
\end{align}
\end{subequations}
with $a_{j}$ and $b_{j}$ being newly introduced local Majorana operators. Then direct substitution leads to the combined Hamiltonian
\begin{align}
H_+ + H_- &= t \sum_j (-1)^j [c^\dagger_{j+1} c_j + c^\dagger_{j}c_{j+1}],
\end{align}
where $c_j =  \frac{a_j+ib_j}{2}$ is a legitimate complex fermion. This lattice Hamiltonian is manifestly $U(1)$ symmetric with conserved particle number $N = \sum_j c^\dagger_j c_j$. Lastly, the fact that 
\begin{align}
\mathcal{T} \lambda \mathcal{T}^{-1} = \sigma^x \lambda, \qquad 
\mathcal{T} \lambda' \mathcal{T}^{-1} = - \sigma^x \lambda'
\end{align}
can be shown to lead to $a_j$ and $b_j$ transforming by opposite signs under time reversal. 

\begin{figure*}\begin{center}
\includegraphics[width=0.95\textwidth]{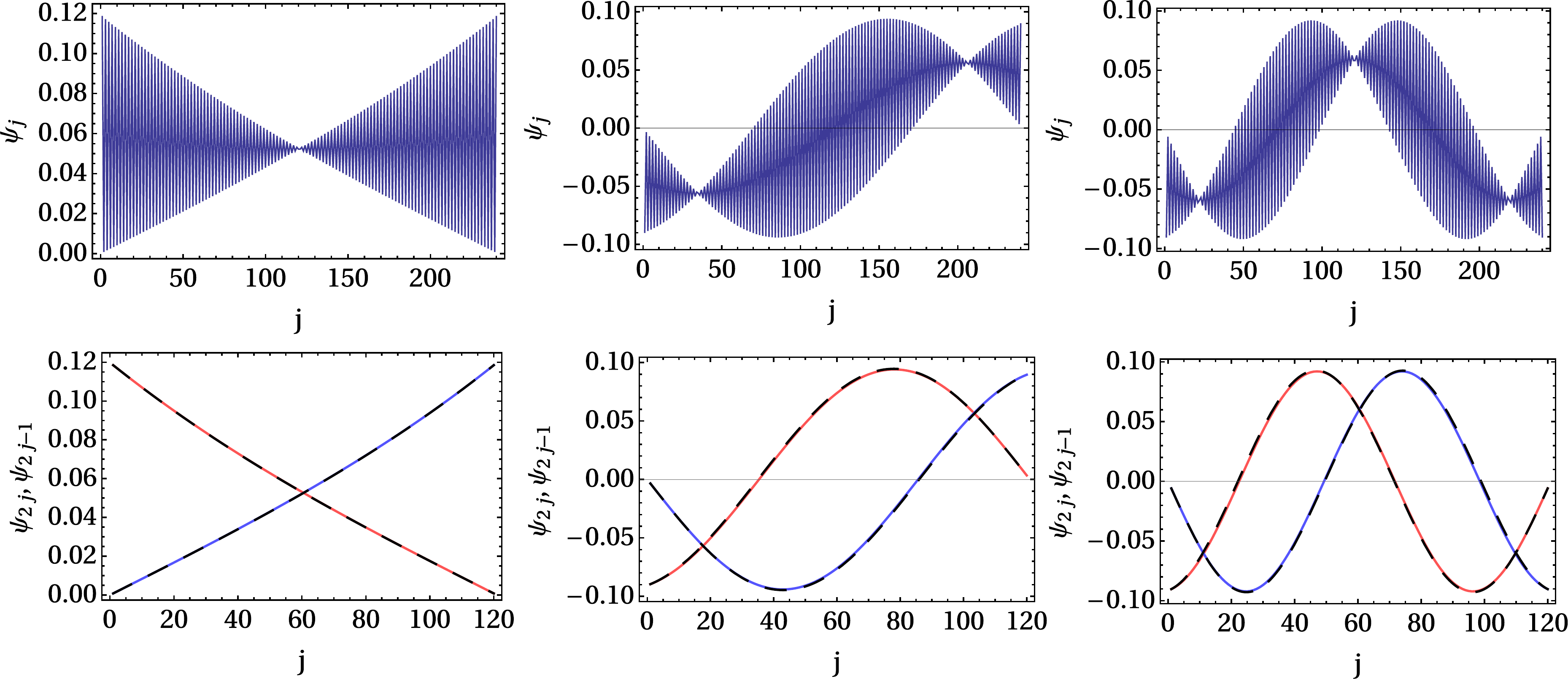}
\end{center}
\caption{Comparison between the numerically determined eigenfunctions of an open Majorana chain [Eqn.~(\ref{eqn:H_Maj_lattice}a)] and exact solutions to the linearized continuum model [Eqn.~(\ref{eqn:H_lambdas}b)]. On display are the first three eigenfunctions with positive energy. The state corresponding to the smallest energy (leftmost panel) corresponds to the topological mode. The parameters used here are $\Delta= 0.011$ and $v_F/a = 0.9955$, giving a coherence length of $v_F/\Delta= 90.5$ unit cells or 181 sites. The chain length is $\LL = 120$ unit cells. The top panels show the numerically obtained eigenfunctions as a function of site number. The apparent fast oscillations are due to the combined plot of the wavefunction from both sublattices site types. The bottom panels contain the same data but separated into sublattice type, with red curves taken from odd sites and blue curves taken from even sites. Dashed black lines are the analytical expressions Eqn.~(\ref{eqn:Phi_soln}) [$a(x)$ and $b(x)$] and Eqn.~(\ref{eqn:phi_soln}) after resolving the 2-spinor fields into sublattice type (real and imaginary parts), cf.  Eqn.~(\ref{eqn:sublattice}).}
\label{fig:eigenfunction}
\end{figure*}

\begin{figure*}
\begin{center}
\includegraphics[width=0.95\textwidth]{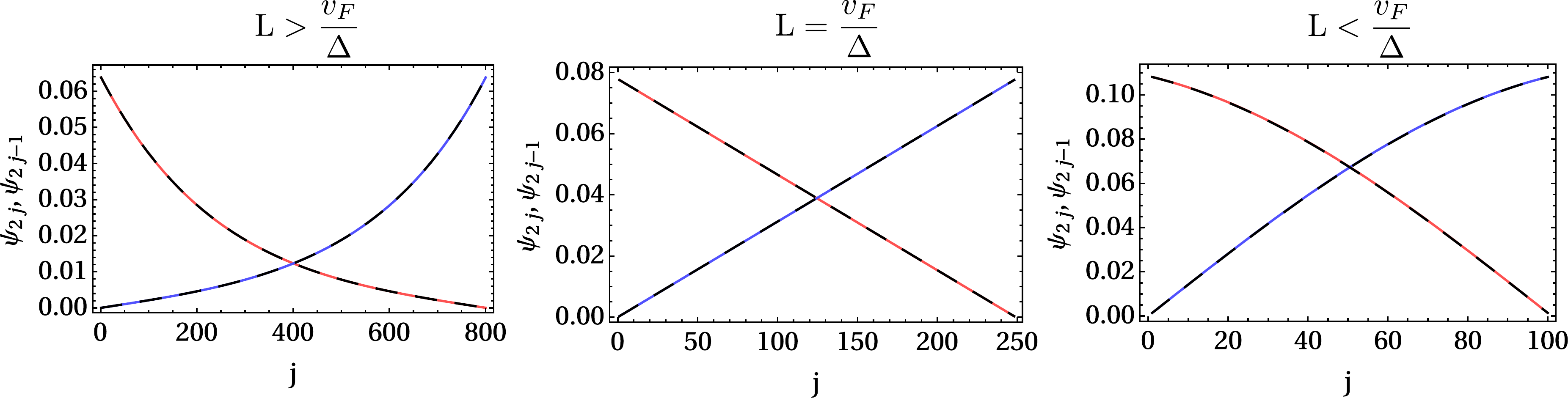}
\end{center}
\caption{Comparison between the numerically computed eigenfunctions (solid lines) and analytical expressions (dashed lines) for the state with the smallest positive energy at different lengths. Like in the bottom panels of Fig.~\ref{fig:eigenfunction}, the eigenfunctions are plotted according to sublattice type. The parameters $\Delta=0.004, v_F/a=0.992$ are chosen such that the coherence length is $v_F/\Delta= 248$ unit cells. (Left) The long wire limit where $\LL = 800$ is well described by the in-gap solution $\Phi^{(+)}(x)$ from Eqn.~(\ref{eqn:Phi_soln}). (Right) The short wire limit with $\LL= 100$, where the dashed curve is the eigenfunction $\phi^{(+)}_{k_{0,1}}(x)$ obtained by an $n=0$ solution of Eqn.~(\ref{eqn:phi_soln}). (Middle) The critical case where $\mathrm{L}=248$ is exactly equal to the coherence length. The eigenfunction now has no curvature and is a piecewise linear function that is described by either $\phi^{(+)}_{k_{0,1}}(x)$ or $\Phi^{(+)}(x)$ in the limit that $k_{0,1},\kappa\rightarrow 0$.} 
\label{fig:lengths}
\end{figure*}

Moving on, we proceed to determine the appropriate boundary conditions for the low-energy effective fields $\lambda,\lambda'$ in an open system. An arbitrary lattice fermionic operator expanded as 
\begin{align}
\psi = \sum_{j=0}^{N+1} \sum_{\alpha=0,1} \psi_{j,\alpha} \gamma_{2j+\alpha} 
\end{align}
will undergo a time evolution according to the Heisenberg equation of motion 
\begin{align}
\partial_t \psi = i[H_+, \psi]. 
\end{align}
This requires that $\psi_{0,1}=0$, $\psi_{N+1,0}=0$ such that the action of $H_+$ cannot introduce new particles into or out of the finite system. Equivalently this means that the slow fields $\lambda_1(x)$ and $\lambda_2(x)$ should obey
\begin{align}
\lambda_1(0)+ \lambda_2(0)=0, \qquad \lambda_1(\LL) - \lambda_2(\LL) =0, 
\end{align}
because 
\begin{align}
\begin{pmatrix} \gamma_0(x) \\ \gamma_1(x) \end{pmatrix} = \frac{1}{\sqrt{2}}
\begin{pmatrix}
\lambda_1(x) -\lambda_2(x) \\ \lambda_1(x) +\lambda_2(x)
\end{pmatrix}. \label{eqn:sublattice}
\end{align}
This boundary condition may be identically expressed as 
\begin{align}
\sigma^x \lambda(0) = - \lambda(0), \qquad
\sigma^x \lambda(\LL) = + \lambda(\LL).
\end{align}
Identical considerations also produce
\begin{align}
\sigma^x \lambda'(0) = - \lambda'(0), \qquad
\sigma^x \lambda'(\LL) = + \lambda'(\LL).
\end{align}
Next, the defining relations of $\lambda,\lambda'$ in Eqns.~(\ref{eqn:lambdas}) translate these boundary conditions to 
\begin{align}
R(0) + L(0) = 0 ,\qquad R(\LL) - L(\LL) =0 ,
\end{align}
which are our previously stated boundary conditions of Sec.~\ref{sec:exact_fermion}. Hence we have verified that the open chain boundary conditions will agree with those of our previous exact continuum analysis. 

Shown in Fig.~\ref{fig:Egap} are exact numerical results from tight-binding model calculations based on Eqn.~(\ref{eqn:H_Maj_lattice}a). In this instance, the Hamiltonian $H_+$ is topological with boundary modes whenever $\Delta>0$. The other Majorana chain model $H_-$ does not display in-gap states in this parameter regime. The roles of $H_+$ and $H_-$ are reversed whenever $\Delta$ changes sign, such that the total model always contains topological boundary modes. Lastly, one may also compare the form of the eigenfunctions, both localized and extended, as is done in Figs.~\ref{fig:eigenfunction} and \ref{fig:lengths}. In Fig.~\ref{fig:lengths} we observe the transition from a localized (convex-profiled) in-gap mode to an extended (concave-profiled) mode  with decreasing $\LL$. In the short length regime when $\LL < v_F/\Delta$, the analytical eigenmode is described by $\phi^{(+)}_{k_{0,1}}(x)$ from Eqn.~(\ref{eqn:Phi_soln}), where the quantization condition for $k_{n,1}$ in Eqn.~(\ref{eqn:kn12_quantization}a) now has a valid solution for $n=0$. 

\raggedleft

\bibliography{references.bib}

\end{document}

%% file: preamble.tex
\usepackage{relsize,mathtools} 

\usepackage{amsmath,amssymb,amsthm,amsfonts}
\usepackage{listings}
\usepackage{enumerate}
\usepackage{latexsym}

\usepackage{mathtools} 

\usepackage[mathscr]{euscript} 
\usepackage{upgreek,mathrsfs}

\usepackage[colorlinks=true,citecolor=blue,linkcolor=blue,urlcolor=blue]{hyperref} 

\usepackage{verbatim}

\usepackage{psfrag}

\usepackage{bm}

\usepackage{bbm} 

\usepackage{graphics}
\usepackage{graphicx}

\usepackage{color}

\usepackage{xr} 

\definecolor{darkred}{RGB}{139,0,0}
\definecolor{crimson}{RGB}{220,20,60}

\newcommand{\e}{\mathrm{e}}

\newcommand{\LL}{{\mathrm{L}}}

\DeclareFontEncoding{LS1}{}{}
\DeclareFontSubstitution{LS1}{stix}{m}{n}
\DeclareSymbolFont{symbols2}{LS1}{stixfrak}{m}{n}
\DeclareMathSymbol{\typecolon}{\mathbin}{symbols2}{"25}